\documentclass[a4paper,11pt]{article}
\pdfoutput=1 
\usepackage{verbatim}

\usepackage{jheppub}
\usepackage{amsmath,amsfonts,amssymb,amsthm,graphicx,color}
\usepackage[T1]{fontenc} 
\usepackage{enumerate}
\usepackage{tikz, tikz-3dplot, pgfplots}
\usepackage[utf8]{inputenc}
\usepackage{slashed}
\usepackage{subcaption}

\usepackage{changepage}
\usepackage{float}
\usepackage{tikz}
\usetikzlibrary{decorations.text}
\usepackage{enumitem}
\setlist{itemsep=0pt}
\usetikzlibrary{math}

\usepackage{tikz}
\usepackage{pgfplots}
\usetikzlibrary{decorations.text,intersections, pgfplots.fillbetween}
\usetikzlibrary{math}
 \usetikzlibrary{snakes,3d,shapes.geometric,shadows.blur}
\usepackage{tikz-3dplot}

\newcommand{\comm}[1]{} 

\def\({\left(}
\def\){\right)}
\def\[{\left[}
\def\]{\right]}

\def\One{{\hbox{ 1\kern-.8mm l}}}

\def\barray{\begin{array}}
\def\earray{\end{array}}
\def\be{\begin{equation}}
\def\ee{\end{equation}}
\def\bea{\begin{eqnarray}}
\def\eea{\end{eqnarray}}
\def\bal{\begin{align}}
\def\eal{\end{align}}

\def\-{\,-\,}
\def\={\,=\,}
\def\+{\,+\,}
\def\equi{\,\equiv\,}

\numberwithin{equation}{section} 

\definecolor{cardinal}{rgb}{0.6,0,0}
\definecolor{darkgreen}{rgb}{0,0.4,0}
\definecolor{golden}{rgb}{0.92, 0.7, 0}
\definecolor{midnight}{rgb}{0, 0, 0.5}
\definecolor{darkblue}{rgb}{0, 0, 0.7}
\definecolor{purple}{rgb}{0.5, 0, 0.5}

\def\IR{\mathbb{R}}

\def\cB{{\cal B}}

\def\cF{{\cal F}}
\def\cG{{\cal G}}

\def\cL{{\cal L}}
\def\cM{{\cal M}}
\def\cN{{\cal N}}

\def\cO{{\cal O}}
\def\cQ{{\cal Q}}

\def\cS{{\cal S}}
\def\cT{{\cal T}}

\makeatletter
\newcommand\footnoteref[1]{\protected@xdef\@thefnmark{\ref{#1}}\@footnotemark}
\makeatother

\usetikzlibrary{positioning,calc,fadings,decorations.pathreplacing}

\tikzset{
 diffuse color/.initial = black,                       
}

\tikzset{
 linear opacity/.initial=0.5,                          
 linear stroke/.style = {                              
   preaction={                                         
     draw=\pgfkeysvalueof{/tikz/diffuse color},        
     line width = (2.0-#1)*\pgflinewidth,              
     opacity=\pgfkeysvalueof{/tikz/linear opacity},white}},  
 diffuse gradient/.style={                             
   draw = none,                                        
   linear opacity=#1,                                  
   linear stroke/.list={0.0,#1,...,1.0}},              
 diffuse gradient/.default=1,                          
}

\tikzset{
 non-linear stroke/.style = {                          
   preaction={                                         
     draw=\pgfkeysvalueof{/tikz/diffuse color},        
     line width = (2.0-#1)*\pgflinewidth,              
     opacity=#1,white}},                                     
 diffuse falloff/.style={                              
   draw = none,                                        
   non-linear stroke/.list={0.0,#1,...,1.0}},          
 diffuse falloff/.default=1,                           
}
 
\tikzset{%
  >=latex, 
  inner sep=0pt,%
  outer sep=2pt,%
  mark coordinate/.style={inner sep=0pt,outer sep=0pt,minimum size=3pt,
    fill=black,circle}%
}

\title{\boldmath Smooth Bubbling Geometries Without Supersymmetry
}

\author{Ibrahima Bah and}
\author{Pierre Heidmann} 
\affiliation{Department of Physics and Astronomy, Johns Hopkins University, 3400 North Charles Street, Baltimore, MD 21218, USA}

\emailAdd{iboubah@jhu.edu}
\emailAdd{pheidma1@jhu.edu}

\abstract{We construct the first smooth bubbling geometries using the Weyl formalism. The solutions are obtained from Einstein theory coupled to a two-form gauge field in six dimensions with two compact directions. We classify the charged Weyl solutions in this framework. Smooth solutions consist of a chain of Kaluza-Klein bubbles that can be neutral or wrapped by electromagnetic fluxes, and are free of curvature and conical singularities. We discuss how such topological structures are prevented from gravitational collapse without struts. When embedded in type IIB, the class of solutions describes D1-D5-KKm solutions in the non-BPS regime, and the smooth bubbling solutions have the same conserved charges as a static four-dimensional non-extremal Cvetic-Youm black hole.}

\preprint{}
\begin{document}

\maketitle
\flushbottom

\section{Introduction}
\label{sec:Intro}

A great challenge in theories of gravity is the construction of physical solutions with multiple sources that are smooth and free from any singularities.  In supergravity theories, such a challenge is surmountable. With the help of supersymmetry, the non-linear higher order partial differential equations of Einstein's equations can be replaced with linear, and often first order, systems that provide broad and diverse families of solutions.  For example, there is a non-exhaustive list of multicenter BPS black holes \cite{Tod:1983pm,Sabra:1997yd}, and large families of smooth and horizonless microstate geometries with interesting topologies have been obtained in higher dimensions \cite{Bena:2004de,Bena:2006kb,Bena:2007qc,Bena:2007kg,Bena:2015bea,Heidmann:2017cxt,Bena:2017fvm}.  An important objective is to understand how to systematically generalize these solutions in general theories of gravity without the use of supersymmetry.

Much effort has been devoted to developing techniques to treat Einstein's equations in four dimensions (see \cite{Stephani:2003tm,Bonnor} for an extensive review). The Weyl formalism is one of the oldest method \cite{Weyl:book}, and has allowed to derive linear equations for static axisymmetric backgrounds in vacuum. Generic solutions correspond to black holes on a line that are held apart by struts.  These are cosmic strings with negative tension that counterbalance the attractive force of gravity. They appear in spacetime as regions with conical excess and hence are singular.  Struts a priori do not have any consistent UV description and a way must be found to replace them with smooth objects in order to build physical solutions.

Emparan and Reall initiated the classification of vacuum Weyl solutions, that is static and axisymmetric geometries, in higher dimensions where extra compact dimensions are added to the four-dimensional spacetime \cite{Emparan:2001wk}. The greater diversity of gravity in higher dimensions allows the solutions to be sourced not only by black objects but also by smooth Kaluza-Klein (KK) bubbles corresponding to a region where extra compact dimensions degenerates. In this way, Elvang and Horowitz were able to show that the struts separating the chain of four-dimensional black holes can be classically replaced by KK bubbles in five dimensions \cite{Elvang:2002br}.  In these solutions, the gravitational attraction of the black holes is counterbalanced by the instability of vacuum KK bubbles to expand \cite{Witten:1981gj}. 

An interesting question arises: can the method of Elvang and Horowitz be used to construct smooth bubbling geometries without black holes?  This is particularly challenging since any construction with KK bubbles only will have some that suffer from their inherent quantum instability to forever expand \cite{Witten:1981gj}.  However, KK bubbles can be stabilized by electromagnetic fluxes \cite{Stotyn:2011tv}.  In a  previous paper, we successfully generalized the Weyl formalism by adding non-trivial gauge-field degrees of freedom while preserving the linearity structure of the equations of motion \cite{Bah:2020pdz}. 

In this paper, we derive a Weyl ansatz of linear equations of motion that allows for the construction of entirely-smooth static solutions made of multiple charged KK bubbles. Due to the presence of non-trivial gauge-field degrees of freedom, the ansatz will have a specific BPS limit where the equations reduce to the supergravity equations of motion.  Generic solutions, however, will be non-BPS and in the non-extremal regime of black-hole charges. 

Our construction is an important step forward in the microstate geometry program, and in understanding the available mechanisms to construct non-trivial smooth geometries with interesting topology beyond supersymmetry.

In what follow, we start with a summary of the main results.  In section \ref{sec:strutGen}, we review various aspects of Weyl solutions in four and five dimensions with struts.  In section \ref{sec:6dFrame}, we describe our general six-dimensional framework and discuss the species of bubbles and black objects that can be used to construct solutions.  In section \ref{sec:chainBoN}, we analyze smooth Weyl solutions in vacuum and then generalize these systems by turning on various electromagnetic gauge fields in section \ref{sec:chainTS}.  We end with a discussion on how to embed our system in type IIB supergravity in section \ref{sec:STembedding}.

\subsection{Summary}

In this paper, we classify static axisymmetric solutions using the Weyl formalism in six dimensions with two compact directions \cite{Weyl:book,Emparan:2001wk}. We extend the method of \cite{Bah:2020pdz} to turn on a two-form gauge field and a Kaluza-Klein gauge field along one of the compact dimensions. 

Generic Weyl solutions consist of bound states of \emph{black branes}, and \emph{two species of smooth bubbles} stacked on a line, wrapped by non-trivial electromagnetic fluxes. If the sources do not touch, they are separated by \emph{struts}, i.e. segments with a conical excess \cite{Costa:2000kf}. A strut is a singularity that represents the repulsive force needed to compensate for the self-attraction between the sources. However, they disappear when the sources touch. Motivated by the work of \cite{Elvang:2002br}, we show that struts can be classically replaced by smooth bubbles in six dimensions. The physical interpretation is that bubbles, whether neutral or charged, are reluctant to be squeezed, and can provide the same repulsive force as a strut when compressed. Moreover, we argue that the role of the electromagnetic fluxes is to support the overall structure from instability of the vacuum bubbles \cite{Witten:1981gj}. Indeed, a single KK bubble is stabilized when wrapped by fluxes \cite{Stotyn:2011tv}. Therefore, Weyl solutions in six dimensions can be prevented from collapsing by pure topology without the aid of conical singularities and from expanding by electromagnetic fluxes.

We focus on the specific subclass of solutions that are completely smooth and horizonless. These are bubbling solutions consisting of a chain of different bubble species. We detail their construction and regularity, which involves as many \emph{bubble equations} as the number of bubbles in the chain. Once these equations are satisfied, the geometry of the bubbles is frozen as all internal degrees of freedom, except the number of bubbles and some gauge field parameters, are fixed. In this paper, we explicitly construct smooth and horizonless solutions with one and three bubbles, both charged and neutral. This allows us to better illustrate the classical replacement of struts by a smooth degeneracy of an extra dimension and to better discuss their stability. In a subsequent paper \cite{bubblebagend}, we construct and analyze bubbling Weyl solutions with a large number of bubbles, nicknamed Weyl stars.

Finally, we embed our Weyl construction into type IIB string theory on T$^4$. Our generic Weyl solutions correspond to static D1-D5-KKm solutions, and the smooth bubbling geometries correspond to a chain of non-BPS D1-D5-KKm and KKm bubbles. The latter have the same conserved charges as a static non-BPS non-extremal D1-D5-KKm black hole. 

Moreover, in a certain limit, our Weyl construction gives the BPS equations of motion for axisymmetric D1-D5-KKm multicenter solutions. It thus encompasses the ansatz of the three-charge BPS floating branes and offers a non-trivial extension into the non-BPS, non-extremal black hole regime. It is the first linear ansatz that allows the construction of smooth horizonless geometries in such a regime that treats only the Einstein's equations.

\section{The strut problem in four and five dimensions}
\label{sec:strutGen}

We first review the state of art about classification of static and axisymmetric Einstein solutions in four and five dimensions. From its initial construction \cite{Weyl:book}, the Weyl formalism has allowed to treat Einstein's equations as a succession of linear differential equations for static solutions \cite{Israel1964,Gibbons:1979nf,Costa:2000kf,Emparan:2001wk,Elvang:2002br,Emparan:2008eg,Charmousis:2003wm}. In \cite{Papapetrou:1953zz,Bah:2020pdz}, the Weyl formalism has been generalized to also include Maxwell fields of various kinds in such a way to preserve the linear structure of  equations. This allows for solutions with superposition of multiple charged sources along a line consistent with the axisymmetry present in the Weyl formalism.

\subsection{The four-dimensional paradigm}

\begin{figure}[h]
\centering
\includegraphics[height=0.5\textwidth]{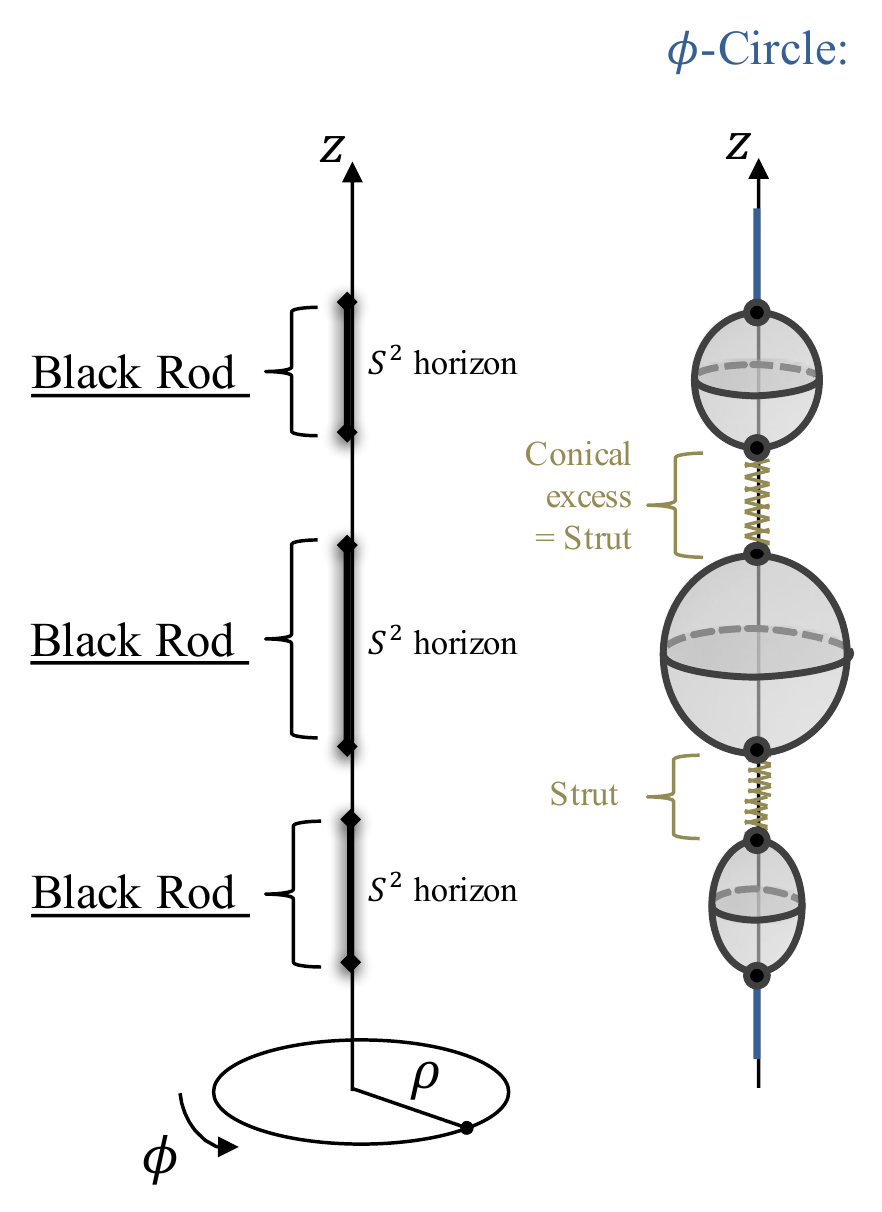}
\caption{Schematic description of a static axisymmetric Weyl solution in four dimensions and the behavior of the $\phi$-circle on the $z$-axis. Solutions are sourced by black rods with a S$^2$ horizon that can be wrapped by electromagnetic fluxes. The multiple black holes are separated by struts.}
\label{fig:Israel4d}
\end{figure}

The static charged black hole is the unique spherically symmetric physical source in asymptotically flat spacetime in four dimensions. Vacuum solutions of multiple four-dimensional black holes stacked on a line have been pioneered in \cite{Israel1964,Gibbons:1979nf}. These are static systems that break the spherical symmetry of individual black holes.  In order to support such structures from collapse, the black holes are welded together by a series of \emph{struts}, i.e. \emph{strings with negative tension} (see Fig.\ref{fig:Israel4d}), between them.  In cylindrical coordinates parametrized by $(\rho,z,\phi)$, a strut on the $z$-axis is a segment for which the axisymmetry circle $\phi$ with period $2\pi$ shrinks in the three-dimensional space as
\begin{equation}
ds_\text{strut}^2 \= d\rho^2 \+ dz^2 \+ \frac{\rho^2}{d^2}\,d\phi^2\,,
\label{eq:conicalexcessGen}
\end{equation}  
where $0<d<1$. Therefore, the local angle, $\phi_\text{loc} = \frac{\phi}{d}$,  has period $\frac{2\pi}{d}>2\pi$ and the local geometry has a \emph{conical excess}.  This can be contrasted to when $d>1$ and integral, where there is a conical deficit corresponding to an orbifold fixed point.  These can be resolved in various ways.  Such resolutions are not available for the strut.  A non-trivial stress tensor can be associated to these objects since they provide sources for the Ricci scalar \cite{Regge:1961px,Costa:2000kf}.  The strut will have a negative energy density which captures the binding energy for the chain of black holes that they hold together from gravitational attraction.  This is reviewed below in section \ref{sec:twobubbles5d} and appendix \ref{App:StrutEnergy}.  

\subsection{The five-dimensional paradigm}
\label{sec:5dparadigm}

The classification of Weyl solutions in five dimensions with one compact dimension, that we will review in this section, can be found in \cite{Costa:2000kf,Emparan:2001wk,Elvang:2002br,Emparan:2008eg,Charmousis:2003wm,Bah:2020pdz}. 
We are interested in static five-dimensional solutions that are asymptotic to $\IR^{1,3}\times$S$^1$ of the following Einstein-Maxwell action\footnote{The norm is defined as follows: $\left| \cF \right|^2 = \cF \wedge \star_5 \cF$.}
\begin{equation}
\left(16 \pi G_5 \right) \,S_5=\int \mathrm{d}^5 x \sqrt{-\det g}\left(R-\frac{1}{2} \left|F^{(m)}\right|^2-\frac{1}{2} \left|F^{(e)}\right|^2\right)\,,
\label{eq:ActionGen5d}
\end{equation}
where $G_5$ is the five-dimensional Newton's constant, $F^{(m)}$ and $F^{(e)}$ are field strengths for a one-form and a two-form gauge fields.  We refer to them as magnetic two-form and electric three-form field strengths respectively.  This naming will be clear below.   

There can exist a richer and more diverse family of solutions of the five-dimensional theory in \eqref{eq:ActionGen5d} than in four dimensions.  In particular, static Weyl solutions can be constructed from two types of physical objects: smooth Kaluza-Klein (KK) bubbles, dubbed topological stars when they are wrapped by fluxes \cite{Bah:2020ogh,Bah:2020pdz} and black strings.  These objects individually preserve spherical symmetry and can be uniformly described by the two-parameter family of metrics:
\begin{equation}\begin{split}
d s_{5}^{2}&\=-\left(1- \frac{r_\text{S}}{r} \right) dt^{2}+\left(1- \frac{r_\text{B}}{r} \right)d y_1^{2}+\frac{r^2\,d r^{2}}{(r-r_\text{B})(r-r_\text{S})}+r^{2}\left(d \theta^{2}+\sin ^{2} \theta d \phi^{2}\right)\,,\\
F^{(e)}&\=\frac{Q}{r^{2}}\, d r \wedge d t \wedge d y, \qquad F^{(m)}\=P \sin \theta\, d \theta \wedge d \phi\,,\qquad P^{2}+Q^{2}= 3 r_{\mathrm{S}} r_{\mathrm{B}}\,.
\label{eq:TS5d}
\end{split}
\end{equation} 
If $r_\text{B}>r_\text{S}$, the solution is a topological star, and the extra dimension, $y_1$-circle with period $2\pi R_{y_1}$, shrinks at $r=r_\text{B}$.  This region corresponds to a smooth end of space with a finite size S$^2$ bubble sitting at the origin of an $\IR^2$-plane composed of $(r,y_1)$.  One can also allow for a conical singularity by having $\IR^2/\mathbb{Z}_k$ with regularity condition
\begin{equation}
R_{y_1}^{2}\=\frac{4 r_{\mathrm{B}}^{3}}{k^{2}\left(r_{\mathrm{B}}-r_{\mathrm{S}}\right)}\,\qquad k \in \mathbb{N}\,.
\end{equation}
Note that when the topological star is free from conical defect, $k=1$, the radius of the bubble given by $r_\text{B}$ is of order the size of the extra dimension, $R_{y_1}$. This gives another motivation to construct stack of topological stars on a line in order to build an object whose size can be parametrically larger than the Kaluza-Klein scale.

When $r_\text{S}\geq r_\text{B}$, the solution corresponds to a black string with a S$^1\times$S$^2$ horizon at $r=r_\text{S}$. This black string solution does not have any curvature singularities but admits a bubble behind the horizon \cite{Bah:2020ogh}.  

Finally there are vacuum solutions when one of the parameters of $(r_\text{S},r_\text{B})$ vanishes.  There is the bubble of nothing solution when $r_\text{S}=0$ \cite{Witten:1981gj} and a four-dimensional Schwarzschild solution times S$^1$ when $r_{\text{B}} =0$.

\begin{figure}[h]\centering
\begin{subfigure}{0.46\textwidth}\centering
 \includegraphics[width=\textwidth]{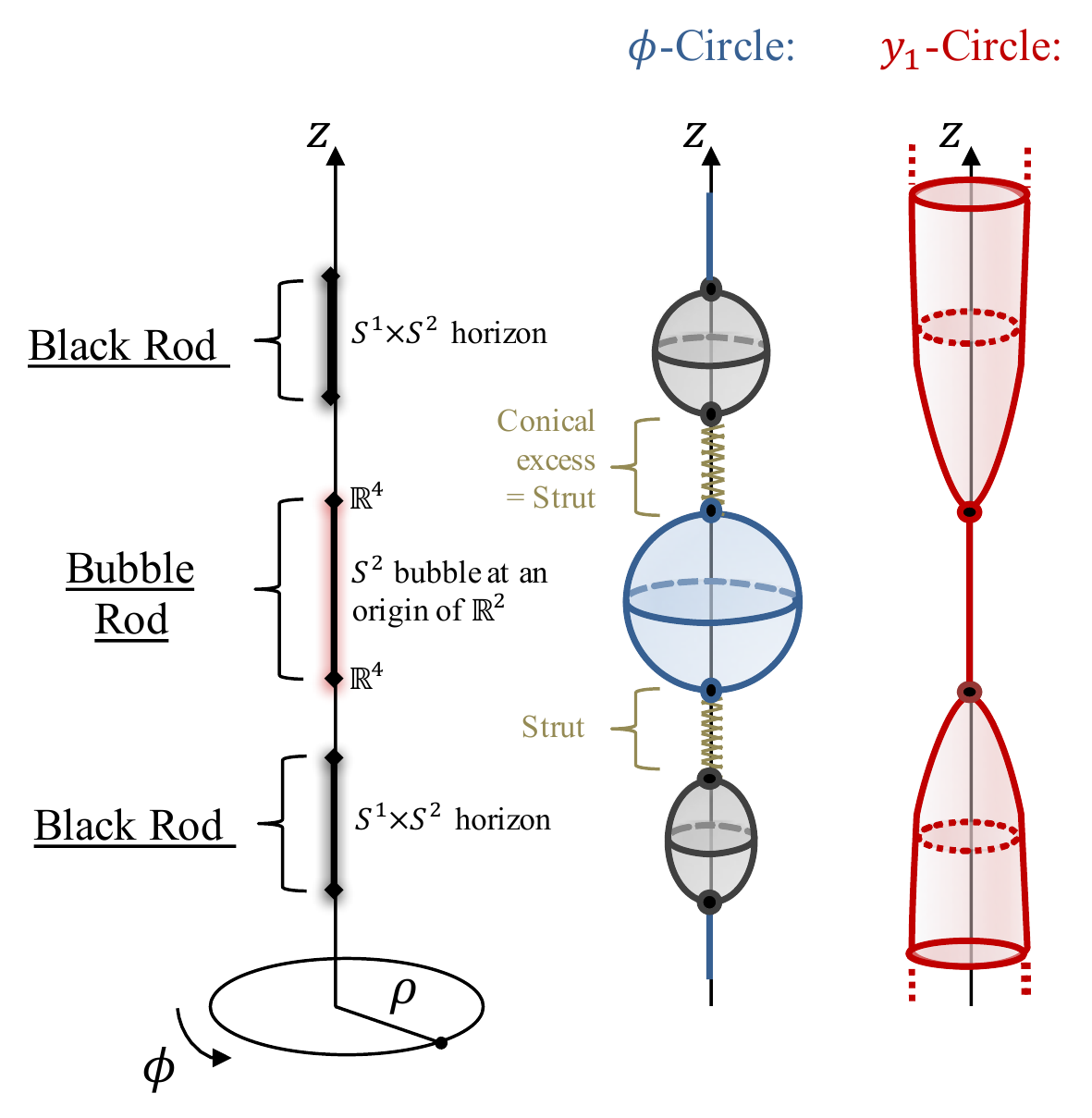}
 \caption{Weyl solutions with disconnected rods.}
 \end{subfigure}
 \hspace{0.5cm}
 \begin{subfigure}{0.46\textwidth}\centering
   \includegraphics[width=\textwidth]{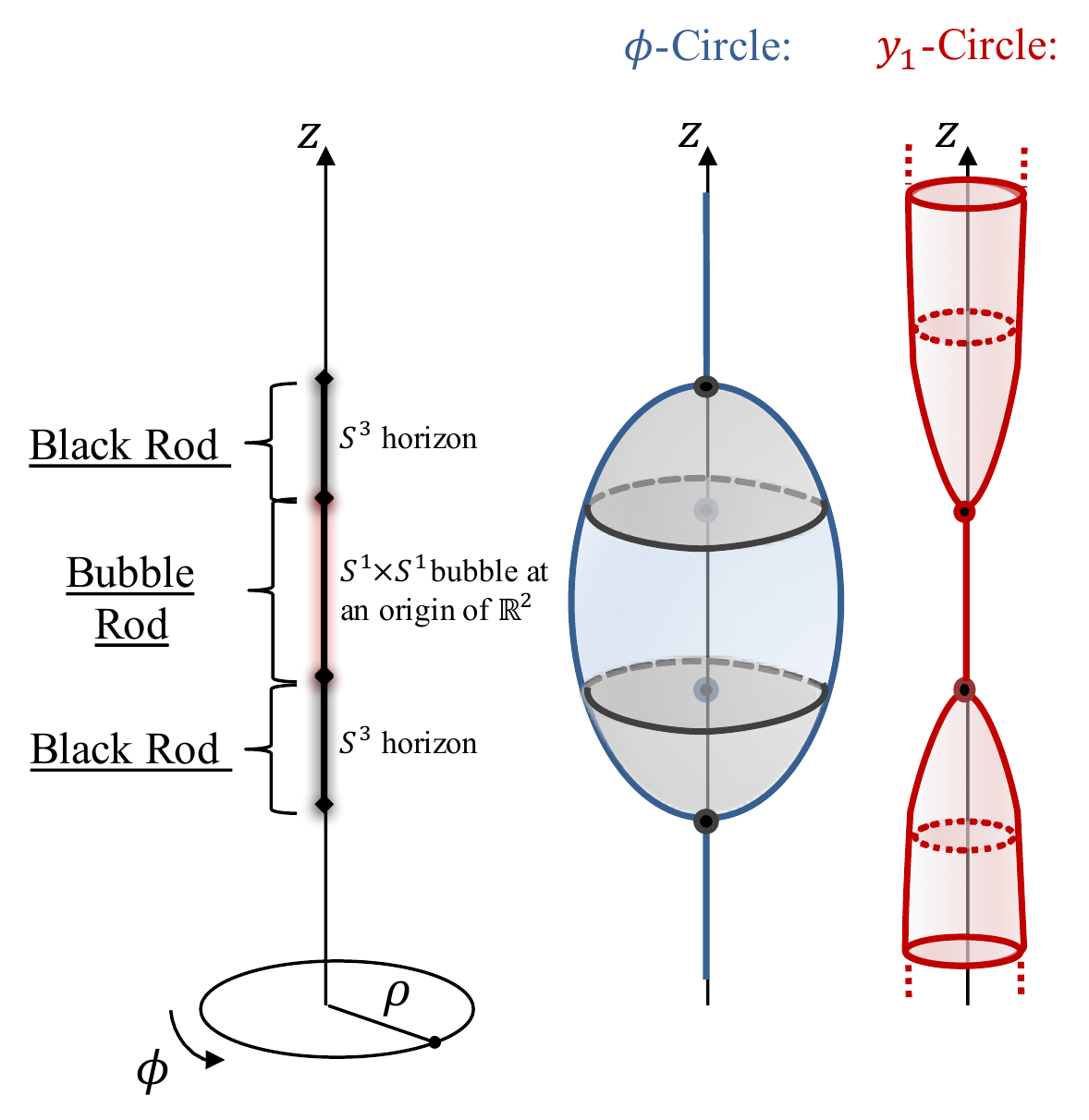}
    \caption{Weyl solutions with connected rods.}
 \end{subfigure}
 \caption{Schematic description of static axisymmetric Weyl solutions in five dimensions and the behavior of the $\phi$ and $y_1$ circles on the $z$-axis. Solutions can be sourced by black rods or bubble rods (topological stars) that can be wrapped by electromagnetic fluxes. At each rod, the $\phi$ circle has a finite size on the $z$-axis. The multiple objects are separated by struts (a) except if two rods of different nature are connected (b).}
 \label{fig:5dWeyl}
\end{figure}

Generic static axisymmetric Weyl solutions can be obtained by stacking topological stars and black strings along a line as depicted in Fig.\ref{fig:5dWeyl}. Vacuum solutions in this class have been classified in \cite{Costa:2000kf,Emparan:2001wk,Elvang:2002br,Emparan:2008eg,Charmousis:2003wm} while generalization with non-trivial gauge fields has been obtained in \cite{Bah:2020pdz}.

As in four dimensions, two disconnected objects are necessarily separated by a strut (see Fig.\ref{fig:5dWeyl}(a)). However, since there are rods of different nature, one can connect two different rods without forming a unique rod (see Fig.\ref{fig:5dWeyl}(b)). The $\phi$-circle does not shrink anymore in between and the solutions are free from struts. The physics of connected black holes and bubbles as in Fig.\ref{fig:5dWeyl}(b) has been studied in details in \cite{Elvang:2002br}.

These solutions can be seen as classical resolutions of the struts appearing for the four-dimensional Weyl solutions of Fig.\ref{fig:Israel4d}. In five dimensions, the segment that separates two black holes is replaced by a smooth region in Fig.\ref{fig:5dWeyl}(b), where the $y_1$-circle shrinks as a KK bubble, free from conical excess and its associate curvature singularity. Under KK reduction along $y_1$, the metric is singular there and the component along the $y_1$-circle gives a singular dilaton.  Therefore, the work in \cite{Elvang:2002br} points out something interesting in that black holes in four-dimensions along a line can be supported with a bubble instead of a struts.  Excitingly this implies that struts can be replaced by bubbles by going to higher dimensions.  


The main issue with the strut-free five-dimensional solutions of Fig.\ref{fig:5dWeyl}(b) is that they are not entirely smooth solutions as they require a succession of black holes with bubbles. One needs an other species of smooth bubbles that can alternate with our present five-dimensional topological stars. This can be ultimately done by considering not one but two extra dimensions, that is a six-dimensional Einstein-Maxwell theory.

Before going through the main construction in six dimensions, we briefly discuss the physics of Weyl solutions in five dimensions and the physics of struts with a simple example of two smooth topological stars on a line separated by a strut. This example will be useful later in the paper when we construct solutions that substitute the strut that separates the two bubbles with another species of bubbles that come from the degeneracy of a sixth dimension in a manner similar to the two black holes described in Fig.\ref{fig:5dWeyl}(b) \cite{Elvang:2002br}. The interested reader can find a summary of generic charged Weyl solutions in five dimensions as constructed in \cite{Bah:2020pdz} in the Appendix \ref{App:Weyl5d}.

\subsection{Strut between two bubbles}
\label{sec:twobubbles5d}

\begin{figure}[b]
\centering
\includegraphics[height=0.5\textwidth]{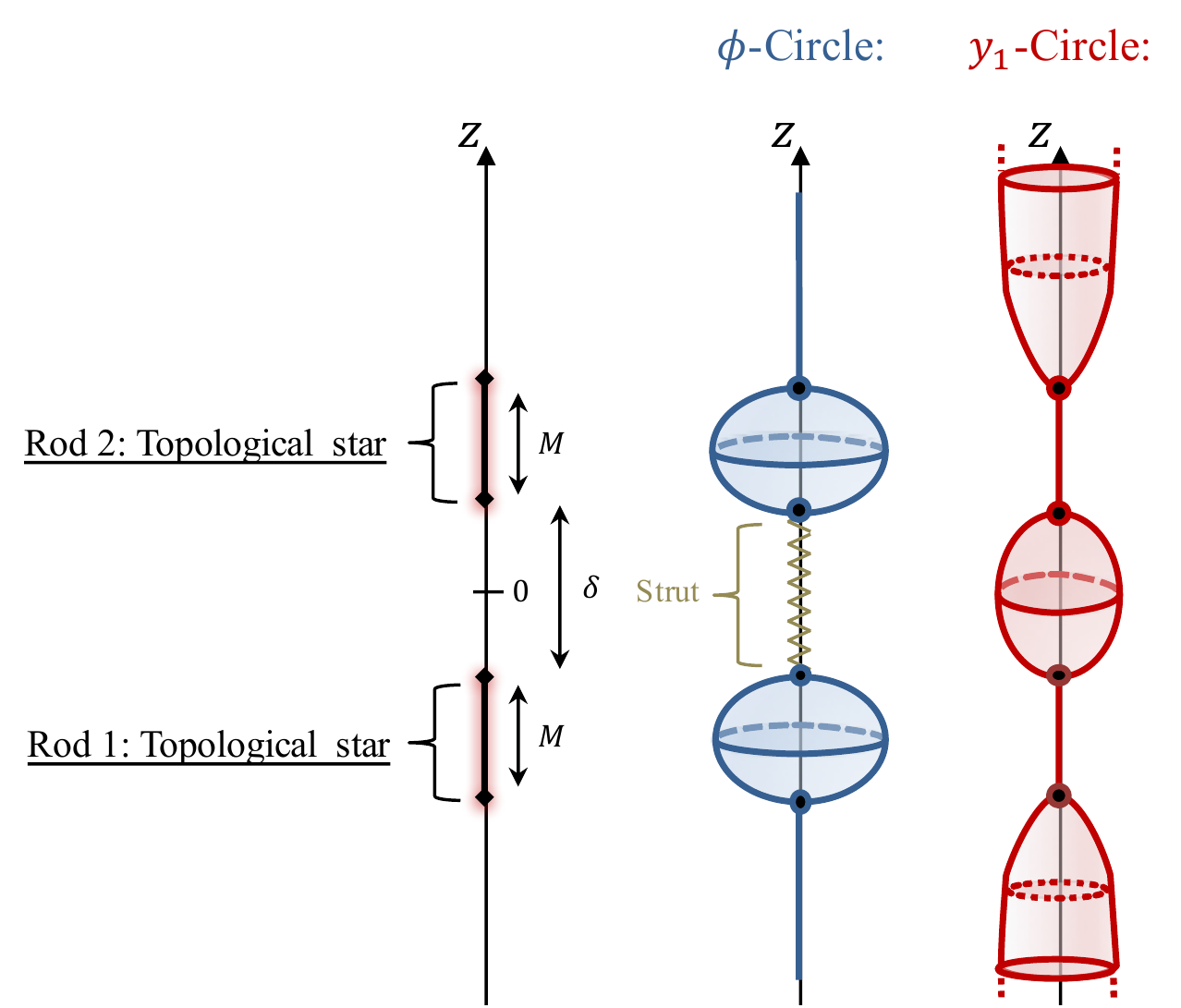}
\caption{Rod profile of a solution with two topological stars and the behavior of the circles on the $z$-axis.}
\label{fig:2TS5d}
\end{figure}

We consider a Weyl solution of the five-dimensional action \eqref{eq:ActionGen5d} that is sourced by two identical topological stars of size $M$ on the $z$-axis separated by a distance $\delta$ (see Fig.\ref{fig:2TS5d}). We keep the possibility of having conical defects at the bubbles, parametrized by an orbifold parameter $k$. The Weyl formalism is usually presented in cylindrical coordinates $(\rho,z,\phi)$ for the three-dimensional base as in \eqref{eq:conicalexcessGen} (See appendix \ref{App:Weyl5d} for more details). However, due to the symmetry of the configuration, it is convenient to consider spherical coordinates centered on the middle segment:
\begin{equation}
\rho \equi \sqrt{r(r-\delta)}\,\sin \theta\,,\qquad z \equi \left(r-\frac{\delta}{2} \right) \,\cos \theta\,,
\end{equation}
with $r\geq\delta$ and $ 0\leq \theta \leq \pi$.  The metric and gauge fields can be given as 
\begin{equation}
\begin{split}
ds_5^2  &\=\frac{1}{\bar{Z}} \,\left[ - \,dt^2 \+ U_1\, dy_1^2 \right] \+ \frac{\bar{Z}^2}{U_1}\left[e^{2\bar{\nu}}\,\left(dr^2+r(r-\delta)d\theta^2 \right) +r(r-\delta)\sin^2\theta\, d\phi^2\right]\,,\\
F^{(e)} &\= dT \wedge dt \wedge dy_1\,,\qquad F^{(m)} \= dH \wedge d\phi\,.
\label{eq:5d2bubble}
\end{split}
\end{equation} 
From the general expressions \eqref{eq:WarpfactorsRods}, we have for the rod configuration in figure \ref{fig:2TS5d}
\begin{align}
U_1 \=& \left( 1- \frac{M}{r_+}\right)\left( 1- \frac{M}{r_-}\right) \,,\qquad \bar{Z} \= \frac{e^b-e^{-b}\,U_1}{2 \sinh b} \,, \nonumber\\
e^{2\bar{\nu}} \=& \frac{1}{2} \left(1 +\frac{(M+r)(r-\delta-M)-\frac{\delta^2}{4}\sin^2\theta}{\left(2 r_- - (M+r)+\delta \sin^2\frac{\theta}{2} \right)\left(2 r_+ - (M+r)+\delta \cos^2\frac{\theta}{2} \right)} \right) \label{eq:2TS5d}\\
& \times  \frac{ \left(r-\delta \sin^2\frac{\theta}{2}\right) r_- - M (r-\delta) \sin^2\frac{\theta}{2}}{ \left(r-\delta \cos^2\frac{\theta}{2}\right) r_- - \left(M  r +\delta^2 \cos^2 \frac{\theta}{2}\right) \sin^2\frac{\theta}{2}} \frac{\left(r-\delta \cos^2\frac{\theta}{2}\right) r_+ - M (r-\delta) \cos^2\frac{\theta}{2} }{ \left(r-\delta \sin^2\frac{\theta}{2}\right) r_+ - \left(M  r +\delta^2 \sin^2 \frac{\theta}{2}\right) \cos^2\frac{\theta}{2}  } \,, \nonumber\\
H \=& \frac{\sqrt{3}}{\sqrt{1+q^2}}\,\frac{r_--r_+-\delta\,\cos \theta}{\sinh b}\,,\quad 
T\=  - \frac{\sqrt{3}\,q\,\sinh b}{\sqrt{1+q^2}} \,\frac{e^b +e^{-b}U_1}{e^b -e^{-b} U_1},\nonumber
\end{align}
where we have defined the distance to the upper and lower topological stars such as
\begin{equation}
\begin{split}
r_+ &\equi \frac{1}{2} \left( r+M-\delta \cos^2\frac{\theta}{2} \+ \sqrt{\left( r \cos\theta-M-\delta \cos^2\frac{\theta}{2} \right)^2 +r(r-\delta)\sin^2\theta} \right)\,,\\
r_- & \equi \frac{1}{2} \left( r+M-\delta \sin^2\frac{\theta}{2} \+ \sqrt{\left( r \cos\theta+M+\delta \sin^2\frac{\theta}{2} \right)^2 +r(r-\delta)\sin^2\theta} \right) \,,
\label{eq:distanceupperlower}
\end{split}
\end{equation}
and $(q,b)$ are gauge field parameters with $b\geq 0$. The quantity $q$ gives the ratio between the electric charges and the magnetic charges carried by $T$ and $H$ respectively. The vacuum limit is obtained by taking $b\to + \infty$. 
The four-dimensional ADM mass, electric and magnetic charges of the configuration are given by\footnote{We used the conventions of \cite{Myers:1986un} to derive the charges: $\cQ_e = \frac{1}{\sqrt{16\pi G_4}}\int_{S^2_\infty} \star_4( F^{(e)}|_{y_1})$ and $\cQ_m = \frac{1}{\sqrt{16\pi G_4}}\int_{S^2_\infty} F^{(m)}$.}
\begin{equation}
\cM \= \frac{1}{4 G_4}\, M\,(3 \,\coth b -1 )\,,\qquad \cQ_e \= q\,\cQ_m \= \frac{\sqrt{3}\,q}{4\sqrt{(1+q^2)\pi G_4}} \,\frac{M}{\sinh b}\,,
\label{eq:Mass2TS5d}
\end{equation}
where $G_4$ is the four-dimensional Newton's constant given by $G_5 \= \frac{G_4}{2\pi R_{y_1}}$.

The $y_1$-circle degenerates when $r_\pm=M$ ($r\leq M+\delta$, $\cos \theta = \pm1$) defining the loci of the two topological stars. The $(r,y_1)$-subspace corresponds to an origin of a smooth discrete quotient $\IR^2/\mathbb{Z}_{k}$ providing that $R_{y_1}$ satisfies \eqref{eq:Ry5dApp},
\begin{equation}
R_{y_1} \=\frac{2 M \,(2 M+\delta)}{k\,\left( 1-e^{-2b}\right)^\frac{3}{2}\,(M+\delta)}\,, \qquad k \in \mathbb{N}\,.
\label{eq:Ry2TS5d}
\end{equation}
The $(\theta,\phi)$-subspace at this loci defines the surface of the topological stars with a S$^2$ topology.

\subsubsection*{The strut region}
The region $r=\delta$ describes the section between the topological stars, and the $\phi$-circle shrinks with a conical excess corresponding to a strut.  The metric in this region can be written as
\begin{align}
ds^2_5 &=ds^2(M_3) +  \frac{d^2\,\bar{Z}^2}{U_1} \left[ d\rho^2 + \frac{\rho^2}{d^2} d\phi^2 \right]   ,\nonumber\\
ds^2(M_3) &= \frac{1}{\bar{Z}} \left[ - dt^2 + U_1\, dy_1^2 + \frac{\bar{Z}^3}{U_1} \frac{d^2 \delta^2 }{4} d\tau^2  \right],  \qquad d= \frac{(2M + \delta) \delta}{(M+\delta)^2} . \\
U_1 &= \frac{\delta^2( 1- \tau^2)}{[2M + \delta(1+ \tau)][2M+\delta(1-\tau)]}, \qquad \bar{Z} = \frac{e^b - e^{-b} U_1}{2\sinh b}. \nonumber
\end{align} The $M_3$ space is the world-volume directions of the strut with coordinate $\tau = \frac{2z}{\delta}=\cos \theta$ ranging in $|\tau| \leq 1$.  We consider $d$ as an independent parameter and solve the regularity condition \eqref{eq:Ry2TS5d} as 
\begin{equation}
M \= \frac{1-\sqrt{1-d}}{2 d}\,\left( 1-e^{-2b}\right)^\frac{3}{2} k\,R_{y_1},\qquad \delta \= \frac{\left( 1-\sqrt{1-d}\right)^2}{2 d\,\sqrt{1-d}}\,\left( 1-e^{-2b}\right)^\frac{3}{2} k\,R_{y_1}.  
\label{eq:inversionMdelta5d}
\end{equation}
It is important to note that both $M$ and $\delta$ are increasing functions of $d$, and in particular the mass parameter $M$ increases with the separation distance $\delta$.  The mass difference when the bubbles are far away compared to when they are close to each other can be interpreted as the binding energy of the final state when two bubbles merge and transition to a single bubble.  This mass difference also corresponds to the energy released when two bubbles collide and transition to a single bubble.  At a fixed separation, the binding energy is measured by the strut itself.

The conical excess is measured by $d$ and it corresponds to the tension of the strut.  In appendix \ref{App:StrutEnergy} we describe the general stress tensor for conical deficits.  We specialize these results for the two-bubble system to obtain a stress tensor for the strut as 
\begin{equation}
T_{mn} = - \frac{ d-1 }{4G_5d  } ~ \frac{ U_1}{\bar{Z}^2}   ~ g_{mn} ~ \frac{\delta(\rho) }{2\pi \rho~ d}, \qquad T_{\mu\nu} =0
\end{equation} where $(m,n)$ correspond to world-volume directions on $M_3$ with metric $g_{mn}$, while $(\mu,\nu)$ correspond to $(\rho,\phi)$ directions.  The stress tensor is localized on the strut as indicated by the delta-function $\delta(\rho)$.    First, we observe that the overall tension of the strut is negative and given by 
\begin{equation}
\text{Strut Tension} = \frac{ d-1 }{4G_5d  }.
\end{equation} The tension vanishes when $d\to 1$ corresponding to the limit when the bubbles are infinitely far away $(\delta \to \infty)$.  When the bubbles approached each other $(\delta \to 0)$, the deficit angle vanish and the tension goes to infinity.  However the stress energy tensor is actually vanishing due to the warp factor $U_1$.

We can also consider the total energy of the strut by integrating the energy density (see appendix \ref{App:StrutEnergy} for details) to obtain
\begin{equation}
E =  - \frac{M^2\delta}{(M+\delta)^2} \frac{2\pi R_{y_1}}{4G_5 } =-  \frac{M^2\delta}{4G_4 (M+\delta)^2 }.  
\label{eq:StrutEnergy}
\end{equation} 

 In the large $\delta$ limit, the energy of the strut approximates the Newtonian potential between two particles of mass, $\frac{M}{2G_4}$, in four dimensions.  From this perspective, the strut measures the binding energy between the two bubbles, or rather the potential energy needed to keep the two bubbles from collapsing on each other.  

An important observation is that the effective masses of the bubbles, from the Newtonian point of view, depend on $b$, which is associated to the charge of the bubbles as given in \eqref{eq:inversionMdelta5d}.  This implies the binding energy as measured by the strut also accounts for effects due to the electromagnetic fields of the bubbles.     

It is interesting to consider the vanishing limit of charges $(b\to \infty)$ in order to make a more precise statement on the dynamics of the bubbles.  The radius of the neutral bubbles and ADM mass in four dimensions are respectively, $M$ and $\mathcal{M} = \frac{M}{4G_4}$.  Immediately we observe that the masses that appear in the Newtonian potential for interacting two bubbles is twice the individual ADM masses of the bubbles.  The physical reason for this is that from a four dimensional perspective, the dynamics of the bubbles and their effective masses depend greatly on the scalar cloud due to the dilaton in the KK reduction \cite{1975ApJ,Damour:1992we,Mirshekari:2013vb,Julie:2017rpw}.  The dilaton seems to make the bubbles more attractive. 

The neutral bubble with ADM mass, $\mathcal{M} = \frac{M}{4G_4}$ in four dimensions, has a radius twice of its Schwarzschild radius.  Since its effective mass in the Newtonian potential is twice that of the ADM mass, its apparent Schwarzschild radius is actually located at the bubble surface.

In general when $b$ is finite, the effective mass from the Newtonian potential depends on the charges of the bubble in a non-trivial way. This accounts for the electromagnetic interactions of the bubbles.  For finite separation between the two bubbles, the gravitational potential between them deviates significantly from that of the Newtonian limit.  In particular, the gravitational potential between two bubbles vanish when $\delta \to 0$.  In this limit, the two rod sources that make the bubbles merge into one and the two bubbles system becomes a single bubble.  A similar phenomenon occurs when one considers two black holes separated by a strut as discussed in \cite{Costa:2000kf}.

\section{Six-dimensional framework}
\label{sec:6dFrame}

In this section, we discuss our six-dimensional framework used to construct and classify Weyl solutions. We work with the following six-dimensional theory
\begin{equation}
\left(16 \pi G_6 \right) \,S_{6} \= \int d^6x \sqrt{-\det g}\,\left(R \-  \frac{1}{2} \,\left|F_3\right|^2\,\right)\,,
\label{eq:Action6d}
\end{equation}
where $F_3= dC_2$\footnote{Its norm is given as \begin{equation}
\left|F_3\right|^2 =  \frac{1}{3!}\,F_{3\,\mu\nu\sigma}\,F_3^{\,\mu\nu\sigma}\,.\nonumber
\end{equation}} is a three-form field strength for a two-form gauge field $C_2$.  The equations of motion are
\begin{equation}
d\star_6 F_3 \= 0,\quad R_{\mu \nu} \= \frac{1}{2} \left(T_{\mu \nu} \- \frac{g_{\mu\nu}}{4} \,T_{\alpha}^{\,\,\alpha} \right), \quad T_{\mu\nu} = \frac{1}{2} \left[F_{3\,\mu\alpha\beta} F_{3\,\nu}^{\,\,\,\,\alpha\beta}- g_{\mu \nu}\,\left|F_3\right|^2\right]
\end{equation}

Our objective is to build solutions that generalize the five-dimensional Weyl system in \ref{sec:5dparadigm} and in Appendix \ref{App:Weyl5d} that are asymptotic to T$^2 \times\IR^{1,3}$.  The general ansatz that can accommodate these structures in six dimensions is
\begin{equation}
\begin{split}
ds_6^2 &\=  - U_0 \,dt^2 \+ U_1\, dy_1^2 \+ U_2 \left( dy_2 + H_0 \,d\phi\right)^2 \+  \frac{1}{U_0 U_1 U_2}\,ds(\cB)^2\,,\\
F_3& \= F^{(m1)} \wedge (dy_2+H_0 \,d\phi) \+ F^{(e1)}\wedge dy_1\,.
\end{split}
\label{eq:metricGen}
\end{equation} The coordinates $\{y_a\}_{a=1,2}$ parametrize the T$^2$ with $2\pi R_{y_a}$ periodicities, $ds(\cB)^2$ is the metric of the asymptotically-$\IR^3$ three-dimensional base.  We assume that the base metric is axially symmetric and admits a $U(1)$ action whose orbits are parametrized with the coordinate $\phi$.  The extra KK circle also admits a KK vector parametrized by $H_0$.  All warp factors and gauge fields depend on the base metric. It may seem intriguing at first sight to impose an asymmetry between the two  compact directions by allowing a KK vector field in the $y_2$ fiber. This is motivated by the observation, which will be clarified later, that one can freely add such a vector field without compromising the linearization of the equations of motion. This vector field is therefore a free degree of freedom that can be turned on and does not couple to the other fields. Therefore, we could have perfectly well considered two symmetric fibers but this would not have made the equations any simpler and would have turned off one degree of freedom of the gauge fields.

\subsection{Kaluza-Klein reduction}
\label{sec:KKreduction}

In this section, we describe the truncation of the six-dimensional theory in \eqref{eq:Action6d} to five and four dimensions that will provide lower-dimensional descriptions of our constructions.  In particular we will also derive the four-dimensional conserved quantities: the ADM mass and charges associated to the various gauge fields.

\subsubsection*{Reduction to five dimensions}

The KK reduction of the action \eqref{eq:Action6d} on $y_2$ gives the following five-dimensional theory 
\begin{align}
S_{5}^\text{KK} &= \frac{1}{16 \pi G_5 } \int d^5x \sqrt{-\det g}\, \left(R - \mathcal{L}_5 \right) \\
\mathcal{L}_5 &= \frac{1}{2}\,\partial_\mu \Phi_2 \,\partial^\mu \Phi_2+ \frac{e^{\frac{-2\sqrt{2}}{\sqrt{3}}\,\Phi_2}}{2}\,\left| F^{(m0)}\right|^2+ \frac{e^{\frac{\sqrt{2}}{\sqrt{3}}\,\Phi_2}}{2}\,\left| F^{(m1)}\right|^2  + \frac{e^{-\frac{\sqrt{2}}{\sqrt{3}}\,\Phi_2}}{2}\,\left| F^{(e)}\right|^2 \nonumber 
\end{align}
with $G_5 \equiv \frac{G_6}{2\pi R_{y_2}}$.  The five-dimensional theory contains a dilaton, $\Phi_2$, a pair of one-form gauge fields with field strength $(F^{(m0)},F^{(m1)})$, and a two-form gauge fields with field strength $F^{(e)}$.  These are identified with the six-dimensional frame as
\begin{equation}
F^{(m0)}\= dH_0 \wedge d\phi \,,\qquad F_3 \= F^{(m1)} \wedge (dy_2+H_0 \,d\phi) \+ F^{(e)} .
\end{equation}
In this framework, our ansatz reduces to 
\begin{equation}
ds_5^2 \= U_2^{\frac{1}{3}} \,\left[ - U_0 \,dt^2 \+ U_1\, dy_1^2\+  \frac{1}{U_0 U_1 U_2}\,ds(\cB)^2\right], \qquad e^{\frac{1}{\sqrt{6}}\Phi_2}  \= U_2^{-\frac{1}{3}} .\\
\label{eq:5dRed}
\end{equation}
Note that we retrieve the five-dimensional action \eqref{eq:ActionGen5d} if we consider the subclass of solutions that have $\Phi_2=0$ and the equations of motion for $\Phi_2$ requires in addition
\begin{equation}
\left| F^{(m0)} \right|^2 \= \frac{1}{2} \left(\left| F^{(m1)} \right|^2 -\left| F^{(e)} \right|^2 \right)\,.
\label{eq:cond5din6d}
\end{equation}

\subsubsection*{Reduction to four dimensions}

After compactification on $y_1$, we restrict to a truncation of the KK theory to the Einstein-Maxwell-Dilaton theory
\begin{align}
S_{4}^\text{KK} &=  \frac{1}{\left(16 \pi G_4\right) } \int d^4x \sqrt{-\det g}\,\left( R - \mathcal{L}_4 \right), \\
\mathcal{L}_4 &= \frac{1}{2}\,\partial_\mu \Phi_2 \,\partial^\mu \Phi_2 + \,\frac{1}{2}\,\partial_\mu \Phi_1 \,\partial^\mu \Phi_1 - \frac{e^{\frac{-2\sqrt{2}}{\sqrt{3}}\,\Phi_2-\frac{1}{\sqrt{3}}\,\Phi_1}}{2}\,\left| F^{(m0)}\right|^2  \nonumber \\
&+ \frac{e^{\frac{\sqrt{2}}{\sqrt{3}}\,\Phi_2-\frac{1}{\sqrt{3}}\,\Phi_1}}{2}\,\left| F^{(m1)}\right|^2 + \frac{e^{-\frac{\sqrt{2}}{\sqrt{3}}\,\Phi_2+\frac{1}{\sqrt{3}}\,\Phi_1}}{2}\,\left| F^{(e1)}\right|^2. \nonumber 
\end{align}
with $G_4 \equiv \frac{G_6}{(2\pi)^2 R_{y_1}R_{y_2}} $. The gauge fields are identified as 
\begin{equation}
 F^{(m0)} \= dH_0 \wedge d\phi,\qquad F_3 \= F^{(m1)} \wedge (dy_2+H_0 \,d\phi) \+ F^{(e1)}\wedge dy_1.
\end{equation} Note that we have turned off a KK vector associated to $y_1$ and the components of $F_3$ orthogonal to $y_1$ and $y_2$ in the truncation.  Our ansatz is 
\begin{align}
ds_4^2 &= \sqrt{U_1 U_2} \,\left[ - U_0 \,dt^2  \+ \frac{1}{U_0 U_1 U_2}\,ds(\cB)^2\right], \label{eq:4dFrameworkcharged}\\
e^{\frac{1}{\sqrt{3}}\,\Phi_1} &= U_1^{-\frac{1}{2}}\,U_{2}^{-\frac{1}{6}}, \qquad e^{\frac{1}{\sqrt{6}}\Phi_2}  = U_2^{-\frac{1}{3}}. \nonumber
\end{align}

For the specific solutions we will construct, $F^{(m1)}$ and $F^{(m0)}$ will be magnetically sourced and $F^{(e1)}$ will be electrically sourced. Moreover, the scalars will be asymptotic to zero. More concretely, the asymptotic behaviors of the relevant quantities are
\begin{equation}
\begin{split}
U_I &\,\underset{r\to\infty}{\sim}\, 1 \- \frac{\cM^{(I)}}{r}\,,\qquad F^{(mI)} \=-P_I \,\sin\theta \, \left( 1\+ \cO\left(\frac{1}{r}\right) \right) \,d\theta \wedge d\phi \+ \ldots\,,\\
 F^{(e1)} &\=  \frac{Q_1}{r^2} \,\left( 1\+ \cO\left(\frac{1}{r}\right) \right)\,dt\wedge dr \+ \ldots\,.
 \label{eq:AsymptoticExpGen}
\end{split}
\end{equation}
Therefore, the solutions correspond to four-dimensional three-charge static solutions with non-trivial scalar fields. The four-dimensional ADM mass, $\cM$, the electric and magnetic charges, $(\cQ^{(1)}_e,\cQ^{(0)}_m,\cQ^{(1)}_m)$, are given by\footnote{We used the conventions of \cite{Myers:1986un} to derive the charges: $\cQ^{(1)}_e = \frac{1}{\sqrt{16\pi G_4}}\int_{S^2_\infty} \star_4 F^{(e1)}$ and $\cQ^{(I)}_m = \frac{1}{\sqrt{16\pi G_4}}\int_{S^2_\infty} F^{(mI)}$.}
\begin{equation}
\cM \= \frac{1}{4 G_4} \,\left(2\cM^{(0)}+\cM^{(1)}+\cM^{(2)} \right), \quad \cQ^{(1)}_e \= \frac{Q_1}{\sqrt{16 \pi G_4}},\quad \cQ^{(I)}_m \= \frac{P_I}{\sqrt{16 \pi G_4}}\,.
\label{eq:conservedchargesGen}
\end{equation}

\subsection{Spherically symmetric solutions}
\label{sec:sphericallySymCharged}

We consider spherically symmetric static solutions of the action \eqref{eq:Action6d} where $F^{(mI)}$ and $F^{(e1)}$ are magnetically and electrically sourced. This allows to describe the basic physical sources of the theory before constructing multi-body solutions with the Weyl formalism. 

\noindent The spherical symmetry constrains the three-form field strength and the KK gauge field such that
\begin{equation}
F_3 \= \frac{Q_1}{r^2} \, dt\wedge dr\wedge dy_1 \- P_1 \,\sin\theta \,d\theta \wedge d\phi\wedge dy_2 \,,\qquad H_0\, d\phi \= P_0 \,\cos\theta \, d\phi\,,
\end{equation}
where $(r,\theta,\phi)$ are the spherical coordinates of the three-dimensional base. We find three types of physical solutions that we can decompose in two categories.

\subsubsection{Species-1 bubble and black brane}

The equations of motion are solved by considering a metric given by three parameters $(r_\text{S},r_\text{B},r_\text{C})$:
\begin{equation}
\begin{split}
ds_6^2 \=&- \left( 1-\frac{r_\text{S}}{r}\right)\,dt^2 \+ \left( 1-\frac{r_\text{B}}{r}\right)\,dy_1^2\+ \left( 1-\frac{r_\text{C}}{r}\right)^{-1}\,(dy_2+P_0\cos\theta d\phi )^2 \\
&\+ \left(1-\frac{r_\text{C}}{r} \right) \,\left[\frac{r^2\,dr^2}{(r-r_\text{S})(r-r_\text{B})} \+ r^2 \,d\Omega_2^2 \right],
\label{eq:BBchargedsingle}
\end{split}
\end{equation}
and with the charges fixed to satisfy  
\begin{equation}
P_1^2 \+ Q_1^2 \=2 \,r_\text{B}\,r_\text{S}\,,\qquad P_0^2 \= (r_\text{B}-r_\text{C})\,(r_\text{S}-r_\text{C})\,.
\label{eq:chargeRelation6d}
\end{equation}
In the four-dimensional framework \eqref{eq:4dFrameworkcharged}, the solutions have two magnetic and one electric charges, and the conserved charges are \eqref{eq:conservedchargesGen}
\begin{equation}
\cM \= \frac{1}{4 G_4}\,\left( 2 r_\text{S}+r_\text{B}-r_\text{C} \right)\,,\qquad\cQ^{(1)}_e \= \frac{Q_1}{\sqrt{16 \pi G_4}}\,,\qquad \cQ^{(I)}_m \= \frac{P_I}{\sqrt{16 \pi G_4}}\,.
\end{equation}
The domain of validity requires $r_\text{B}r_\text{S}\geq 0$ and $(r_\text{B}-r_\text{C})\,(r_\text{S}-r_\text{C})\geq 0$. Without restriction, we assume that $r_\text{B}$ and $r_\text{S}$ are positive and $r_\text{C} \leq r_\text{S/B}$ with the possibility of being negative. Therefore, we have two types of solutions:
\begin{itemize}
\item[-] When $r_\text{B} > r_\text{S}$:

The outermost coordinate singularity is $r=r_\text{B}$, where the $y_1$-circle shrinks forming an end to spacetime. The local geometry is given by
\begin{equation}
\begin{split}
ds_6^2 \,\underset{\rho \to 0}{\sim}\,& - \frac{r_\text{B} - r_\text{S}}{r_\text{B}}\,dt^2 \+  d\rho^2 \+ \frac{\rho^2 \,(r_\text{B} - r_\text{S})}{4 r_\text{B} ^2(r_\text{B} - r_\text{C})} dy_1^2 \\
& \+ \frac{r_\text{B}}{r_\text{B} - r_\text{C}}\, (dy_2+P_0\cos\theta \,d\phi)^2\+ r_\text{B}(r_\text{B} - r_\text{C} )d\Omega_2^2 \,,
\end{split}
\label{eq:localBubblecharged}
\end{equation}
with $\rho^2 \equi \frac{4 r_\text{B} (r_\text{B} - r_\text{C})}{r_\text{B} - r_\text{S}}\,(r-r_\text{B} )$. The line element $d\Omega_2^2$ describes a round S$^2$ of radius $\sqrt{r_\text{B}(r_\text{B} - r_\text{C})}$ while the $(\rho,y_1)$-subspace corresponds to an origin of $\IR^2/\mathbb{Z}_k$, $k\in \mathbb{N}$, providing that
\begin{equation}
R_{y_1} \= \frac{2r_\text{B}}{k} \sqrt{\frac{r_\text{B} - r_\text{C}}{r_\text{B} - r_\text{S}}}\,.
\end{equation} 
The geometry at the coordinate singularity corresponds to a \emph{bolt}, a smooth S$^{y_2}\times$S$^2$ bubble sitting at an origin of a $\IR^2$ with a conical defect of order $k$. One can check that the gauge fields are regular at this locus and that the geometry is regular elsewhere. 

If $k=1$, the bubble is free from conical defect and we have $2r_\text{B} \leq R_{y_1}$. Therefore, the mass, the conserved charges of the solution and the area of the bubble are bounded by the KK scale.

Finally, by taking $r_\text{C}=0$, the metric along the $y_2$ circle is trivial and one has a six-dimensional embedding of the single-center topological star \eqref{eq:TS5d}, studied in \cite{Bah:2020ogh,Bah:2020pdz}.

\item[-] When $r_\text{S} \geq  r_\text{B}$:

The outermost coordinate singularity is now $r=r_\text{S}$, where the timelike Killing vector $\partial_t$ vanishes. It corresponds to a horizon with a S$^2\times$T$^2$ topology. The local geometry can be obtained from \eqref{eq:localBubblecharged} by Wick rotation of $t$ and $y_1$ and changing the role of $r_\text{B}$ and $r_\text{S}$. The Bekenstein-Hawking entropy, $S = \frac{A}{4G_6}$, and the temperature, $\cT$,  are
\begin{equation}
S \= \frac{ \pi}{G_4}\,r_\text{S}\,\sqrt{(r_\text{S}-r_\text{C})(r_\text{S}-r_\text{B})}\,, \qquad \cT\= \frac{1}{4\pi r_\text{S}} \sqrt{\frac{r_\text{S} - r_\text{B}}{r_\text{S} - r_\text{C}}}\,. 
\end{equation} 
The second critical radius $r=r_\text{B}$ is part of the interior and corresponds to an origin of a Milne space \cite{Bah:2020ogh,Bah:2020pdz}. The extremal limit is obtained by taking $r_\text{S}=r_\text{B}>r_\text{C}$, and we recognize a two-dimensional BPS extremal black brane with an AdS$_3\times$S$^2\times$S$^1$ near-horizon geometry.
\end{itemize}

\subsubsection{Species-2 bubble}

The remaining spherically symmetric solutions are given by two parameters $(r_{B1},r_{B2})$:
\begin{equation}
\begin{split}
ds_6^2 \=& -dt^2 \+ dy_1^2 \+ \frac{(r-r_{B1})(r-r_{B2})}{r^2-r_{B1}r_{B2}}\left( dy_2 \+ P_0 \cos \theta\,d\phi \right)^2 \\
& \+  \frac{r^2-r_{B1}r_{B2}}{(r-r_{B1})(r-r_{B2})}\,dr^2 \+ \left(r^2-r_{B1}r_{B2} \right) \,d\Omega_2^2.
\end{split}
\end{equation} 
Without loss of generality, we take $r_{B1} \geq r_{B2}$.  The charges are fixed as
\begin{equation}
P_1 \= Q_1 \=0 \,,\qquad P_0^2 \= 4 r_{B1}r_{B2}\,,
\end{equation}

In the four-dimensional framework \eqref{eq:4dFrameworkcharged}, the solutions have one magnetic charge only, and the conserved charges are \eqref{eq:conservedchargesGen}
\begin{equation}
\cM \= \frac{1}{4 G_4}\,\left(  r_{B1}+ r_{B2} \right)\,,\qquad\cQ^{(1)}_e \= \cQ^{(1)}_m \=0\,,\qquad \cQ^{(0)}_m \= \frac{P_0}{\sqrt{16 \pi G_4}}\,.
\end{equation}
If $r_{B1}=r_{B2}=r_B$, the solutions correspond to BPS solutions with a Taub-NUT base of charge $2r_B$. The space ends at $r=r_\text{B}$ as a S$^{y_1}$ fibration over an origin of $\IR^4 / \mathbb{Z}_{\frac{2r_B}{R_{y_2}}}$. If $r_{B1} \neq r_{B2} $, the $y_2$-circle shrinks at $r=r_{B1}$, while the other circles have finite size. The local geometry corresponds to a S$^{y_1}\times$S$^2$ bubble sitting at an origin of $\IR^2$ with potential conical defect if
\begin{equation}
R_{y_2} \= \frac{2 \,r_{B1}}{k}\,,\qquad k\in \mathbb{N}\,.
\end{equation}

Now we have described the three types of physical sources of the theory, we can discuss the axisymmetric generalization corresponding to such sources stacked on a line.

\section{Vacuum Weyl solutions}
\label{sec:chainBoN}

Motivated by the work of Elvang and Horowitz \cite{Elvang:2002br} we propose to use bubbles from extra dimensions as substitutes of struts to construct bound states of black holes and bubble geometries. This is akin to providing a resolution for struts by using degrees of freedom from extra dimensions.  To construct entirely smooth and regular spacetimes, we need to consider Weyl solutions that are made of connecting bubbles of different species, which requires two extra compact dimensions. As a warm-up and to illustrate the physics of the resolution, we consider vacuum solutions of our six-dimensional theory first. More precisely, we aim to construct neutral axisymmetric static Weyl solutions of the six-dimensional action \eqref{eq:Action6d} with the gauge field turned off, $F_3 = 0$.

\subsection{Weyl ansatz}
\label{sec:WeylVac}

We consider canonical Weyl coordinates $(\rho,z,\phi)$ for the three-dimensional base and the  ansatz of metric for neutral solutions is then
\begin{equation}
ds_6^2  \= - U_0 \,dt^2 \+ U_1\, dy_1^2 \+ U_2 \,dy_2^2 \+ \frac{1}{U_0 U_1 U_2}\left[e^{2\nu}\,\left(d\rho^2+dz^2 \right) +\rho^2 d\phi^2\right]\,,
\label{eq:metricVac}
\end{equation}
where the warp factors, $(U_0,U_1,U_2,\nu)$, are functions of $(\rho,z)$. We define the two-dimensional cylindrical Laplacian
\begin{equation}
\cL \equi \frac{1}{\rho} \,\partial_\rho \left( \rho \,\partial_\rho \right) + \partial_z^2\,,
\label{eq:Laplacian}
\end{equation}
and the Einstein equations can be decomposed into
\begin{equation}
\begin{split}
&\cL \log U_I  \= 0 \,,\qquad I=0,1,2 \,,\qquad \frac{1}{\rho} \,\partial_z \nu \=  \frac{1}{4} \,\sum_{I<J} \partial_\rho \log (U_I U_J )\,\partial_z \log (U_I U_J )  \,,\\
&\frac{1}{\rho} \,\partial_\rho \nu  \=   \frac{1}{8} \,\sum_{I<J} \left( \partial_\rho \log (U_I U_J )\right)^2- \left(\partial_z \log (U_I U_J )\right)^2 \,.
\end{split}
\label{eq:EQforUI}
\end{equation}
The system of equations is entirely integrable. One can source the warp factors, $U_I$, by rod segments  on the $z$-axis,\footnote{Sourcing $\log U_I$ by point particles does not lead to any physical solutions.} and the equations for $\nu$ are simple integral equations for which the integrability is guaranteed by the harmonicity of $\log U_I$. 

\subsubsection{Rod sources: generic solutions}
\label{sec:RodSolGenVac}

\begin{figure}[h]
\centering
\includegraphics[width=0.23\textwidth]{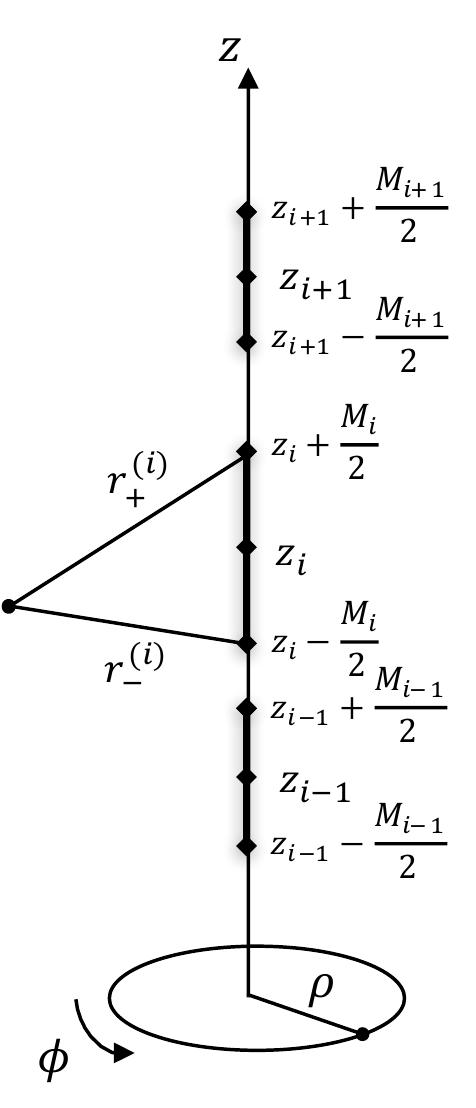}
\caption{Schematic description of axisymmetric rod sources.}
\label{fig:RodsSources}
\end{figure}

 We consider $n$ distinct rods of length $M_i>0$ along the $z$-axis centered on $z=z_i$. Without loss of generality we can order them as $z_i < z_j$ for $i<j$. Our conventions are illustrated in Fig.\ref{fig:RodsSources}. The coordinates of the endpoints of the rods on the $z$-axis are given by
\begin{equation}
z^\pm_i \equi z_i \pm \frac{M_i}{2}\,.
\label{eq:coordinatesEndpoints}
\end{equation}
We define the distances to the endpoints $r_\pm^{(i)}$, the distances $R_\pm^{(i)}$ and the generating functions $E_{\pm \pm}^{(i,j)}$ such as
\begin{align}
&r_\pm^{(i)} \equiv \sqrt{\rho^2 + \left(z-z^\pm_i\right)^2}\,, \qquad R_\pm^{(i)} \equiv r_+^{(i)}+r_-^{(i)}\pm M_i\,, \label{eq:Rpmdef}\\
&E_{\pm \pm}^{(i,j)} \equi r_\pm^{(i)} r_\pm^{(j)} + \left(z-z_i^{\pm}  \right)\left(z-z_j^{\pm}  \right) +\rho^2\,, \qquad \nu_{ij} \equi \log \frac{E_{+-}^{(i,j)}E_{-+}^{(i,j)}}{E_{++}^{(i,j)}E_{--}^{(i,j)}}\,.\nonumber
\end{align}
The warp factors associated to such sources are 
\begin{equation}
U_I = \prod_{i=1}^n \left( \frac{R_+^{(i)}}{R_-^{(i)}} \right)^{G^{(I)}_i}\,, \qquad \nu \= \frac{1}{8} \,\sum_{I<J} \sum_{i,j=1}^n \left(G_i^{(I)}+G_i^{(J)}\right)\left(G_j^{(I)}+G_j^{(J)}\right)\,\nu_{ij}\,,
\label{eq:ExprUI}
\end{equation}
where $G^{(I)}_i$ defines the weight of the $i^\text{th}$ rod on the warp factor $U_I$. 
The solutions are directly asymptotic to T$^2\times\IR^{1,3}$ as $\frac{R_+^{(i)}}{R_-^{(i)}} \to 1$ and $\nu_{ij} \to 1$ at large distance.

\subsubsection{Regularity}
\label{sec:RegVac}

Potential singularities arise on the $z$-axis: at the rod sources and elsewhere where the $\phi$-circle shrinks. We discuss the regularity analysis in details in the appendix \ref{App:WeylVac}. Regularity at the rods fixes the weights $G_i^{(I)}$ to three choices corresponding to the three categories of physical sources highlighted in section \ref{sec:sphericallySymCharged} (see Fig.\ref{fig:RodCategories} for a schematic description). To elucidate the regularity conditions, we define 
 \begin{equation}
 \alpha_{jk} \equi \frac{1}{2}\,\sum_{I<J}\left(G_j^{(I)}+G_j^{(J)}\right)\left(G_k^{(I)}+G_k^{(J)} \right). 
  \end{equation} This appears in the solution of $\nu$ in \eqref{eq:ExprUI} and in the various regularity constraints below.  These parameters take specific values depending on the three species of object that can exist at the rods:
\begin{equation}
\alpha_{jk} \= \begin{cases} 
1 \qquad &\text{if the }j^\text{th}\text{ and }k^\text{th}\text{ same object species,}  \\
\frac{1}{2} \qquad &\text{otherwise.}
\end{cases}
\label{eq:alphaDef}
\end{equation}
We also introduce the parameters $d_i$ appearing in the regularity conditions
\begin{equation}
\begin{split}
d_1 &\equi 1\,,\qquad d_i \equi  \prod_{j=1}^{i-1} \prod_{k=i}^n \left(\dfrac{(z_k^- - z_j^+)(z_k^+ - z_j^-)}{(z_k^+ - z_j^+)(z_k^- - z_j^-)}  \right)^{\alpha_{jk}}\quad \text{when } i=2,\ldots n\,.
\end{split}
\label{eq:dialphaDef}
\end{equation}
The three categories of possible objects that can live on the rods are (see appendix \ref{App:AttheRod} for more details):
\begin{itemize}
\item \underline{\bf Black rod:} The $i^\text{th}$ rod corresponds to the horizon of a static two-dimensional black brane or black string\footnote{\label{footnote1}The horizon topology depends on the close environment of the black rod. If the rod is disconnected as in Fig.\ref{fig:RodCategories}, it corresponds to the T$^2\times$S$^2$ horizon of a two-dimensional black brane. If the rod is connected on one side only as in Fig.\ref{fig:TouchingRods}, it corresponds to the S$^1\times$S$^3$ horizon of a black string.} providing that 
\begin{equation}
G^{(0)}_i =-1 \,,\qquad  G^{(1)}_i = G^{(2)}_i=0\,.
\label{eq:BlackBraneGi}
\end{equation} 
At the rod, $\rho= 0$ and $z_i^- \leq z \leq z_i^+$, the timelike Killing vector $\partial_t$  vanishes and the local geometry defines the locus of a horizon with a T$^2\times$S$^2$ or S$^1\times$S$^3$ topology (see Fig.\ref{fig:RodCategories} and Fig.\ref{fig:TouchingRods}).\footnoteref{footnote1} The surface gravity and horizon area associated to this black brane or black string is
\begin{equation}
\kappa_i  \= \frac{1}{2d_i\,M_i}\, \prod_{j\neq i} \left(\frac{z_j^+ - z_i^-}{z_j^- -z_i^-} \right)^{\text{sign}(i-j)\,\alpha_{ij}}\,,\qquad A_i \=  \frac{(2\pi)^3}{\kappa_i}\,M_i\,R_{y_1}R_{y_2}\,,
\label{eq:surfaceGrav&AreaVac}
\end{equation}
\item \underline{\bf Species-1 bubble rod:} The $i^\text{th}$ rod corresponds to a static KK bubble where the $y_1$-circle shrinks providing that
\begin{equation}
G^{(1)}_i = -1 \,,\qquad  G^{(0)}_i = G^{(2)}_i=0\,.
\label{eq:1bubbleGi}
\end{equation} 
At the rod, $\rho= 0$ and $z_i^- \leq z \leq z_i^+$, the orbits of the spacelike Killing vector $\partial_{y_1}$ shrink. The local geometry corresponds to a bubble with a S$^1\times$S$^2$ or S$^3$ topology\footnote{\label{footnote2}As for a black rod, the local topology of the bubble depends if the bubble rod is disconnected (Fig.\ref{fig:RodCategories}) or connected (Fig.\ref{fig:TouchingRods}).} at the origin of a $\IR^2$ space parametrized by $(\rho,y_1)$ (see Fig.\ref{fig:RodCategories} and Fig.\ref{fig:TouchingRods}), and the internal parameters must be related to the radius of the extra dimension:
\begin{equation}
R_{y_1} \= 2d_i\,M_i\, \prod_{j\neq i} \left(\frac{z_j^+ - z_i^-}{z_j^- -z_i^-} \right)^{\text{sign}(j-i)\,\alpha_{ij}}\,.
\label{eq:RegRy1Vac}
\end{equation}
One can also consider that the $\IR^2$ has a conical defect by introducing a positive integer $k_i\in \mathbb{N}$ and replacing $R_{y_1} \to k_i \,R_{y_1}$  in the above expression.

\item \underline{\bf Species-2 bubble rod:}  The last category of physical source is identical to the previous one by inverting the role of $y_1$ and $y_2$. If we consider
\begin{equation}
G^{(2)}_i = -1 \,,\qquad  G^{(0)}_i = G^{(1)}_i=0\,,
\label{eq:2bubbleGi}
\end{equation} 
then the $i^\text{th}$ rod corresponds to a S$^1\times$S$^2$ or S$^3$ bubble\footnoteref{footnote2} at the origin of a $\IR^2$ space  parametrized by $(\rho,y_2)$(see Fig.\ref{fig:RodCategories} and Fig.\ref{fig:TouchingRods}). The regularity fixes similarly
\begin{equation}
R_{y_2} \= \frac{2d_i\,M_i}{k_i}\,\prod_{j\neq i} \left(\frac{z_j^+ - z_i^-}{z_j^- -z_i^-} \right)^{\text{sign}(j-i)\,\alpha_{ij}}\,,
\label{eq:RegRy2Vac}
\end{equation}
where $k_i \in \mathbb{N}$ is an orbifold parameter defining a potential conical defect at the bubble.
\begin{figure}[h]
\centering
\includegraphics[width=0.8\textwidth]{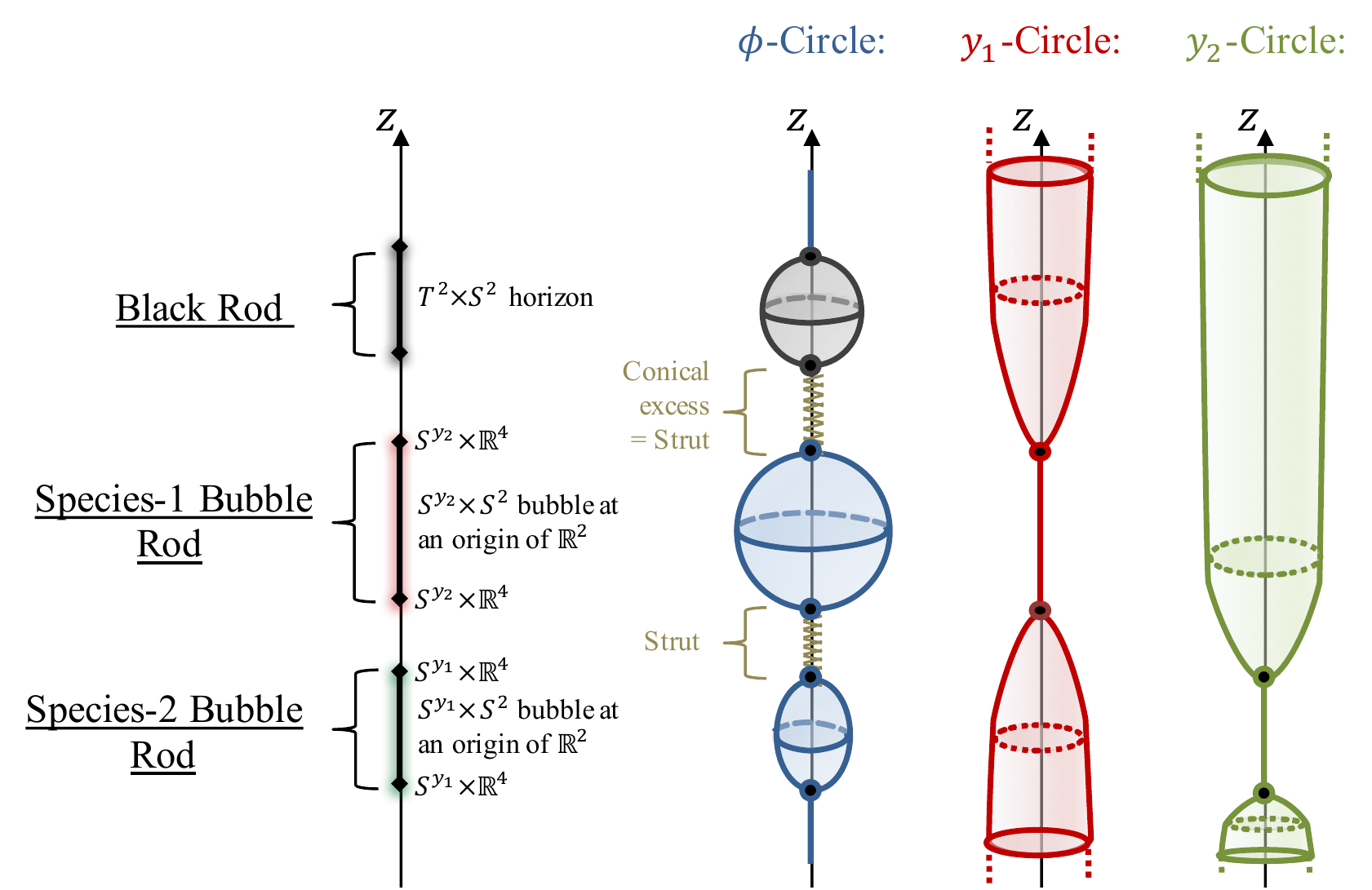}
\caption{Schematic description of the three possible categories of physical rods and the behavior of the circles on the $z$-axis when the rod are disconnected. The black rod sources a two-dimensional static black brane with a T$^2\times$S$^2$ horizon. The species-1 bubble rod corresponds to the degeneracy of the $y_1$-circle inducing a S$^{y_2}\times$S$^2$ bubble and reversely for the species-2 bubble rod. Each section between sources has a conical excess.}
\label{fig:RodCategories}
\end{figure}
\end{itemize}

\paragraph{\underline{Regularity elsewhere on the $z$-axis:}} Out of the rods but still on the $z$-axis, the $\phi$-circle degenerates as the cylindrical coordinate degeneracy \eqref{eq:metricVac}. The warp factor $e^{2\nu}$ can induce conical singularities. As discussed in the appendix \ref{App:OutoftheRod}, we found that:
\begin{itemize}
\item The semi-infinite segments above and below the rod configuration, $z > z_n^+$ and $z<z_1^-$, are free from conical singularity and the geometry is smooth.
\item If the two successive $(i-1)^\text{th}$ and $i^\text{th}$ rods, with $i=2...n$, are disconnected, then they are separated by a segment with a conical excess given by the orbifold parameter $d_i <1$ \eqref{eq:dialphaDef}. More concretely, the local metric for $z_{i-1}^+ < z < z_{i}^-$ and $\rho \sim 0$, behaves as
\begin{equation}
ds_6^2 \sim -f_t \,dt^2 + f_1 \,dy_1^2 +f_2 \,dy_2^2 + f_z \,\left(d\rho^2 + dz^2 + \frac{\rho^2}{d_i^2} \,d\phi^2 \right)\,,
\end{equation}
where the $f$'s are regular functions of $z$. Therefore, the local angle, $\phi_i \= \frac{\phi}{d_i}$, has a periodicity of $\frac{2\pi}{d_i}>2\pi$.
\item If the two successive $(i-1)^\text{th}$ and $i^\text{th}$ rods, with $i=2...n$, are connected, then their separation is reduced to a point on the $z$-axis,\footnote{Note that two touching rods are necessarily of a different nature. Indeed, two touching rods of the same nature form a single rod.} $z=z_{i-1}^+= z_i^-$. We have shown in the appendix \ref{App:OutoftheRod} that the $\phi$-circle keeps a finite size there and the local geometry is free from struts. 
The local geometry depends on the nature of the connecting rods. If a black rod is touching a bubble rod, the intersection corresponds to a pole of the horizon \eqref{eq:intersectionBB&Bu} (see Fig.\ref{fig:TouchingRods}). If we have two touching bubbles of different species, the intersection corresponds to a S$^\phi$ fibration over an origin of $\IR^4$ \eqref{eq:intersectionBu&Bu}. The $\IR^4$ has the same conical defects as the individual bubbles. If the bubbles are free from conical defect, we obtain a smooth S$^\phi\times\IR^4$ local geometry.

\begin{figure}[h]
\centering
\includegraphics[width=0.8\textwidth]{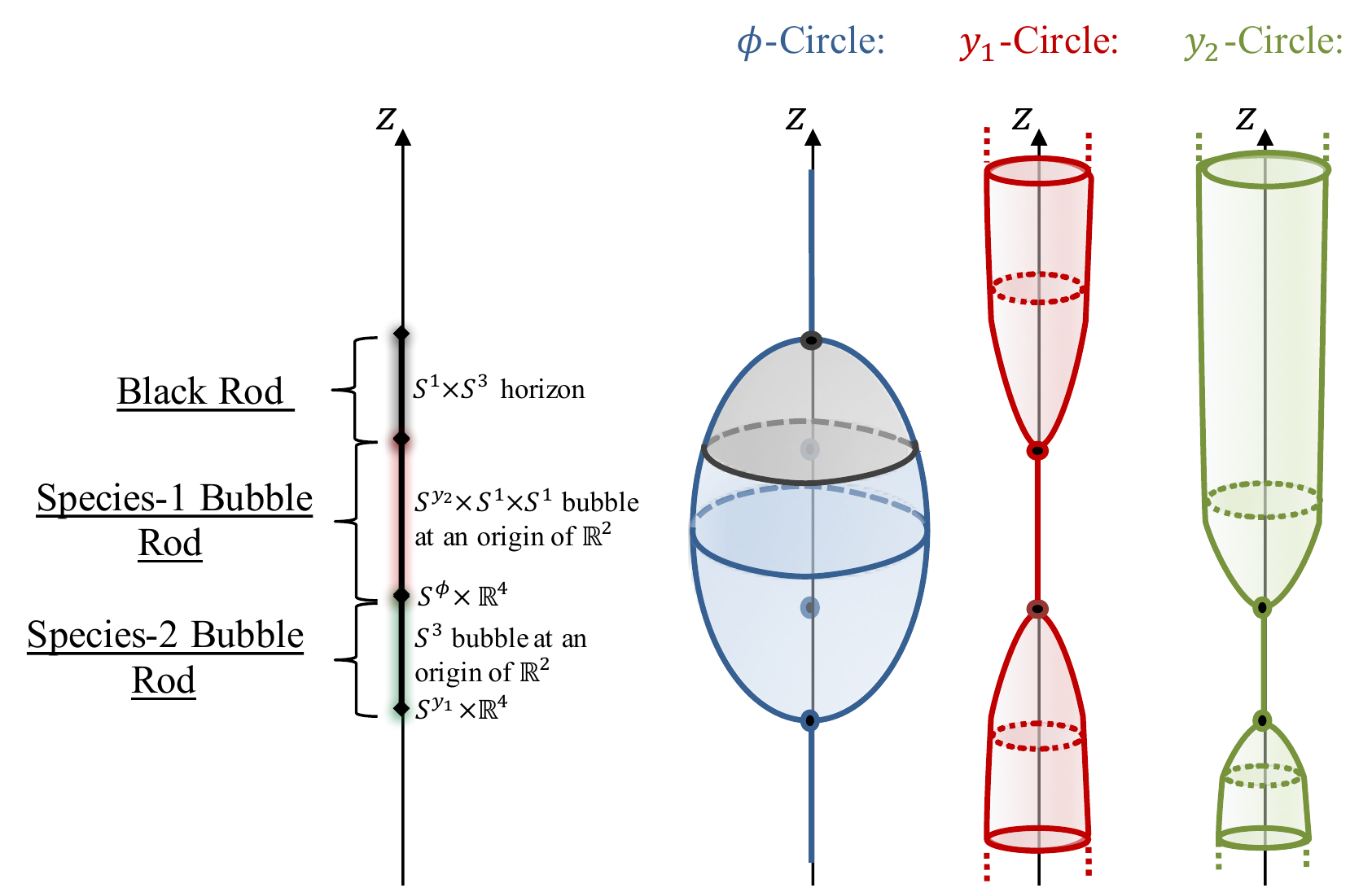}
\caption{Schematic description of connected rods by making the rods of Fig.\ref{fig:RodCategories} to touch. The $\phi$-circle describes now a single bubble without struts. The black rod sources a black string where the $\phi$-circle and the $y_1$-circle degenerate at its north and south poles respectively. At the intersection of the bubble rods, both the $y_1$ and $y_2$ circles shrink defining an origin of $\IR^4$.}
\label{fig:TouchingRods}
\end{figure}

\end{itemize}

We have seen that the struts between the rods can be replaced with bubble rods. The resulting geometry will be free from conical excess. This geometric replacement has been discussed in section \ref{sec:5dparadigm} and in \cite{Elvang:2002br} for five-dimensional solutions. However, fully smooth solutions can be constructed by building a succession of species-1 and species-2 KK bubbles in six dimensions. We will demonstrate such a solution in detail by an explicit construction in the next section.

Note also that if we consider multiple KK bubbles, their regularity conditions \eqref{eq:RegRy1Vac} and \eqref{eq:RegRy2Vac} constrain non-trivially the geometry and the size of the rods. These constraints are the equivalent of the \emph{bubble equations} for BPS multicenter solutions \cite{Denef:2000nb,Denef:2002ru,Bates:2003vx}.

After reduction to four dimensions, the class of solutions correspond to massive and neutral solutions with two scalars \eqref{eq:4dFrameworkcharged}. The black rods reduce to the horizons of regular black holes with non-trivial scalar profiles while the bubble rods correspond to singularities where the scalars diverge. The four-dimensional ADM mass \eqref{eq:conservedchargesGen} gives
\begin{equation}
\cM \= \frac{1}{4G_4}\,\sum_{i=1}^n M_i \left( 1 - G_i^{(0)} \right)\,.
\label{eq:ADMmassVacregRod}
\end{equation}
Thus, black rods ($G_i^{(0)}=-1$) contributes more to the mass than bubble rods ($G_i^{(0)}=0$). More concretely, a bubble rod of length $M_i$ contributes to the total mass as half of a black rod with the same length.\footnote{However, note that the physical size of the objects in six dimensions, $\int_\text{rod} \sqrt{-g}|_\text{rod}$, can be very different from the rod lengths, $\int_\text{rod}1 = M_i$, due to warping effects in six dimensions.}

\subsection{Smooth bubbling Weyl solutions}
\label{sec:smoothchainVac}

In this section, we apply the generic construction to an explicit illustrative strut-free example of six-dimensional Weyl solutions. As already said, this requires to deal with configurations of connected bubble rods. 

The solution here is the six-dimensional counterpart of the five-dimensional two-bubble solutions with a strut constructed in section \ref{sec:twobubbles5d}, and the strut will be replaced by a smooth species-2 bubble. This simple example gives a nice illustration of the classical replacement of struts by the degeneracy of extra dimensions. 

\subsubsection{Three connected bubbles}
\label{sec:threeconnectedVac}

\begin{figure}[h]
\centering
\includegraphics[width=0.8\textwidth]{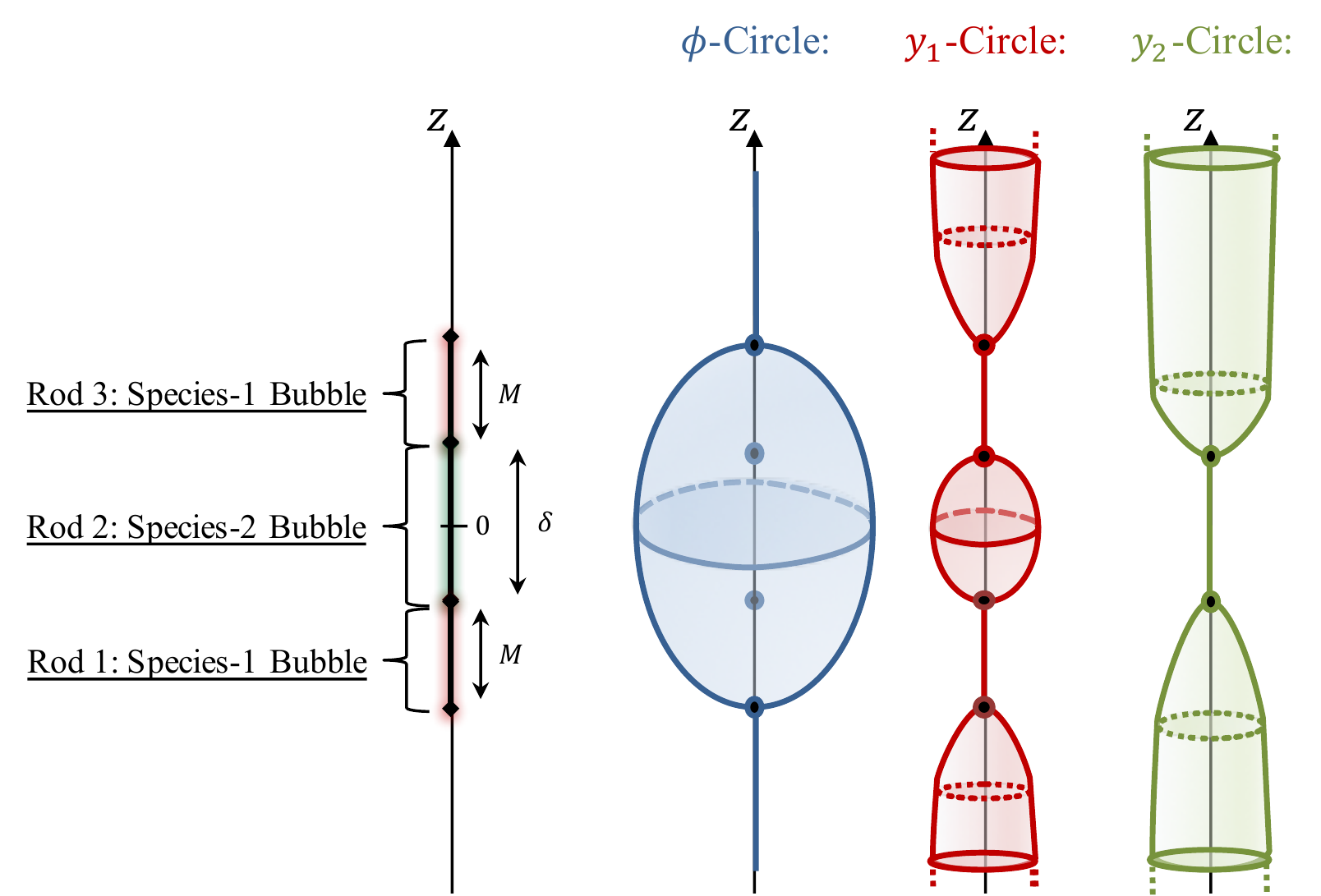}
\caption{Rod profile of the three-bubble solutions and the behavior of the circles on the $z$-axis.}
\label{fig:ThreeTouchingRods}
\end{figure}

We consider the same rod configuration as in section \ref{sec:twobubbles5d}, that is two identical species-1 bubble rods of size $M$ separated by a distance $\delta$, but with an additional species-2 bubble rod of size $\delta$ in between (see Fig.\ref{fig:ThreeTouchingRods}). We keep the possibility of having conical defects at the bubbles by considering the same integer $k$ for the three bubbles. We also use the spherical coordinates centered on the species-2 bubble:
\begin{equation}
\rho \equi \sqrt{r(r-\delta)}\,\sin \theta\,,\qquad z \equi \left(r-\frac{\delta}{2} \right) \,\cos \theta\,,
\end{equation}
with $r\geq \delta$ and $ 0\leq \theta \leq \pi$. The six-dimensional metric \eqref{eq:metricVac} is now given by
\begin{align}
ds_6^2  &= - U_0 \,dt^2 \+ U_1\, dy_1^2 \+ U_2 \,dy_2^2 \\ &+ \frac{1}{U_0 U_1 U_2}\left[e^{2\bar{\nu}}\,\left(dr^2+r(r-\delta)d\theta^2 \right) +r(r-\delta)\sin^2\theta\, d\phi^2\right], \nonumber
\end{align}
where $$e^{2\bar{\nu}} \equi \frac{\left(r-\delta \sin^2\frac{\theta}{2}\right)\left(r-\delta \cos^2\frac{\theta}{2}\right)}{r (r-\delta)} \, e^{2\nu}\,. $$
From the generic expressions \eqref{eq:ExprUI} with the rod configuration considered, we have
\begin{equation}
\begin{split}
U_0 \=& 1 \,,\qquad U_1 \= \left( 1- \frac{M}{r_+}\right)\left( 1- \frac{M}{r_-}\right)\,,\qquad U_2 \= 1 \- \frac{\delta}{r}\,, \\
e^{2\bar{\nu}} \=& \frac{1}{2} \left(1 +\frac{(M+r)(r-\delta-M)-\frac{\delta^2}{4}\sin^2\theta}{\left(2 r_- - (M+r)+\delta \sin^2\frac{\theta}{2} \right)\left(2 r_+ - (M+r)+\delta \cos^2\frac{\theta}{2} \right)} \right) \\
& \times \sqrt{ \frac{ \left(r-\delta \sin^2\frac{\theta}{2}\right) r_- - M (r-\delta) \sin^2\frac{\theta}{2}}{ \left(r-\delta \cos^2\frac{\theta}{2}\right) r_- - \left(M  r +\delta^2 \cos^2 \frac{\theta}{2}\right) \sin^2\frac{\theta}{2}}} \\
& \times \sqrt{ \frac{\left(r-\delta \cos^2\frac{\theta}{2}\right) r_+ - M (r-\delta) \cos^2\frac{\theta}{2} }{ \left(r-\delta \sin^2\frac{\theta}{2}\right) r_+ - \left(M  r +\delta^2 \sin^2 \frac{\theta}{2}\right) \cos^2\frac{\theta}{2}  }} \,,
\end{split}
\label{eq:WarpFact3BubblesVac}
\end{equation}
where  the distances to the upper and lower species-1 bubble, $r_\pm$, are sill given by \eqref{eq:distanceupperlower}.

The $y_2$-circle degenerates at $r=\delta$ and the $(r,y_2)$-subspace corresponds to an origin of a smooth discrete quotient $\IR^2/\mathbb{Z}_{k}$ providing that the radius $R_{y_2}$ satisfies \eqref{eq:RegRy2Vac}
\begin{equation}
R_{y_2} \= \frac{2\delta (2 M +\delta)}{k\,\left(M+\delta\right)}\,.
\label{eq:BE3connecVac1}
\end{equation}
The $(y_1,\theta,\phi)$-subspace at this locus defines the surface of the species-2 bubble with a finite S$^{\phi}\times$S$^2$. 

Moreover, the $y_1$-circle degenerates when $r_\pm=M$ ($r\leq M+\delta$, $\cos \theta = \pm1$) defining the loci of both species-1 bubbles. The $(r,y_1)$-subspace corresponds to origins of smooth discrete quotients $\IR^2/\mathbb{Z}_{k}$ providing that $R_{y_1}$ satisfies \eqref{eq:RegRy1Vac},
\begin{equation}
R_{y_1} \= \frac{2\sqrt{M}\,(2M+\delta)}{k\,\sqrt{M+\delta}}\,.
\label{eq:BE3connecVac2}
\end{equation}
The $(y_2,\theta,\phi)$-subspace at this loci defines the surface of the species-1 bubbles with a  finite S$^3$. By inverting both bubble equations, we find
\begin{equation}
M \= \frac{1}{2} \,\frac{k\,R_{y_1}^2}{\sqrt{4R_{y_1}^2 +R_{y_2}^2}}\,,\qquad  \delta \= \frac{k\,R_{y_2}}{4} \,\left( \frac{R_{y_2}}{\sqrt{4R_{y_1}^2 +R_{y_2}^2}}+1\right)\,.
\label{eq:M&deltainverse}
\end{equation}

\begin{figure}[h]
\begin{adjustwidth}{-1cm}{-1cm}
\centering
\includegraphics[width=1.1\textwidth]{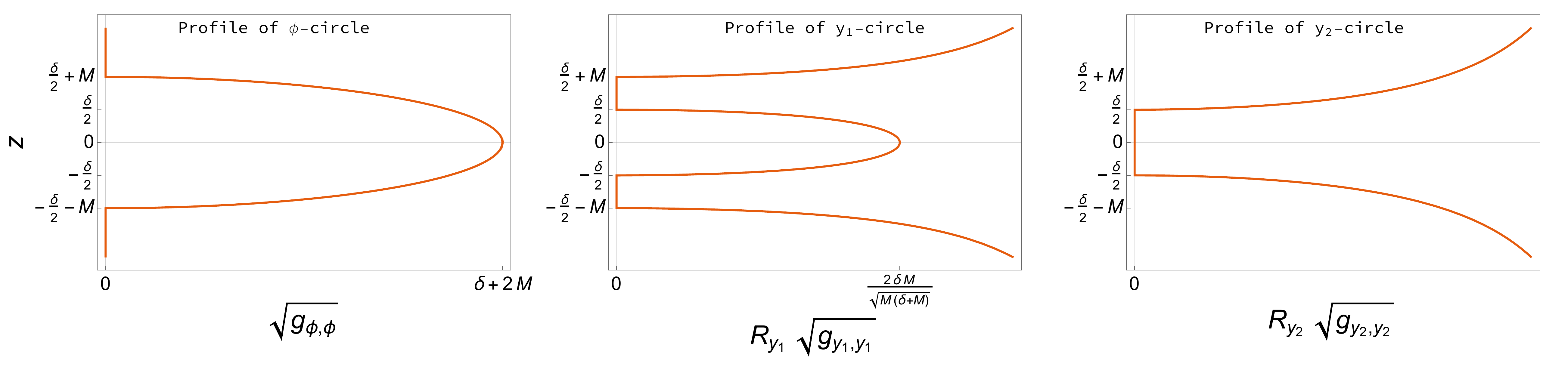}
\end{adjustwidth}
\caption{Sizes of the $\phi$, $y_1$ and $y_2$ circles on the $z$-axis.}
\label{fig:sizethreeVac}
\end{figure}

The bubbles are connected at the north and south poles of the species-2 bubble, $\theta= 0,\,\pi$ and $r=\delta$, as depicted in Fig.\ref{fig:ThreeTouchingRods}. Both the $y_1$ and $y_2$ circle degenerate and the $(r,\theta,y_1,y_2)$-subspace defines origins of $\IR^4/(\mathbb{Z}_k\times \mathbb{Z}_k)$. For the interested reader, we have plotted the size of the $\phi$, $y_1$ and $y_2$ circles on the $z$-axis in Fig.\ref{fig:sizethreeVac}.

We have constructed a smooth horizonless Weyl solution without conical singularities if $k=1$. If $k>1$, the solutions is free from conical excess but have conical defects at the bubbles. However, conical defects are rather benign singularities that are nicely resolved in string theory, and such geometries are usually viewed as regular. On one hand, conical defects have a well-known classical resolution by blowing up smooth Gibbons-Hawking cycles \cite{GIBBONS1978430,Bena:2007kg}. On the other hand, in AdS/CFT the corresponding worldsheet conformal field theory is completely well-defined and there are many examples, for instance in AdS$_3$/CFT$_2$ \cite{Jejjala:2005yu,Giusto:2004id,Giusto:2012yz}, where regular CFT states have bulk duals with conical defects.

In the four-dimensional framework \eqref{eq:4dFrameworkcharged}, the solution is singular at the bubble loci since the dilatons diverge and the four-dimensional metric degenerates there. The four-dimensional ADM mass \eqref{eq:ADMmassVacregRod} gives
\begin{equation}
\cM \= \frac{1}{4G_4}\,\left( 2M +\delta \right) \= \frac{k}{16 G_4}\,\left( R_{y_2} + \sqrt{4 R_{y_1}^2+R_{y_2}^2} \right) \,.
\end{equation}
The mass is bounded by the sum of the extra-dimension radii if $k=1$. As for the solutions of \cite{Bah:2020ogh,Bah:2020pdz} and the single-bubble solutions in section \ref{sec:sphericallySymCharged}, one needs large conical defects at the bubbles to have macroscopic objects. These defects highlight a much richer space of states via the classical resolution of the orbifold singularity by blowing up Gibbons-Hawking cycles at the poles of the bubbles \cite{Bah:2020ogh,Bah:2020pdz}.

The strut of the neutral two-bubble solution in five dimensions, obtained from the solution of section \ref{sec:twobubbles5d},\footnote{The two-bubble solution  of section \ref{sec:twobubbles5d} has non-trivial charges, so one needs to take their neutral limit $b\to \infty$.} has been replaced by a smooth bubble where the sixth dimension degenerates. It is rather surprising that the strut, which gives the necessary pressure to keep the two bubbles apart, can be replaced by a vacuum KK bubble that does not interact at first sight and does not carry energy in six dimensions. Our interpretation can be made in the line of \cite{Elvang:2002br}. The bubble in the middle has an inherent nature to expand since it is vacuum bubble \cite{Witten:1981gj}. Therefore, this gives the pressure that allows the two bubbles to stay away from each other and compensates for their attraction. The nucleation of a smooth KK bubble at the location of a strut singularity is then a perfectly viable scenario.

Under compactification along $y_2$, our six-dimensional solutions are given by \eqref{eq:5dRed}, and have a singularity at the place of the species-2 bubble $r=\delta$. In order to evaluate which of the two solutions with two bubbles and a strut or three bubbles are preferred, one should a priori compute the free energy of both euclideanized versions in six dimensions and compare in the manner of \cite{Gibbons:1978ji,Gross:1982cv,York:1986it,Brown:2014rka}. If the three-bubble solution has a lower free energy than the two-bubble one, it would suggest that a phase transition from the latter to the former can occur where the third bubble nucleates. We reserve this for future projects and focus in this article on the simple replacement of the struts with smooth bubbles.


As a final comment, a major drawback of the viability of neutral smooth Weyl solutions is related to their stability (which also undermines any free energy analysis). As we already argued, a single KK bubble in vacuum, or bubble-of-nothing, has a quantum instability that forces the bubble to grow in size \cite{Witten:1981gj,Brown:2014rka}. Studying similar instability with our axisymmetric configurations is a difficult task, but, because the main ingredients are KK bubbles stacked on a line, it is expected that the whole geometry is also suffering from such an instability and decays to nothing. The expansion of the bubble in the middle can be compensated by the attraction of the two species-1 bubbles, but the two species-1 bubbles have nothing to stop them from expanding, and hence the expected instability.

 In \cite{Miyamoto:2006nd,Stotyn:2011tv}, it has been argued that KK bubbles are stabilized by adding suitable electromagnetic fluxes wrapping the bubbles. This has allowed the construction of topological stars and charged Weyl solutions with struts in five dimensions as reviewed in section \ref{sec:5dparadigm} and appendix \ref{App:Weyl5d} \cite{Bah:2020ogh,Bah:2020pdz}. In the next section, we will therefore construct charged Weyl solutions in six dimensions and replace the struts by a smooth degeneracy of the sixth dimension.

\section{Three-charge Weyl solutions}
\label{sec:chainTS}

We now apply the Weyl formalism to construct charged Weyl solutions of the six-dimensional action \eqref{eq:Action6d}. As presented in section \ref{sec:6dFrame}, we  consider that the solutions can carry Kaluza-Klein monopole charges along the $y_2$ fiber and that $F_3$ can carry electric and magnetic line charges along $y_1$ \eqref{eq:metricGen}.

\subsection{Weyl ansatz}
\label{sec:Weyl}

We consider Weyl canonical coordinates $(\rho,z,\phi)$ for the three-dimensional base and the following ansatz for the metric and gauge fields
\begin{align}
ds_{6}^2 = &\frac{1}{Z_1} \left[- W_0\,dt^2 + \frac{ dy_1^2}{W_0} \right] + \frac{Z_1}{Z_0}\, \left(dy_2 +H_0 \,d\phi\right)^2 + Z_0 Z_1\,\left( e^{2\nu} \left(d\rho^2 + dz^2 \right) +\rho^2 d\phi^2\right) ,\nonumber\\
 F_3 \= & d\left[ H_1 \,d\phi \wedge dy_2 \+ T_1 \,dt \wedge dy_1 \right]\,,\label{eq:WeylSol2circle}
\end{align}
where the warp factors $(Z_0,Z_1,W_0,\nu)$ and the gauge potentials $(H_0,H_1,T_1)$ are functions of $\rho$ and $z$. Moreover, we restrict $F_3$ to have two electromagnetic dual contributions
\begin{equation}
dT_1 \wedge dt \wedge dy_1 \= q\,\star_6 \left( dH_1 \wedge d\phi \wedge dy_2 \right) \quad \Rightarrow \quad dT_1 \= \frac{1}{\rho \,Z_1^2}\,\star_2 dH_1\,,
\end{equation}
and $q$ is an overall constant corresponding to the ratio between both charges. This will restrict the sources of the solutions to carry electric and magnetic charges for which the ratio is fixed and given by $q$. This restriction simplifies the form of the Einstein and Maxwell equations. Moreover, the technique of linearization of these equations, that we will review in a moment,  also requires this ``almost'' self-duality of $F_3$. We are therefore anticipating a constraint in our ansatz that is necessary for the linearity of the equations.

The choice of warp factors have been appropriately made for solving the equations of motion: $Z_I$ are sourced by $H_I$ while $W_0$ is governed by the same equations as in vacuum. The equations of motion can be decomposed into three almost-linear layers:
\begin{align}
&\text{\underline{Vacuum layer:}} \quad \cL \log W_0  \= 0\,,\nonumber\\
&\text{\underline{Maxwell layer:}} \quad  \cL \log Z_I \= \- \frac{\gamma_I^{-2}}{\rho\, {Z_I}^2}\,\left[ (\partial_\rho H_I)^2 + (\partial_z H_I)^2 \right] \,,\quad \left(\gamma_0,\gamma_1 \right)\=\left( 1, \sqrt{\frac{2}{1+q^2}}\right),\nonumber\\
&\hspace{2.90cm} \partial_\rho \left( \frac{1}{\rho\,{Z_I}^2}\,\partial_\rho H_I \right)\+\partial_z \left( \frac{1}{\rho\,{Z_I}^2}\,\partial_z H_I\right)  \=0\,,\label{eq:EOMWeyl}\\
&\text{\underline{Base layer:}}  \quad \frac{1}{\rho}\,\partial_z \nu  \=  \cS_z \left(W_0,Z_I,H_I\right)\,, \qquad \frac{1}{\rho}\,\partial_\rho\nu \=  \cS_\rho \left(W_0,Z_I,H_I\right) \nonumber\,,
\end{align}
where $\cL$ is the cylindrical Laplacian \eqref{eq:Laplacian} and $\cS_x$ are source functions that are quadratic in $ \left(W_0,Z_I,H_I\right)$ \eqref{eq:EOMWeylApp}. The equations can be treated in a similar fashion as in vacuum except for the Maxwell layer which is a non-trivial set of non-linear coupled differential equations. In \cite{Bah:2020pdz}, a procedure have been found to extract closed-form solutions. The solutions are determined by three arbitrary axisymmetric functions that solve a cylindrical Laplace equation
\begin{equation}
\cL \left(\log W_0 \right) \=\cL \left(L_0 \right)\=  \cL \left(L_1 \right) \= 0\,.
\end{equation}
The scalars $(Z_I, H_I,T_1)$ are given by
\begin{equation}
Z_I \= \cG^{(I)}_\ell\left(L_I\right) \,,\qquad \star_3 d\left(H_I\,d\phi\right) \= \gamma_I\,dL_I\,, \qquad dT_1 \=  q\,\gamma_1\, \frac{dL_1}{ \cG^{(1)}_\ell(L_1)^2} \,,
\label{eq:WGF&H}
\end{equation}
where $\cG^{(I)}_\ell$ is one of the four generating functions of one variable given by
\begin{align}
\cG_1 (x) &\= \frac{\sinh(a x+b)}{a} \,, \qquad \cG_2 (x) \= i\,\frac{\cosh(a x+b)}{a}\,,\qquad \cG_3 (x) \= x +b, \nonumber \\
\cG_4 (x) &\= \frac{\sin(a x+b)}{a} \,,\qquad \cG_5(x) \=\frac{\cos(a x+b)}{a}\,,\qquad a \in \mathbb{R}_+, \quad b\in \IR \,.
\label{eq:DefFi}
\end{align}
The equations for $\nu$ are integrable and given as
\begin{align}
& \frac{1}{\rho}\partial_z \nu = \frac{1}{2}\, \partial_\rho \log W_0\,\partial_z \log W_0+ \frac{\epsilon_\ell^{(0)}a_0^2}{2}\,\partial_\rho L_0\,\partial_z L_0 + \epsilon_\ell^{(1)}a_1^2\,\partial_\rho L_1\,\partial_z L_1, \nonumber \\
& \frac{1}{\rho}\partial_\rho \nu = \frac{1}{4} \left( \left(\partial_\rho \log W_0\right)^2- \left(\partial_z \log W_0 \right)^2\right)+ \frac{\epsilon_\ell^{(0)}a_0^2}{4} \left( \left( \partial_\rho L_0 \right)^2- \left(\partial_z L_0 \right)^2\right) \label{eq:nuEq}\\
&\hspace{1.3cm} + \frac{\epsilon_\ell^{(1)}a_1^2}{2} \left( \left( \partial_\rho L_1\right)^2- \left(\partial_z L_1 \right)^2\right),\nonumber
\end{align}
where $\epsilon_\ell^{(I)}$ is a constant that depends on the choice of generating functions, $\cG_\ell^{(I)}$, for the pair $(Z_0,H_0)$ and $(Z_1,H_1,T_1)$ in \eqref{eq:WGF&H}
\begin{equation}
\epsilon_\ell \= \begin{cases}
1 \,,\qquad \text{if } \ell=1,2\,,\\
0 \,, \qquad \text{if } \ell=3\,,\\
-1\,, \qquad \text{if } \ell=4,5\,.
\end{cases}
\label{eq:epsilonchoice}
\end{equation}
The integrability of the equations for $\nu$ is guaranteed by the harmonicity of the functions on the right-hand sides. These equations can be integrated in a case-by-case manner depending on the choice of sources for the harmonic functions.

The different choices of generating functions $\cG^{(I)}_\ell$ \eqref{eq:DefFi} for the warp factors $Z_I$ drastically change the solutions. The choices $\ell=4$ and $\ell=5$ seem to lead to unphysical solutions. Indeed, the $\sin$ and $\cos$ has too many zeroes to avoid, and this constrains too much the harmonic functions. The choice $\ell=2$ is imaginary but the metric can be still real by taking $W_0$ imaginary. However, it changes the signature of the metric. We will therefore focus on the two other branches $\ell=1$ and $\ell=3$.

\subsubsection{Rod sources: generic solutions}
\label{sec:RodSolGen}

The harmonic functions $(\log W_0,L_0,L_1)$ are sourced by $n$ distinct rods of size $M_i$ on the $z$-axis as depicted in Fig.\ref{fig:RodsSources}. The construction is similar to the neutral solutions in section \ref{sec:RodSolGenVac}. The warp factors are given according to distance functions $R_\pm^{(i)}$, that are defined in \eqref{eq:Rpmdef}, and weights $(G_i,P_i^{(0)},P_i^{(1)})$ associated to each rod such that
\begin{equation}
\log W_0 = \sum_{i=1}^n G_i \,\log  \frac{R_+^{(i)}}{R_-^{(i)}}\,,\qquad L_0 = \sum_{i=1}^n P^{(0)}_i \,\log  \frac{R_+^{(i)}}{R_-^{(i)}}\,,\qquad L_1 = \sum_{i=1}^n P_i^{(1)} \,\log  \frac{R_+^{(i)}}{R_-^{(i)}}\,.
\end{equation}
We restrict our study to the branch of solutions \eqref{eq:WGF&H} with $\ell=1$. 
The metric and gauge fields are then given by
\begin{align}
&Z_I= \frac{1}{2a_I} \left[e^{b_I} \prod_{i=1}^n \left(  \frac{R_+^{(i)}}{R_-^{(i)}}\right)^{a_I P^{(I)}_i}-e^{-b_I} \prod_{i=1}^n \left(  \frac{R_-^{(i)}}{R_+^{(i)}}\right)^{a_I P^{(I)}_i} \right]\,,\qquad W_0 = \prod_{i=1}^n \left(  \frac{R_+^{(i)}}{R_-^{(i)}}\right)^{G_i} \,,\nonumber\\
&H_1 \= \sqrt{\frac{2}{1+q^2}}\sum_{i=1}^n P^{(1)}_i \left(r_-^{(i)}-r_+^{(i)} \right)\,,\quad T_1\=-q a_1 \, \sqrt{\frac{2}{1+q^2}} \coth \left(\sum_{i=1}^n a_1 P^{(1)}_i \,\log  \frac{R_+^{(i)}}{R_-^{(i)}} +b_1\right),\nonumber\\
&H_0 \= \sum_{i=1}^n P^{(0)}_i \left(r_-^{(i)}-r_+^{(i)} \right) \,,\qquad e^{2\nu} \=\prod_{i,j=1}^n\, \left(  \frac{E_{+-}^{(i,j)}E_{-+}^{(i,j)}}{E_{++}^{(i,j)}E_{--}^{(i,j)}}\right)^{\frac{1}{2}\,\alpha_{ij}} \,,
\label{eq:WarpfactorsRods6d}
\end{align}
where the generating functions $E_{\pm \pm }^{(i,j)}$ have been defined in \eqref{eq:Rpmdef} and the exponents $\alpha_{ij}$ are
\begin{equation}
\alpha_{ij} \equi G_i G_j \+ 2 a_1^2 P^{(1)}_i P_j^{(1)}\+ a_0^2 P^{(0)}_i P_j^{(0)} \,.
\label{eq:alphaijDef}
\end{equation}
The solutions are asymptotic to T$^2\times\IR^{1,3}$ if $Z_I \to 1$, which fixes
\begin{equation}
a_I \= \sinh b_I\,.
\end{equation}
Moreover, we can assume without restriction that $b_I \geq 0$. Finally, one can retrieve the neutral Weyl solutions of section \ref{sec:RodSolGenVac} by taking the limit $(b_I \to \infty,P_i^{(I) }\to0)$ while keeping all $P_i^{(I) }~ \sinh b_I$ constant.

\subsubsection{Regularity}

As in vacuum, the regularity on the $z$-axis fixes the weights, $(G_i,P_i^{(0)},P_i^{(1)})$. There will be three types of physical rod sources that have the same nature as the three spherically symmetric solutions detailed in section \ref{sec:sphericallySymCharged}. An exhaustive analysis is given in the appendix \ref{App:Weylcharged}. The Fig.\ref{fig:RodCategories} and Fig.\ref{fig:TouchingRods} used for neutral Weyl solutions also illustrate appropriately the physics of charged solutions with non-trivial electromagnetic fluxes wrapping the S$^2$ and a choice of a circle from the T$^2$. The exponents $\alpha_{ij}$ will simplify for the physical sources and it will be convenient to use the aspect ratios $d_i$ that we remind to be
\begin{equation}
\begin{split}
\alpha_{jk}& \= \begin{cases} 
1 \qquad &\text{if the }j^\text{th}\text{ and }k^\text{th}\text{ rods are of same species.}  \\
\frac{1}{2} \qquad &\text{otherwise,}
\end{cases} \\
d_1 &\equi 1\,,\qquad d_i \equi  \prod_{j=1}^{i-1} \prod_{k=i}^n \left(\dfrac{(z_k^- - z_j^+)(z_k^+ - z_j^-)}{(z_k^+ - z_j^+)(z_k^- - z_j^-)}  \right)^{\alpha_{jk}}\quad \text{when } i=2,\ldots n\,.
\end{split}
\label{eq:dialphaDefcharged}
\end{equation}
The regularity at each rod requires that the sources are in one of these three categories:
\begin{itemize}
\item[•] \underline{\bf Black rod:} The $i^\text{th}$ rod corresponds to the horizon of a static three-charge black brane with a T$^2\times$S$^2$ horizon topology or black string with a S$^1\times$S$^3$ topology\footnote{\label{footnote3}The local topology depends on whether the rod is connected to other rods, as for the Weyl neutral solutions constructed in section \ref{sec:RegVac}.} providing that 
\begin{equation}
P^{(0)}_i \= \frac{1}{2\,\sinh b_0} \,,\qquad P^{(1)}_i \= \frac{1}{2\,\sinh b_1} \,,\qquad G_i \= - \frac{1}{2}\,.
\label{eq:BlackBraneGicharged}
\end{equation} 
The surface gravity and horizon area associated to this black brane is
\begin{align}
A_i &=  \frac{(2\pi)^3}{\kappa_i}\,M_i\,R_{y_1}R_{y_2}, \label{eq:surfaceGrav&Area} \\ \kappa_i &\equi \frac{\sinh b_1}{e^{b_1}}\,\sqrt{\frac{2\sinh b_0}{e^{b_0}}}\frac{1}{d_i\,M_i}\, \prod_{j\neq i} \left(\frac{z_j^+ - z_i^-}{z_j^- -z_i^-} \right)^{\text{sign}(i-j)\,\alpha_{ij}} \nonumber
\end{align}
where we remind that $M_i$ is the size of the rod and $z_j^\pm$ are the rod endpoints \eqref{eq:coordinatesEndpoints}.
\item[•] \underline{\bf Species-1 bubble rod:} The $i^\text{th}$ rod corresponds to a static three-charge bubble where the $y_1$-circle shrinks providing that
\begin{equation}
P^{(0)}_i \= \frac{1}{2\,\sinh b_0} \,,\qquad P^{(1)}_i \= \frac{1}{2\,\sinh b_1} \,,\qquad G_i \=  \frac{1}{2}\,.
\label{eq:1TSGi}
\end{equation} 
At the rod, $\rho= 0$ and $z_i^- \leq z \leq z_i^+$, the orbits of the spacelike Killing vector $\partial_{y_1}$ shrink and the local geometry defines the locus of a bubble with a S$^1\times$S$^2$ or S$^3$ topology\footnoteref{footnote3} at the origin of a $\IR^2$ space. The subspace $(\rho,y_1)$ defines the $\IR^2$ with a potential conical defect of order $k_i\in \mathbb{N}$ if the following bubble equation relating the internal parameters to the radius of the extra dimension is satisfied:
\begin{equation}
k_i\,R_{y_1} \=\frac{e^{b_1}}{\sinh b_1}\sqrt{\frac{e^{b_0}}{2\sinh b_0}}\,d_i\,M_i\, \prod_{j\neq i} \left(\frac{z_j^+ - z_i^-}{z_j^- -z_i^-} \right)^{\text{sign}(j-i)\,\alpha_{ij}}\,.
\label{eq:RegRy1}
\end{equation}
\item[•] \underline{\bf Species-2 bubble rod:} The $i^\text{th}$ rod corresponds to a static one-charge bubble where the $y_2$-circle shrinks providing that
\begin{equation}
P^{(0)}_i \= \frac{1}{\sinh b_0} \,,\qquad P^{(1)}_i \= G_i \=  0\,.
\label{eq:2TSGi}
\end{equation} 
The local geometry defines the locus of a bubble with a S$^1\times$S$^2$ or S$^3$ topology\footnoteref{footnote3} at the origin of a $\IR^2$ space with a potential conical defect of order $k_i\in \mathbb{N}$ if
\begin{equation}
k_i\,R_{y_2} \=\frac{e^{b_0}}{\sinh b_0}\,d_i\,M_i\, \prod_{j\neq i} \left(\frac{z_j^+ - z_i^-}{z_j^- -z_i^-} \right)^{\text{sign}(j-i)\,\alpha_{ij}}\,.
\label{eq:RegRy2}
\end{equation}
Because $P_i^{(1)} =0$, the field strength $F_3$ is not charged at the rod and the bubble carries a magnetic charge in the KK gauge field, $H_0 \,d\phi$, only.

\end{itemize}

Out of the rods, the $\phi$-circle shrinks on the $z$-axis. The regularity analysis is identical to the one led for neutral solutions in section \ref{sec:RegVac}. The semi-infinite segments above and below the rod configuration are free from conical singularities while any segments that separate two rods are struts and have a conical excess of order $d_i<1$ given in \eqref{eq:dialphaDefcharged} (as depicted in Fig.\ref{fig:RodCategories}). If two rods of different nature touch, the strut that separates them disappears and the intersection is free from conical excess (as depicted in  Fig.\ref{fig:TouchingRods}). For instance, the intersection between two touching species-1 and species-2 charged bubbles is free from conical excess and the local topology correspond to a S$^\phi$ at an origin of an $\IR^4$ with the same potential orbifold defects as at each bubble.

As for neutral solutions, if we consider solutions sourced by bubble rods, the regularity conditions \eqref{eq:RegRy1} and \eqref{eq:RegRy2} correspond to bubble equations that strongly constrain the geometry of the bubbles. Moreover, note that the effect of the gauge fields on these equations can be absorbed by considering the rescaled extra-dimensional radii
\begin{equation}
\widetilde{R}_{y_1} \equi \frac{2 \sinh b_1}{e^{b_1}} \sqrt{\frac{2 \sinh b_0}{e^{b_0}} }\,R_{y_1}\,,\qquad \widetilde{R}_{y_2} \equi \frac{2 \sinh b_0}{e^{b_0}}\,R_{y_2}\,,
\label{eq:tildeRy}
\end{equation}
and the bubble equations are identical to the neutral ones \eqref{eq:RegRy1Vac} and \eqref{eq:RegRy2Vac}. Therefore, the gauge fields do not affect the bubble equations but rather act as a flux decoration on top of the neutral backgrounds.

\subsubsection{Profile in four dimensions and conserved charges}

In the four-dimensional framework obtained by compactification along $y_1$ and $y_2$ \eqref{eq:4dFrameworkcharged}, the solutions correspond to asymptotically-flat three-charge static solutions with non-trivial scalar profiles given by
\begin{equation}
\begin{split}
ds_4^2 &\=  - \frac{ \sqrt{W_0}\,dt^2}{Z_1 \sqrt{ Z_0}}  \+ \frac{Z_1 \sqrt{ Z_0}}{ \sqrt{W_0}} \,\left( e^{2\nu} \left(d\rho^2 + dz^2 \right) +\rho^2 d\phi^2\right)\,,\\
e^{\frac{1}{\sqrt{3}}\,\Phi_1} &\=\sqrt{W_0} {Z_1}^\frac{1}{3}{Z_0}^\frac{1}{6} \,,\qquad e^{\frac{1}{\sqrt{6}}\Phi_2}  \= \left( \frac{Z_0}{Z_1}\right)^\frac{1}{3}\,, \\
 F^{(m0)}& \= dH_0 \wedge d\phi\,,\qquad  F^{(m1)} \= dH_1 \wedge d\phi \,,\qquad  F^{(e1)}\= dT_1 \wedge dt\,.
\label{eq:4dFrameworkchargedbis}
\end{split}
\end{equation}
We use the generic expressions, \eqref{eq:AsymptoticExpGen} and \eqref{eq:conservedchargesGen} and the asymptotic spherical coordinates, $ 
\rho \equi r \, \sin \theta \,,\,\, z \equi r \, \cos \theta$, to compute the conserved charges:
\begin{align}
\cM &\= \frac{1}{4 G_4}\,\sum_i \left(-G_i+ P_i^{(0)}\,\cosh b_0 + 2P_i^{(1)}\,\cosh b_1 \right) M_i\,,\label{eq:conservedchargesrod}\\
 \cQ_e^{(1)}&\=q\cQ_m^{(1)} \=  \frac{q}{\sqrt{16 \pi G_4}}\sqrt{\frac{2}{1+q^2}} \sum_i M_i \,P_i^{(1)}\,, \qquad \cQ_m^{(0)}  \=  \frac{1}{\sqrt{16 \pi G_4}} \sum_i M_i \,P_i^{(0)}\,. \nonumber
\end{align}

As for neutral solutions, a black rod \eqref{eq:BlackBraneGicharged} will contribute twice more to the four-dimensional mass than a bubble with the same rod length \eqref{eq:1TSGi} and \eqref{eq:2TSGi}. However, this does not mean that the bubble is twice less compact than a black brane as the physical size in six dimensions can be very different from the rod lengths $M_i$.

Note that the five-dimensional Weyl solutions discussed in section \ref{sec:5dparadigm} and in the appendix \ref{App:Weyl5d} are a subclass of this family of solutions. Assuming 
$$Z_1 = Z_0\quad \Leftrightarrow \quad H_1 \= \sqrt{\frac{2}{1+q^2}} H_0 \quad \Leftrightarrow \quad P_i^{(0)} \= P_i^{(1)}\,,\,\, b_1 \= b_0 \,,$$
the condition \eqref{eq:cond5din6d} is satisfied and the solutions solve Einstein-Maxwell equations of the five-dimensional action \eqref{eq:ActionGen5d} with $ F^{(m)} \= \sqrt{\frac{3+q^2}{1+q^2}} \,F^{(m0)}\,.$

\subsection{Smooth bubbling Weyl solutions}
\label{sec:SmoothBubbleCharged}

In this section, we apply the generic construction of six-dimensional charged Weyl solutions to an explicit strut-free bubbling geometry. We will consider the six-dimensional counterpart of the five-dimensional two-bubble solutions with a strut constructed in section \ref{sec:twobubbles5d}, and the strut will be replaced by a species-2 bubble. This example can also be seen as the charged version of the three-bubble neutral solutions of section \ref{sec:smoothchainVac}. The physics will then be similar and we will discuss the effect of the electromagnetic fluxes.

\subsubsection{Three connected bubbles}
\label{sec:threeconnected}

We consider two identical species-1 bubble rods of size $M$ with a species-2 bubble rod of size $\delta$ in between. Moreover, we allow the three bubbles to have the same conical defects parametrized by the orbifold parameter $k\in \mathbb{N}$. Since the topology is identical to the neutral three-bubble solutions in section \ref{sec:threeconnectedVac}, we refer to Fig.\ref{fig:ThreeTouchingRods} for a schematic description of the topology. The only difference is that the circles are now wrapped by fluxes. We use the spherical coordinates centered on the species-2 bubble, $
\rho \equi \sqrt{r(r-\delta)}\,\sin \theta\,,$ $z \equi \left(r-\frac{\delta}{2} \right) \,\cos \theta\,,
$
with $r\geq \delta$ and $ 0\leq \theta \leq \pi$. The six-dimensional metric and gauge fields \eqref{eq:WeylSol2circle} can be decomposed into
\begin{equation}
\begin{split}
&ds_6^2  \= \frac{1}{\bar{Z}_1} \left[- dt^2 \+ U_1\, dy_1^2 \right] \+ U_2\, \frac{\bar{Z}_1}{\bar{Z}_0}\,(dy_2+H_0\,d\phi)^2  \\
&\hspace{1.05cm} \+ \frac{\bar{Z}_0\bar{Z}_1}{U_1 U_2}\, \left[e^{2\bar{\nu}}\,\left(dr^2+r(r-\delta)d\theta^2 \right) +r(r-\delta)\sin^2\theta\, d\phi^2\right], \\
&F_3 \=dH_1 \wedge d\phi\wedge dy_2 \+  dT_1  \,\wedge\, dt \,\wedge\, dy_1 \,,
\end{split}
\end{equation}
We obtain from \eqref{eq:WarpfactorsRods6d}
\begin{align}
U_1 \=&\left( 1- \frac{M}{r_+}\right)\left( 1- \frac{M}{r_-}\right)\,,\qquad U_2 \=  \left(1 \- \frac{\delta}{r} \right) \,,  \qquad \bar{Z}_0 \= \frac{e^{b_0} -e^{-b_0} \,U_1 \,{U_2}^2}{2 \,\sinh b_0}  \,, \nonumber \\
 \bar{Z}_1 \= &\frac{e^{b_1} -e^{-b_1} \,U_1 }{2 \,\sinh b_1}  \,,\qquad H_0 \= \frac{1}{\sinh b_0}\,\left( r_- - r_+\right)\,,\nonumber \\
 H_1 \=&\sqrt{\frac{2}{1+q^2}}\,\frac{r_--r_+-\delta\,\cos \theta}{\sinh b_1}\,,\qquad T_1 \= -q \,\sinh b_1\,\sqrt{\frac{2}{1+q^2}}\,\frac{e^{b_1} + e^{-b_1} \,U_1 }{e^{b_1} - e^{-b_1} \,U_1}\,,\nonumber\\
e^{2\bar{\nu}} \=& \frac{1}{2} \left(1 +\frac{(M+r)(r-\delta-M)-\frac{\delta^2}{4}\sin^2\theta}{\left(2 r_- - (M+r)+\delta \sin^2\frac{\theta}{2} \right)\left(2 r_+ - (M+r)+\delta \cos^2\frac{\theta}{2} \right)} \right)\label{eq:warpfactors3bubblecharged} \\
& \times \sqrt{ \frac{ \left(r-\delta \sin^2\frac{\theta}{2}\right) r_- - M (r-\delta) \sin^2\frac{\theta}{2}}{ \left(r-\delta \cos^2\frac{\theta}{2}\right) r_- - \left(M  r +\delta^2 \cos^2 \frac{\theta}{2}\right) \sin^2\frac{\theta}{2}}} \nonumber\\
& \times \sqrt{ \frac{\left(r-\delta \cos^2\frac{\theta}{2}\right) r_+ - M (r-\delta) \cos^2\frac{\theta}{2} }{ \left(r-\delta \sin^2\frac{\theta}{2}\right) r_+ - \left(M  r +\delta^2 \sin^2 \frac{\theta}{2}\right) \cos^2\frac{\theta}{2}  }} \,, \nonumber
\end{align}
where the distances to the upper and lower species-1 bubbles, $r_\pm$, have been defined in \eqref{eq:distanceupperlower}. Note that the warp factor $e^{2\bar{\nu}}$ is identical to the neutral solutions \eqref{eq:WarpFact3BubblesVac}, and thus the three-dimensional bases are identical. This was far from certain as the equations of motion for $\nu$ are very different with gauge fields \eqref{eq:EOMWeylApp} than in vacuum \eqref{eq:EQforUI}. An important lesson is that the regularity conditions constrain the effect of gauge fields to a flux decoration on the top of a topology dictates by the base.

The $y_2$-circle degenerates at $r=\delta$ defining a S$^{\phi}\times$S$^2$ bubble wrapped by fluxes while the $y_1$-circle degenerates at $r_\pm=M$ defining the loci of  two S$^3$ bubbles. The bubble equations, obtained from the regularity at each bubble \eqref{eq:RegRy1} and \eqref{eq:RegRy2}, are identical to the neutral ones by considering the rescaled radii \eqref{eq:tildeRy}. Using \eqref{eq:M&deltainverse}, we find 
\begin{equation}
M \= \frac{1}{2} \,\frac{k\,\widetilde{R}_{y_1}^2}{\sqrt{4\widetilde{R}_{y_1}^2 +\widetilde{R}_{y_2}^2}}\,,\qquad  \delta \= \frac{k\,\widetilde{R}_{y_2}}{4} \,\left( \frac{\widetilde{R}_{y_2}}{\sqrt{4\widetilde{R}_{y_1}^2 +\widetilde{R}_{y_2}^2}}+1\right)\,.
\label{eq:rodsizes3bubbles}
\end{equation}
Since $\widetilde{R}_{y_a} \leq R_{y_a}$, the gauge field parameters $b_I$ makes the rod lengths smaller compared to the three neutral bubbles. We also note the physical sizes of the bubbles in six dimensions (see \eqref{eq:AreaBubbleAppcharged}) given as
\begin{equation}
A_\text{B-1} \= 4\pi^2 k\,\sqrt{\frac{e^{b_1}}{2\sinh b_1}}\,M \,R_{y_1} R_{y_2}\,,\qquad A_\text{B-2} \= 4\pi^2 k\, \sqrt{\frac{e^{b_1}}{2\sinh b_1}} \delta\, R_{y_1} R_{y_2}\,.
\end{equation}
We observe that the structure reaches its maximum size in the neutral limit $b_I \to \infty$ and shrinks when $b_I \to 0$.\footnote{Even if the coefficient $\frac{e^{b_1}}{2\sinh b_1}$ is greater than $1$, so greater than the neutral limit $b_I \to \infty$, the scaling of $M$ and $\delta$ in terms of $b_I$ still makes the area of the bubbles smaller than their neutral values.} The latter limit will be specified in the next section.  The overall effect of the charges is to squeeze the bubbles to smaller sizes.  This is consistent with the expectation that the fluxes provide a counterbalance to the expansion of the bubbles.  

It can be verified through this example that the solutions are regular outside the bubble loci as are the gauge fields. We refer the reader to the generic regularity analysis conducted in the appendix \ref{App:Regzaxischarged}.

We have then constructed smooth horizonless three-charge solutions without struts.\footnote{As for neutral solutions, the solutions can be considered smooth even with conical defects at the bubbles, $k>1$.} In the four-dimensional framework \eqref{eq:4dFrameworkcharged}, they correspond to Einstein-Maxwell-dilaton solutions with conserved charges \eqref{eq:conservedchargesrod}
\begin{equation}
\begin{split}
\cM &\= \frac{1}{4G_4}\,\left[(M+\delta) \coth b_0\+ M \left( 2\coth b_1 -1\right)\right]\,,\\
 \cQ^{(1)}_e &\= q\,\cQ^{(1)}_m \= \frac{q\,M}{2\sqrt{2\pi G_4 \left(1+q^2 \right)}\,\sinh b_1}\,,\qquad \cQ_m^{(0)}\= \frac{M+\delta}{4\sqrt{\pi G_4}\,\sinh b_0}\,.
 \label{eq:conservedcharges3bubbles}
\end{split}
\end{equation}

By expressing the mass and charges according to the extra-dimension radii with \eqref{eq:rodsizes3bubbles}, one can check that they are bounded by the sum of the radii in the absence of conical defects at the bubbles, $k=1$, as expected. The conical defects are then necessary for the solutions to be macroscopic.

The strut of the two-bubble solution in five dimensions, constructed in section \ref{sec:twobubbles5d}, has been classically replaced by a third bubble where an extra dimension degenerates. 

Moreover, unlike the smooth neutral bubbling solutions of section \ref{sec:threeconnectedVac}, it is most likely that the present solutions are stable. Indeed, it has been argued that single KK bubbles can be quantumly stabilized  by electromagnetic fluxes \cite{Stotyn:2011tv}. Although we have more than one bubble in our configuration and performing a similar analysis to that of \cite{Stotyn:2011tv} is a challenge with our solutions, the main ingredients are charged KK bubbles that are stable by themselves. Therefore, we do expect that the charged bubbling Weyl solutions are stable.

One might wonder how the strut in five dimensions can be replaced by a stable bubble in six dimensions. Indeed, our argument for neutral solutions was that the inherent instability of the middle KK bubble gives the same pressure as a strut to push the two outer bubbles away. If the bubbles are now stable, one may wonder how the energy of the strut can be replaced by a stable charged bubble. However, even if the charged bubble is stable by itself, it can still induce pressure if it is squeezed to a size smaller than its equilibrium configuration. To make this statement clear, we first isolate the contribution of the species-2 bubble in the three bubble configuration. It carries the following charges
\begin{equation}
q_m^{(0)} \= \frac{\delta}{4\sqrt{\pi G_4}\,\sinh b_0}\,,\qquad q_e^{(1)} \= q_m^{(1)} \=0\,,
\end{equation}
and $\delta$ is expressed in terms of $b_0$ and $R_{y_2}$ in \eqref{eq:rodsizes3bubbles}. If the bubble was alone, that is if we consider a configuration with a single species-2 bubble of size $\bar{\delta}$, the regularity condition \eqref{eq:RegRy2} and the charges would be
\begin{equation}
q_m^{(0)} \= \frac{\bar{\delta}}{4\sqrt{\pi G_4}\,\sinh \bar{b}_0}\,,\qquad q_e^{(1)} \= q_m^{(1)} \=0\,, \qquad \bar{\delta} \= \frac{k\, \sinh \bar{b}_0\,R_{y_2}}{e^{\bar{b}_0}}\,.
\end{equation}
By fixing $\bar{b}_0$ to have the same charges, our middle bubble has then the same charges as a single bubble of size
\begin{equation}
\bar{\delta} \= \delta \,\frac{\sinh \left( b_0 - \log \left[ \frac{1}{2}\left(\frac{\widetilde{R}_{y_2}}{\sqrt{4\widetilde{R}_{y_1}^2 +\widetilde{R}_{y_2}^2}}+1\right) \right]\right)}{\sinh b_0}
\end{equation}
It is clear from the arguments of the $\sinh$ that $\bar{\delta}>\delta$. As a result, the middle bubble is compressed more in the presence of the other two bubbles than if it were alone. This results in a pressure that pushes the two outer bubbles away from each other. Moreover, note that $\bar{\delta} \sim \delta$ when $\widetilde{R}_{y_2} \gg \widetilde{R}_{y_1}$, and the pressure from the middle bubble should be then almost zero. However, in this limit $\delta  \gg M$ \eqref{eq:rodsizes3bubbles} and therefore the outer bubbles are very far from each other. This matches the strut picture where the binding energy goes to zero when the bubbles are infinitely separated \eqref{eq:StrutEnergy}.

In conclusion, a smooth bubble can replace a strut and the artificial binding energy induced by a strut is smoothly substituted by the reluctance of a KK bubble to be compressed and its nature to grow. This is a completely new approach to constructing objects that repel each other without imposing BPS conditions that fix ``Mass=Charge''. In the present construction, objects are repelled by pure topology. Therefore, Weyl's formalism gives the tools to linearly construct smooth bubbling geometries in a non-BPS regime. The solutions require some charges to guarantee their quantum stability.

\subsubsection{A BPS limit}

We have already highlighted that taking $b_I \to \infty$ makes the charges \eqref{eq:conservedcharges3bubbles} to vanish, and one can check more precisely that the solutions \eqref{eq:warpfactors3bubblecharged} indeed converges to the neutral three-bubble solutions given in \eqref{eq:WarpFact3BubblesVac}. An other interesting limit is in the other side of the parameter space, by taking $b_I \to 0$. We consider the limit $b_0 \to 0$ with $b_1 = \beta \,{b_0}^x$, where $x$ and $\beta$ are constant. The three-bubble solutions converge to
\begin{align}
ds_6^2  = &- dt^2 +  dy_1^2+  \frac{1}{1+\frac{Q_0}{r}}\,\left(dy_2+ Q_0 \cos \theta\,d\phi\right)^2 \\
&+\left( 1+\frac{Q_0}{r} \right)\,\left[dr^2+r^2 \left( d\theta^2+\sin^2 \theta \,d\phi^2 \right)\right], \qquad F_3 =0 \,,
\label{eq:met&GFbubbling}
\end{align}
where 
\begin{equation}
Q_0 \= \begin{cases}
k \,R_{y_2}\,, \qquad \text{if } x> \frac{1}{2}\,,\\
\frac{k}{4} \,\sqrt{8 \beta^2 R_{y_1}^2+R_{y_2}^2 } \, \left(1+ \frac{R_{y_2}}{\sqrt{8 \beta^2 R_{y_1}^2+R_{y_2}^2 }} \right)^2\,, \qquad \text{if } x= \frac{1}{2}\,,\\
\cO\left(b_0^{x-\frac{1}{2}}\right) \,, \qquad \text{if } x< \frac{1}{2}\,,
\end{cases}\,.
\end{equation}
We recognize a Taub-NUT space in six dimensions with charge $Q_0$ if $x\geq \frac{1}{2}$, while $Q_0$ diverges and the limit is singular if $x<\frac{1}{2}$. The former corresponds to a BPS solution that is asymptotic to $\IR^{1,3}\times$T$^2$ and the space caps off smoothly as a S$^1\times \IR^4/\mathbb{Z}_{Q_0/R_{y_2}}$. 

One can reverse the argument and interpret the parameters $b_I$ as gauge field parameters that blow up a Taub-NUT center into non-trivial bubbling geometries. It allows one to interpolate from the BPS to the non-BPS regime while keeping the solutions smooth. This is a non-trivial process in the phase space of Einstein solutions. The standard lore for constructing smooth solutions with multiple objects is that one needs strong gauge field potentials to compensate for the gravitational attraction between them. For our present construction, this is not the case as the solutions are supported by topology, and the gauge fields do not play the role of supporting individual bubbles.  The structure is non-collapsing even in the limit where we turn off the gauge fields $(b_I \to \infty)$.  Their main role is to support the overall topological structure and prevent them from expanding.  This further contrast with the standard lore where gauge fields are used to prevent BPS bubbling geometry from collapsing.  Our work shows that one can construct ``floating'' objects from pure topology in classical gravity theories without requiring the ``Mass=Charge'' that has made BPS bubbling geometries so successful \cite{Bena:2004de,Bena:2006kb,Bena:2007qc,Bena:2007kg,Heidmann:2017cxt,Bena:2017fvm}.

\section{Embedding in type IIB}
\label{sec:STembedding}

Until now, we have taken on purpose a bottom-up approach to highlight that smooth bubbling geometries can be constructed from classical theories of gravity without the help of supersymmetry. However, our six-dimensional action \eqref{eq:Action6d} arises as a specific truncation of type IIB string theory on a four-dimensional torus \cite{Chow:2014cca}. In this section, we describe the Weyl solutions from a type IIB supergravity perspective.

\subsection{D1-D5-KKm Weyl ansatz}

The action \eqref{eq:Action6d} can be seen as the minimal pure $\cN = (2,0)$ six-dimensional supergravity with the extra assumption that $F_3$ is self-dual, $\star_6 F_3 = F_3$. This theory arises as a consistent truncation of type IIB supergravity on T$^4$ where the non-trivial ten-dimensional fields are the metric and the Ramond-Ramond two-form $C_2$ only, $F_3 = dC_2$. 

The ansatz for axisymmetric Weyl solutions \eqref{eq:WeylSol2circle} translates to
\begin{align}
ds_{10}^2 \= &\,\frac{1}{Z_1}\, \left[- W_0\,dt^2 + \frac{ dy_1^2}{W_0} \right] \+ \frac{Z_1}{Z_0}\, \left(dy_2 +H_0 \,d\phi\right)^2 \\
&\+ Z_0 Z_1\,\left( e^{2\nu} \left(d\rho^2 + dz^2 \right) +\rho^2 d\phi^2\right) \+ ds(T^4)^2,\nonumber\\
 F_3 \= & \,d\left[ H_1 \,d\phi \wedge dy_2 \+ T_1 \,dt \wedge dy_1 \right]\,,\label{eq:WeylSol2circle}
\end{align}
The equations of motion and the solutions we found in section \ref{sec:Weyl} are then identical with the extra assumption that $q=1$ in order for $F_3$ to  be self-dual. Weyl solutions are then determined by three arbitrary functions that solve a cylindrical Laplace equation
\begin{equation}
\cL \left(\log W_0 \right) \=\cL \left(L_0 \right)\=  \cL \left(L_1 \right) \= 0\,, \qquad \cL \equi  \frac{1}{\rho} \,\partial_\rho \left( \rho \,\partial_\rho \right) + \partial_z^2
\end{equation}
The scalars $(Z_I, H_I,T_1)$ are given by
\begin{equation}
Z_I \= \cG^{(I)}_\ell\left(L_I\right) \,,\qquad \star_3 d\left(H_I\,d\phi\right) \=\,dL_I\,, \qquad dT_1 \=   \frac{dL_1}{ \cG^{(1)}_\ell(L_1)^2} \,,
\label{eq:WGF&HtypeIIB}
\end{equation}
where $\cG^{(I)}_\ell$ is one of the four generating functions of one variable given in \eqref{eq:DefFi}. The equations for $\nu$ are still given by \eqref{eq:nuEq}.

The harmonic functions $(\log W_0,L_0,L_1)$ can be freely sourced by point or rod sources on the $z$-axis. The sources for $L_0$ induce Kaluza-Klein monopoles (KKm) along the $y_2$-fiber. The sources for $L_1$ induce equal electric and magnetic charges in $F_3$. From the form of $F_3$, the magnetic charges correspond to D5-branes wrapping the T$^4$ and the $y_1$-circle, while the electric charges correspond to D1-branes along the $y_1$-circle. As for the sources of $\log W_0$, they correspond to pure vacuum sources that modify the topology of the solutions. To conclude, our Weyl solutions correspond to axisymmetric static D1-D5-KKm solutions when embedded in type IIB.

\subsection{A BPS limit}

So far in this paper, we have restricted our analysis to the branch $\ell=1$ for $\cG^{(I)}_\ell$ \eqref{eq:DefFi} and to rod sources. In this section, we investigate the branch $\ell=3$, $\cG^{(I)}_3(x)=x+b_I$. For such a choice, the $Z_I$ are also harmonic functions, and the equations of motion simplify to
\begin{equation}
\cL \left(\log W_0 \right) \=\cL \left(Z_0 \right)\=  \cL \left(Z_1 \right) \= 0\,, \qquad \star_3 d\left(H_I\,d\phi\right) \=\,dZ_I\,, \qquad T_1 \=  - \frac{1}{Z_1}\,.
\end{equation}
We recognize partially the BPS equations of motion for axisymmetric D1-D5-KKm systems. More precisely, one can show that the sources leading to physical solutions correspond to point particles and require that we fix $W_0=1$. This also fixes $\nu=0$ \eqref{eq:nuEq}.  This choice for $\cG^{(I)}_\ell$ then corresponds to the ansatz for axisymmetric  static BPS multicenter solutions with D1-D5-KKm charges on a flat three-dimensional base \cite{Bena:2007kg}.

Note that in this limit, the electric potential of the D1-branes $T_1$ is proportional to the induced gravitational potential $Z_1^{-1}$, and a similar argument can be made for the D5-branes and KKm charges. This is a specific property of the BPS system that has allowed the construction of non-collapsing solutions with multiple sources via the floating-brane ansatz \cite{Bena:2004de,Bena:2007kg}. Moreover, it was also thought that having $T_I= Z_I^{-1}$ should remain to have a valid non-BPS floating-brane ansatz \cite{Bena:2009fi}. Our specific Weyl ansatz contradicts this assertion; $T_1$ can be very different from $Z_1^{-1}$ for the other choices of generating functions $\cG_\ell^{(1)}$ \eqref{eq:WGF&HtypeIIB}. As we have seen in previous sections, Weyl solutions can always have multiple sources that do not collapse under their own gravitational attraction with the gauge fields playing the role of stabilizing the overall structure. We will examine this class of solutions and their physics from a type IIB perspective in the next section.

\subsection{D1-D5-KKm Weyl solutions}

Generic D1-D5-KKm Weyl solutions can be derived from the six-dimensional Weyl solutions  of section \ref{sec:RodSolGen} sourced by $n$ rods on the $z$-axis with $q=1$, that is 
\begin{align}
&Z_I= \frac{1}{2\sinh b_I} \left[e^{b_I} \prod_{i=1}^n \left(  \frac{R_+^{(i)}}{R_-^{(i)}}\right)^{\sinh b_I P^{(I)}_i}-e^{-b_I} \prod_{i=1}^n \left(  \frac{R_-^{(i)}}{R_+^{(i)}}\right)^{\sinh b_I P^{(I)}_i} \right] \,,\label{eq:WarpfactorsRods6dtypeIIB}\\
& W_0 = \prod_{i=1}^n \left(  \frac{R_+^{(i)}}{R_-^{(i)}}\right)^{G_i} \,,\qquad e^{2\nu} \=\prod_{i,j=1}^n\, \left(  \frac{E_{+-}^{(i,j)}E_{-+}^{(i,j)}}{E_{++}^{(i,j)}E_{--}^{(i,j)}}\right)^{\frac{1}{2}\,\left(G_i G_j + 2 \sinh^2 b_1\,P^{(1)}_i P_j^{(1)}+  \sinh^2 b_0\, P^{(0)}_i P_j^{(0)} \right)} \,\nonumber\\
&H_I \= \sum_{i=1}^n P^{(I)}_i \left(r_-^{(i)}-r_+^{(i)} \right)\,,\quad T_1\=-\sinh b_1 \,  \coth \left(\sum_{i=1}^n \sinh b_1 P^{(1)}_i \,\log  \frac{R_+^{(i)}}{R_-^{(i)}} +b_1\right).\nonumber
\end{align}
where the $(\rho,z)$-dependent functions $R_\pm^{(i)}$ and $E_{\pm\pm}^{(i,j)}$ are defined in \eqref{eq:Rpmdef}, and $b_I \geq 0$. The weights $(P_i^{(0)},P_i^{(1)},G_i)$ are fixed such that the $i^\text{th}$ rod corresponds to either the horizon of a D1-D5-KKm black hole (if $ \sinh b_I P_i^{(I)} =-G_i = \frac{1}{2} $), or the locus of a D1-D5-KKm bubble where the $y_1$ circle shrinks (if $ \sinh b_I P_i^{(I)} = G_i = \frac{1}{2} $) or the locus of a KKm bubble where the $y_2$ circle shrinks (if $P_i^{(1)} = G_i = 0 $ and $\sinh b_0 P_i^{(0)} =1$).

As said earlier, $T_1$ is not proportional to $Z_1^{-1}$ for generic Weyl solutions. More precisely,  we have
\begin{equation}
T_1 \= - \frac{1}{Z_1} \,\sqrt{1+ \sinh^2 b_1 \, {Z_1}^2}\,.
\end{equation}
It is more difficult to read the gravitational potential induced by the D1-branes than in the BPS regime since part of it should come from $e^{2\nu}$. One would a priori have to do a D1-probe calculation to extract this quantity. However, we know for sure that the electric potential does not compensate the gravitational attraction. Indeed, we have already highlighted a neutral limit $b_I\to \infty$ where the solutions become neutral without collapsing. One can therefore ask what are the mechanisms that allow the brane sources to stay away from each other in the Weyl construction. There are essentially two, which have been the main topics of this paper:
\begin{itemize}
\item[•] If the brane sources are separated from each other (as in Fig.\ref{fig:RodCategories}), the solutions develop struts between them, i.e. strings with a negative tension. As discussed in the section \ref{sec:strutGen} and in \cite{Costa:2000kf}, the strut compensates for the repulsion deficit between the sources. This can be rightly seen as an artificial mechanism to avoid solutions to collapse.  It is not well understood what UV degrees of freedom for string theory can account for struts.  One possible set of objects can be O-planes, however this has yet to be studied in these cases.  The configurations are still physically interesting to describe as they capture aspects of the interaction between non-trivial string theory objects. For example, one can study the interaction between two non-extremal static D1-D5-KKm black holes by considering a configuration of two separate black rods and analyzing the strut that emerges between them in the manner of \cite{Costa:2000kf}.
\item[•] If the brane sources are touching (as in Fig.\ref{fig:TouchingRods} or Fig.\ref{fig:ThreeTouchingRods}), the solutions have no struts and are not collapsing. For such configuration, the sources balance each other from their desire to expand. One can therefore study and construct D1-D5-KKm black holes supported by D1-D5-KKm bubbles or KKm bubbles in the manner of \cite{Elvang:2002br}, or chains of D1-D5-KKm bubbles and KKm bubbles, which we have done in section \ref{sec:SmoothBubbleCharged}. For the latter, we have already argued that the solutions are supported by pure topology and stabilized by electromagnetic fluxes. Our interpretation is that Kaluza-Klein bubbles are tempted to expand \cite{Witten:1981gj}. Therefore, even in the presence of electromagnetic fluxes that stabilize them, each bubble in the chain is further compressed from its stable radius and exert sufficient pressure to prevent the collapse of the whole structure. It would be interesting in future projects to analyse how the pressure of the bubbles, their gravitational and electromagnetic interactions prevent the whole structure from collapsing or expanding.
\end{itemize}

The smooth bubbling Weyl solutions constructed in section \ref{sec:SmoothBubbleCharged} correspond to the first large family of non-BPS bubbling geometries that have the same conserved charges as non-extremal static D1-D5-KKm black holes. The three-bubble examples constructed can be easily embedded in type IIB from \eqref{eq:warpfactors3bubblecharged} by taking $q=1$. The conserved charges are given in \eqref{eq:conservedcharges3bubbles}, where the D1 and D5 charges are $\sqrt{16 \pi G_4} Q_e^{(1)}$ and $\sqrt{16 \pi G_4} Q_m^{(1)}$ respectively and the KKm charges is $\sqrt{16 \pi G_4} Q_m^{(0)}$. Note that these specific examples have a large sizes compared to the extra-dimension radii and then correspond to a macroscopic black hole if and only if we impose a large conical defect at the three bubbles. It would be interesting to investigate further more sophisticated bubbling solutions in this new non-BPS floating brane ansatz \cite{bubblebagend}.

It is also questionable whether the constructed geometries can correspond to classical D1-D5-KKm non-BPS black hole microstates. At this level, the solutions in this paper and in \cite{bubblebagend} do not have a black hole throat, i.e. an AdS region, and they do not have internal parameters to be fine-tuned so that the bubble structure scales towards the horizon of a black hole of the same charges. Therefore, they may correspond at best to very atypical microstates. However, they possess the same fundamental ingredients as BPS microstate geometries, and are therefore good prototypes of non-BPS microstate geometries, if they exist. We believe that turning on more degrees of freedom, which will allow for example Chern-Simons interactions, could yield more relevant bubbling geometries that could correspond to classical microstate geometries. Another direction we wish to investigate is to try to construct bubbling geometries that are asymptotic to the near-horizon region of a non-BPS D1-D5-KKm black hole with the present ansatz.

\section*{Acknowledgments}
The work of IB and PH is supported in part by NSF grant PHY-1820784.

\vspace{1cm}

\appendix
\leftline{\LARGE \bf Appendices}

\section{Charged Weyl solutions in five dimensions}
\label{App:Weyl5d}

We review the construction of \cite{Bah:2020pdz} and more details can be found there. Weyl solutions of the five-dimensional action \eqref{eq:ActionGen5d} are obtained by the following ansatz of metric and gauge fields
\begin{align}
ds_5^2 &\= Z^{-1} \left[ -W_0 \,dt^2 +W_0^{-1} \, dy_1^2 \right]+ Z^2 \left[ e^{2\nu} \left(d\rho^2 + dz^2 \right) + \rho^2 d\phi^2 \right]\,,\nonumber \\
F^{(m)} &\= d H \wedge d\phi\,,\qquad F^{(e)} \= dT \wedge dt\wedge dy_1\,,
\label{eq:metricAxisym}
\end{align}
where $(\rho,z,\phi)$ correspond to the Weyl's canonical coordinates of the three-dimensional base, $y_1$ parametrizes the extra dimension with $2\pi R_{y_1}$ periodicity and the warp factors and gauge potential are functions of $(\rho,z)$. 

\subsection{The solutions}

The solutions are determined by two arbitrary functions that solve a Laplace equation on the three-dimensional base
\begin{equation}
\cL \left(\log W_0 \right) \=0 \,,\qquad \cL \left(L \right) \= 0\,, \qquad \text{with} \quad \cL \equi \frac{1}{\rho} \,\partial_\rho \left( \rho \partial_\rho \right) \+ \partial_z^2\,.
\end{equation}
The warp factor and the gauge potentials $(Z, H,T)$ are given by
\begin{equation}
Z \= \cG_\ell \left(L \right) \,,\quad \star_3 d(H\,d\phi) \= \sqrt{\frac{6}{2(1+q^2)}}~dL\,,\quad dT \= q\, \sqrt{\frac{6}{2(1+q^2)}}~ \frac{dL}{\cG_\ell (L)^2} ,
\label{eq:WGF&HApp}
\end{equation}
where $q$ is a constant giving the ratio between the magnetic and electric charges and $\cG_\ell$ is one of the following generating functions of one variable 
\begin{align}
\cG_1 (x) &\= \frac{\sinh(a x+b)}{a} \,, \qquad \cG_2 (x) \= i\,\frac{\cosh(a x+b)}{a}\,,\qquad \cG_3 (x) \= x +b, \nonumber \\
\cG_4 (x) &\= \frac{\sin(a x+b)}{a} \,,\qquad \cG_5(x) \=\frac{\cos(a x+b)}{a}\,,\qquad a \in \mathbb{R}_+, \quad b\in \IR \,.
\end{align}
The base scalar $\nu$ is obtained by integrating
\begin{align}
& \frac{1}{\rho}\partial_z \nu = \frac{1}{2}\, \partial_\rho \log W_0\,\partial_z \log W_0+ \frac{3\epsilon_\ell\,a^2}{2}\,\partial_\rho L\,\partial_z L, \nonumber \\
& \frac{1}{\rho}\partial_\rho \nu = \frac{1}{4} \left( \left(\partial_\rho \log W_0\right)^2- \left(\partial_z \log W_0 \right)^2\right)+ \frac{3\epsilon_\ell\,a^2}{4} \left( \left( \partial_\rho L \right)^2- \left(\partial_z L \right)^2\right) \label{eq:EqfornuGF}
\end{align}
where $\epsilon_\ell$ is a constant that depends on which choice of generating functions, $\cG_\ell$, has been made:
\begin{equation}
\epsilon_\ell \= \begin{cases}
1 \,,\qquad \text{if } \ell=1,2\,,\\
0 \,, \qquad \text{if } \ell=3\,,\\
-1\,, \qquad \text{if } \ell=4,5\,.
\end{cases}
\end{equation}
These integrals are simple to integrate but must be treated in a case-by-case manner depending on the type of sources chosen for $\log W_0$ and $L$.

One can retrieve vacuum Weyl solutions by considering $\ell=1$, $L = \lambda \bar{L}$ and the limit $\lambda \to 0$, $a \lambda \to 1$ and $a^{-1}\,e^{b}\to 1$. If we take $\ell=3$, physical sources must be point particles and the gauge field part will not modify the base scalar $\nu$. Physical solutions correspond to well-known BPS multicenter solutions in five dimensions with a flat base. 

Our interest relies on non-supersymmetric solutions and therefore on the choice $\cG_1$ for a generating function. For such a  function, physical sources are necessarily rod sources. 
We therefore consider $n$ distinct rods of length $M_i>0$ along the $z$-axis centered on $z=z_i$. Without loss of generality we can order them such that $z_i < z_j$ for $i<j$. The conventions are illustrated in Fig.\ref{fig:RodsSources}. The coordinates of the endpoints of the rods on the $z$-axis are given by
\begin{equation}
z^\pm_i \equi z_i \pm \frac{M_i}{2}\,.
\label{eq:coordinatesEndpoints5d}
\end{equation}
We define the distances to the endpoints $r_\pm^{(i)}$ and the distances $R_\pm^{(i)}$ as
\begin{equation}
r_\pm^{(i)} \equi \sqrt{\rho^2 + \left(z-z^\pm_i\right)^2}\,, \qquad R_\pm^{(i)} \equi r_+^{(i)}+r_-^{(i)}\pm M_i\,\,.
\label{eq:Rpmdef5d}
\end{equation}
Generating solutions of Laplace equation associated to such sources are given by $\log \frac{R_+^{(i)}}{R_-^{(i)}}$ and solutions are given by
\begin{equation}
\log W_0 = \sum_{i=1}^n G_i \,\log  \frac{R_+^{(i)}}{R_-^{(i)}}\,,\qquad L = \sum_{i=1}^n P_i \,\log  \frac{R_+^{(i)}}{R_-^{(i)}}\,,
\end{equation}
where $(G_i,P_i)$ corresponds to the weights of the $i^\text{th}$ rod on the harmonic functions. The warp factors and the gauge field potentials \eqref{eq:metricAxisym} are then given by
\begin{align}
&Z = \frac{1}{2a} \left[e^b \prod_{i=1}^n \left(  \frac{R_+^{(i)}}{R_-^{(i)}}\right)^{a P_i}-e^{-b} \prod_{i=1}^n \left(  \frac{R_-^{(i)}}{R_+^{(i)}}\right)^{a P_i} \right]\,,\qquad W_0 = \prod_{i=1}^n \left(  \frac{R_+^{(i)}}{R_-^{(i)}}\right)^{G_i} \,,\label{eq:WarpfactorsRods}\\
&H = \sqrt{\frac{6}{2(1+q^2)}} \sum_{i=1}^n P_i \left(r_-^{(i)}-r_+^{(i)} \right)\,,\quad T\=-\frac{\sqrt{6}\,q\,a}{\sqrt{2(1+q^2)}} \,\coth \left(\sum_{i=1}^n a P_i \,\log  \frac{R_+^{(i)}}{R_-^{(i)}} +b\right),\nonumber\\
& \nu \= \frac{1}{4} \, \sum_{i,j=1}^n \left(G_i G_j + 3 a^2 \,P_i P_j \right)\,\nu_{ij}\,,\nonumber
\end{align}
where we have also defined
\begin{equation} 
E_{\pm \pm}^{(i,j)} \equi r_\pm^{(i)} r_\pm^{(j)} + \left(z-z_i^{\pm}  \right)\left(z-z_j^{\pm}  \right) +\rho^2\,, \qquad \nu_{ij} \equi \log \frac{E_{+-}^{(i,j)}E_{-+}^{(i,j)}}{E_{++}^{(i,j)}E_{--}^{(i,j)}}\,.
\label{eq:GeneratingNu5d}
\end{equation}
We have constructed a family of solutions given by $4n+2$ parameters $(M_i,G_i,P_i,z_i,a,b)$. We now have to study the regularity of the solutions that constrains the parameter space. The potential constraints arise from coordinate singularities on the $z$-axis, regularity of the spacetime elsewhere and from conditions on the asymptotics.

\begin{itemize}\setlength{\itemsep}{0pt}\setlength{\parskip}{0pt}
\item[•] The solutions are asymptotic to $\IR^{1,3}\times$S$^1$ at large distance and regular everywhere out of the $z$-axis if
\begin{equation}
a \= \sinh b \,,\qquad P_i > 0\,.
\end{equation}
\item[•] \underline{Two types of physical rod sources:}
\begin{itemize}
\item[-] \underline{Black rods:}
If the weights at the $i^\text{th}$ rod satisfy
\begin{equation}
G_i \= -\frac{1}{2} \,,\qquad \,P_i \= \frac{1}{2 \sinh b} \,,
\end{equation}
the timelike Killing vector $\partial_t$ shrinks at the rod and it corresponds to a regular S$^2\times$S$^1$ horizon of a two-charge black string or a S$^3$ horizon of a two-charge black holes. These two different topologies depend if the rod is connected with other rod or not (see Fig.\ref{fig:5dWeyl}).
Its contribution to the four-dimensional ADM mass after KK reduction along $y_1$, $\cM^{(i)}$, its electric and magnetic charges, $Q_e^{(i)}$ and $Q_m^{(i)}$, are given by
\begin{equation}
\cM^{(i)} \= \frac{M_i}{8 G_4} \left(3 \coth b+1\right)\,,\quad {Q_e^{(i)}}^2 \= q^2\,{Q_m^{(i)}}^2 \= \frac{3\,q^2}{64\pi(1+q^2)G_4} \,\frac{M_i^2}{\sinh^2 b}\,.
\end{equation}
The presence of a black string or black hole induces a temperature to the whole solution, $\cT$, which can be derived from regularity of the Euclidean metric. We find that
\begin{equation}
\cT^{-2} \= \frac{2 \pi^2 \,M_i^2 \,e^{3 b}}{\sinh^3 b}\,d_i^2\,  \prod_{j\neq i} \left(\frac{z_j^+ - z_i^-}{z_j^- -z_i^-} \right)^{\text{sign}(j-i) \,\frac{3 - 2 G_j }{2}} \,,
\end{equation}
where $d_i$ corresponds to the following product of aspect ratios
\begin{equation}
d_1 \equi 1\,,\qquad d_i \equi  \prod_{j=1}^{i-1} \prod_{k=i}^n \left(\dfrac{(z_k^- - z_j^+)(z_k^+ - z_j^-)}{(z_k^+ - z_j^+)(z_k^- - z_j^-)}  \right)^{ \frac{3 + 4 G_j G_k}{4}}\quad \text{when } i=2,\ldots n\,.
\label{eq:diApp5d}
\end{equation}
\item[-] \underline{Bubble rods:} If the weights at the $i^\text{th}$ rod satisfy
\begin{equation}
G_i \= \frac{1}{2} \,,\qquad \,P_i \= \frac{1}{2 \sinh b} \,,
\end{equation}
the spacelike Killing vector $\partial_{y_1}$ shrinks and the rod corresponds to a degeneracy of the $y_1$-circle. The $(\rho,y_1)$ subspace corresponds to $\IR^2 / \mathbb{Z}_{k_i}$ with $k_i \in \mathbb{Z}_+$ if
\begin{equation}
R_{y_1}^2 \=  \frac{M_i^2 \,e^{3 b}}{2\,k_i^2\,\sinh^3 b}\,d_i^2\,  \prod_{j\neq i} \left(\frac{z_j^+ - z_i^-}{z_j^- -z_i^-} \right)^{\text{sign}(j-i) \,\frac{3 + 2 G_j }{2}}\,.
\label{eq:Ry5dApp}
\end{equation}
The rod corresponds to a S$^2$ or S$^1\times$S$^1$ bubble, depending if the rod is connected with a black rod, and is wrapped by electromagnetic fluxes. Its contribution to the four-dimensional ADM mass, $\cM^{(i)}$, its electric and magnetic charges, $Q_e^{(i)}$ and $Q_m^{(i)}$, are given by
\begin{equation}
\cM^{(i)} \= \frac{M_i}{8G_4} \left(3 \coth b-1\right)\,,\quad {Q_e^{(i)}}^2 \= q^2\,{Q_m^{(i)}}^2 \= \frac{3\,q^2}{64\pi(1+q^2)G_4} \,\frac{M_i^2}{\sinh^2 b}\,.
\end{equation}
\end{itemize}
\item[•] On the $z$-axis in between two disconnected rods, the $\phi$-circle shrinks as the usual cylindrical degeneracy. The local three-dimensional base in between the $(i-1)^\text{th}$ and $i^\text{th}$ has a conical singularity given by the metric
\begin{equation}
ds_3^2 \,\sim\,d\rho^2 \+ dz^2 \+ \frac{\rho^2}{d_i^2}\,d\phi^2\,.
\label{eq:strutmetApp}
\end{equation}
One can check that $0<d_i<1$ for $i\geq2$ \eqref{eq:diApp5d}, and therefore corresponds to a conical excess, i.e. a strut. 

\item[•] If two rods of different nature are connected, the intersection is free from strut and the $\phi$-circle has a finite size there. More precisely, such an intersection can only appear between a black rod touching a bubble rod (see Fig.\ref{fig:5dWeyl}).
\end{itemize}

Generic solutions have been depicted in Fig.\ref{fig:5dWeyl}. From far away, the solutions are asymptotic to $\IR^{1,3}\times$S$^1$ and have the following conserved charges in four dimensions after KK reduction along $y_1$
\begin{equation}
\cM \= \frac{1}{8G_4}\sum_{i=1}^n M_i \left(3 \coth b-2G_i\right)\,,\qquad {Q_e} \= q\,{Q_m} \= \frac{\sqrt{3}\,q}{8\sqrt{\pi(1+q^2) G_4}} \,\frac{\sum_{i=1}^n M_i}{\sinh b}\,.
\end{equation}

\subsection{Strut Energy}
\label{App:StrutEnergy}

If two successive rods are separated from each other, the section on the $z$-axis has a strut between them \eqref{eq:strutmetApp}. The strut carries negative energy density \cite{Costa:2000kf}.  In this appendix, we briefly review the computation of its stress tensor.

The strut or conical defects in general can be studied as a point source on a two dimensional plane. In general it can be model by the Liouville system
\begin{equation}
ds^2_\Sigma = e^{2A(x)} \left( dx_1^2 + dx_2^2 \right), \qquad \left( \partial_{x_1}^2 + \partial_{x_2}^2 \right) A +\kappa e^{2A} =0
\end{equation} where $\kappa$ is the curvature of the surface $\Sigma$.  We are interested in the behavior of sources for the Liouville potential, $A$, given as
\begin{equation}
\left( \partial_{x_1}^2 + \partial_{x_2}^2 \right) A = - 2\pi S~ \delta(x_1, x_2).\footnote{We normalize the delta function as $\int \delta(x_1, x_2) ~  dx_1 ~ dx_2 =1$.}
\end{equation} The effect of the curvature can be ignored in the region near the source and we can fix $\kappa=0$ without loss of generality. The solution can be expressed in terms of cylindrical coordinates around the source, $(x_1 = r \cos\phi, x_2 = r \sin\phi)$, as
\begin{equation}
A = - S \log r, \qquad ds^2_{\Sigma} = r^{-2S} \left( dr^2 + r^2 d \phi^2 \right).
\end{equation} For $S<1$, the space near the source has a conical deficit most easily seen with the coordinates 
\begin{equation}
\rho = r^{1-S}, \qquad ds^2_{\Sigma} = \frac{1}{(1-S)^2} d\rho^2 + \rho^2 d\phi^2.  
\end{equation} The local angle is $(1-S) \phi$ and since $\phi$ has period $2\pi$, there is a conical deficit given by the source $S$. We can express the volume form of $\Sigma$ and the $\delta$ function in the new coordinates as
\begin{equation}
dV_\Sigma = e^{2A} dx_1 \wedge dx_2 = \frac{\rho }{1-S} d\rho \wedge d\phi, \qquad  e^{-2A} \delta(x_1,x_2) = \frac{1-S}{2\pi \rho } \delta(\rho).  
\end{equation}  These are such that
\begin{equation}
\int_{\Sigma} e^{-2A} \delta(x_1, x_2) dV_\Sigma = 1.
\end{equation}
The Ricci curvatures of $\Sigma$ with the source is given as
\begin{equation}
R_{ij} = - \left( \partial_{x_1}^2 + \partial_{x_2}^2 \right) A ~ \delta_{ij} = 2\pi S~ \delta(x_1, x_2) ~ \delta_{ij}, \qquad R = 4\pi S~ \delta(x_1, x_2) e^{-2A} 
\end{equation}
Now, we consider a $D$ dimensional spacetime with a conical deficit corresponding to a codimension-two source.  The metric near such source $(\rho =0)$ can be written as 
\begin{equation}
ds^2_{D} = ds^2\left(M_{D-2} \right) + e^{2W} \left( e^{2\nu} d\rho^2 + \rho^2 d\phi^2 \right)
\end{equation} with $\phi \sim \phi + 2\pi$ and $\nu$ constant.  The external spacetime $M_{D-2}$ is the world volume directions of the source.  The warp function $W$ generically depends on $M_{D-2}$, while $\nu$ is a constant.  We can identify the two dimensional space where the source is localized with the Liouville system $\Sigma$ above.  The deficit angle associate to the source is then
\begin{equation}
S = 1 - e^{-\nu}.  
\end{equation} The contribution of the source to the Ricci curvature of the full system is
\begin{equation}
R_{mn} = 0+ \dots, \qquad R_{ij} = 2\pi S~ \delta(x_1, x_2) ~ \delta_{ij} + \dots
\end{equation} where the ellipses correspond to contributions from other sources in the spacetime, $(i,j)$ to the $(\rho,\phi)$ direction and $(m,n)$ to $M_{D-2}$ directions.  By using Einsteins equation, we read off the stress tensor for the co-dimension 2 source as
\begin{equation}
T_{mn} = - \frac{ S}{4G_D} \left[ e^{-2A} \delta(x_1, x_2) \right] e^{-2W} g_{mn}, \qquad T_{ij} =0.
\end{equation}
To obtain the energy density of the source, first we consider the conserved current
\begin{equation}
j^{\mu} = -T^{\mu \nu} k_\nu, \qquad k^\mu = \left(\partial_t \right)^\mu. 
\end{equation}  We can then write
\begin{equation}
j^m = \frac{S }{4G_5} \left[e^{-2A} \delta(x_1, x_2)  \right] e^{-2W} \delta^{m}_{t}, \qquad j^i =0.  
\end{equation} To compute the energy density for an asymptotic observer, we consider the normalized killing vector $v^\mu =  \frac{1}{\sqrt{-k \cdot k}} k^\mu$.  In the static cases of interest, this gives
\begin{equation}
e =  -j^\mu v_\mu=  \frac{S }{4G_5} \left[e^{-2A} \delta(x_1, x_2)   \right] e^{-2W} \sqrt{-g_{tt}} .  
\end{equation} We can integrate over a spacial slice to obtain the energy of the strut as 
\begin{align}
E &= \int e \cdot \left[e^{-2A} \delta(x_1, x_2) \right]  \cdot \sqrt{\hat{g}} ~e^{2W} dV_{\Sigma}  ~dw_1 \cdots dw_{D-3} \\
&=  \frac{2 \pi S }{\kappa_D^2} \int \sqrt{g_{M_{D-2}}} ~ dw_1 \cdots dw_{D-3}.  
\end{align} where the $w$'s are spacial coordinates of $M_{D-2}$ and $\hat{g}$ is the determinant of the induced metric on $M_{D-2}$.

We are now in a position to compute the stress energy of struts in between the rods of a generic Weyl solutions in five dimensions. We consider then a Weyl solution sourced by $n$ rods as in \eqref{eq:WarpfactorsRods} given by the metric \eqref{eq:metricAxisym}.  In the region near the strut in between the $(i-1)^\text{th}$ and $i^\text{th}$ rods, we have \cite{Bah:2020pdz}
\begin{align}
ds^2 (M_{3}) &= Z^{-1} \left[ -W_0 dt^2 + W_0^{-1} dy_1^2 + Z^3 e^{2\nu} dz^2  \right] \,,\qquad e^{2\nu} = {d_i}^2\,,
\end{align} where $0<d_i<1$ is given in \eqref{eq:diApp5d}. The energy of the strut is then
\begin{equation}
E= -\frac{2 \pi R_y}{4G_5} \,(1-d_i) (z_i^- -z_{i-1}^+)\= -\frac{1-d_i}{4 G_4}(z_i^- -z_{i-1}^+)\,.
\end{equation}
For the two-bubble configuration of the section \ref{sec:twobubbles5d}, we find
\begin{equation}
E =  - \frac{M^2\delta}{(M+\delta)^2} \frac{2\pi R_y}{4G_5 } =-  \frac{M^2\delta}{4G_4 (M+\delta)^2 }.  
\end{equation}

\section{Vacuum Weyl solutions in six dimensions}
\label{App:WeylVac}

In this section, we construct in details axisymmetric solutions of the six-dimensional theory \eqref{eq:Action6d} considering that the gauge field is trivial. We remind that the ansatz for the metric is
\begin{equation}
ds_6^2  \= - U_0 \,dt^2 \+ U_1\, dy_1^2 \+ U_2 \,dy_2^2 \+ \frac{1}{U_0 U_1 U_2}\left[e^{2\nu}\,\left(d\rho^2+dz^2 \right) +\rho^2 d\phi^2\right]\,,
\end{equation}
where $U_I$ and $\nu$ are $(\rho,z)$-dependent warp factors governed by the equations \eqref{eq:EQforUI}.

We consider $n$ distinct rods of length $M_i>0$ along the $z$-axis centered on $z=z_i$. Without loss of generality we can order them as $z_i < z_j$ for $i<j$ (see Fig.\ref{fig:RodsSources}). The warp factors are given by
\begin{equation}
U_I = \prod_{i=1}^n \left( \frac{R_+^{(i)}}{R_-^{(i)}} \right)^{G^{(I)}_i}\,, \qquad \nu \= \frac{1}{8} \,\sum_{I<J} \sum_{i,j=1}^n \left(G_i^{(I)}+G_i^{(J)}\right)\left(G_j^{(I)}+G_j^{(J)}\right)\,\nu_{ij}\,,
\end{equation}
where $G^{(I)}_i$ defines the weight of the $i^\text{th}$ rod on the $I^\text{th}$ warp factor that will be fixed by regularity and $(R_\pm^{(i)},\nu_{ij})$ are function of $(\rho,z)$ defined by \eqref{eq:Rpmdef}.

\noindent To study the metric far away from the rods, it is appropriate to use spherical coordinates
\begin{equation}
\rho \= r\,\sin \theta \,,\qquad z \= r \, \cos \theta \,,
\end{equation}
and we have, at large $r$, 
\begin{equation}
U_I \,\sim\, 1 \,,\qquad e^{2\nu} \,\sim \, 1\,.
\end{equation}
Therefore, the solutions are asymptotic to T$^2\times \IR^{1,3}$. 

It is rather straightforward from the expressions of $(R_\pm^{(i)},\nu_{ij})$, \eqref{eq:Rpmdef}, that $U_I$ and $e^{2\nu}$ are finite and strictly positive out of the $z$-axis. The solutions are then regular and have the topology of T$^2\times \IR^{1,3}$ out of the $z$-axis. 

The $\phi$-circle degenerates on the $z$-axis out of the rods. This degeneracy corresponds to the usual cylindrical coordinate singularity on the axis but conical singularities can occur depending on the value of $e^{2\nu}$ in these regions. Moreover, on the $z$-axis but on a rod, $U_I$ is either diverging or blowing and $e^{2\nu}$ goes to zero. This requires a careful analysis constraining the values of $G_i^{(I)}$ to induce either a coordinate singularity corresponding to degeneracies of the extra circles $(y_1,y_2)$ or a horizon. 

\subsection{Regularity on the $z$-axis}

We will first discuss the regularity on the rods before discussing the regularity elsewhere on the $z$ axis.

\subsubsection{At the $i^\text{th}$ rod}
\label{App:AttheRod}

The local spherical coordinates around the i$^{\text{th}}$ rod are given by $r_i \rightarrow 0$ for $0\leq \theta_i \leq \pi$ with
\begin{equation}
\rho = \sqrt{r_i(r_i+M_i)}\,\sin\theta_i \,,\qquad z = \left(r_i+\frac{M_i}{2} \right) \cos\theta_i +z_i\,.
\label{eq:localcoorApp}
\end{equation}
The two-dimensional base behaves as
\begin{equation}
d\rho^2 + dz^2 \sim \frac{M_i \sin^2 \theta_i}{4} \left(\frac{dr_i^2}{r_i}+M_i\, d\theta_i^2 \right)\,.
\end{equation}
Moreover,
\begin{equation}
\frac{R_+^{(i)}}{R_-^{(i)}} \sim \frac{M_i}{r_i} \,,\qquad \frac{R_+^{(j)}}{R_-^{(j)}} \sim \left(\frac{z_j^+-\left(z_i + \frac{M_i}{2} \cos \theta_i \right)}{z_j^--\left(z_i + \frac{M_i}{2}  \cos \theta_i \right)}\right)^{\text{sign}(j-i)}\,,\quad j\neq i\,,
\label{eq:RpRmexp}
\end{equation}
where we remind that $z^\pm_j$ are the rod endpoints \eqref{eq:coordinatesEndpoints}. Thus,
\begin{equation}
U_I \sim\left( \frac{M_i}{r_i} \right)^{ G^{(I)}_i}\,\prod_{j\neq i} \left(  \frac{R_+^{(j)}}{R_-^{(j)}}\right)^{G^{(I)}_j}\,.
\end{equation}
To derive the behavior of $e^{2\nu}$ we first need the limits for $\nu_{jk}$. We have different situations:
\begin{equation}
\begin{split}
e^{\nu_{ii}}&\sim \frac{16 r_i^2}{M_i^2\sin^4 \theta_i}\,,\qquad \quad e^{\nu_{jk}}\ \sim \begin{cases}
1 \qquad &\text{if } i<j\leq k  \text{ or } j\leq k<i  \\
\dfrac{(z_k^- - z_j^+)^2(z_k^+ - z_j^-)^2}{(z_k^+ - z_j^+)^2(z_k^- - z_j^-)^2} \qquad &\text{if } j<i<k 
\end{cases}\,, \\
e^{\nu_{ij}} &\sim \begin{cases} 
\dfrac{(z_j^+ - z_i^-)^2\left(z_j^- - \left(z_i +\frac{M_i}{2} \cos\theta_i \right)\right)^2}{(z_j^- - z_i^-)^2\left(z_j^+ - \left(z_i +\frac{M_i}{2} \cos\theta_i \right)\right)^2} \qquad &\text{if } j>i,  \\
\dfrac{(z_j^- - z_i^+)^2\left(z_j^+ - \left(z_i +\frac{M_i}{2} \cos\theta_i \right)\right)^2}{(z_j^+ - z_i^+)^2\left(z_j^- - \left(z_i +\frac{M_i}{2} \cos\theta_i \right)\right)^2} \qquad &\text{if } j<i, 
\end{cases}\,.
\end{split}
\label{eq:EEEEatrod}
\end{equation}
We gather everything to derive the behavior of $$ 2\nu  = \frac{1}{4} \,\sum_{I<J} \sum_{j,k=1}^n \left(G_j^{(I)}+G_j^{(J)}\right)\left(G_k^{(I)}+G_k^{(J)}\right)\,\nu_{jk} \equi \frac{1}{2} \,\sum_{j,k=1}^n \alpha_{jk}\,\nu_{jk}\,, $$
where we have defined $
\alpha_{jk} \equi \frac{1}{2}\,\sum_{I<J}\left(G_j^{(I)}+G_j^{(J)}\right)\left(G_k^{(I)}+G_k^{(J)}\right)\,.
$
Finally, we have
\begin{equation}
\begin{split}
e^{2 \nu} \sim & \left(\frac{4 r_i}{M_i \sin^2 \theta_i} \right)^{\alpha_{ii}} \,\prod_{j\neq i} \left( \frac{R_+^{(j)}}{R_-^{(j)}}\right)^{-2\alpha_{ij}} \, \\
&\times \prod_{j=1}^{i-1} \prod_{k=i}^n \left(\dfrac{(z_k^- - z_j^+)(z_k^+ - z_j^-)}{(z_k^+ - z_j^+)(z_k^- - z_j^-)}  \right)^{2\alpha_{jk}}  \prod_{j\neq i} \left(\frac{z_j^+ - z_i^-}{z_j^- -z_i^-} \right)^{2\,\text{sign}(j-i) \,\alpha_{ij}} \,,
\end{split}
\end{equation}
and the $\theta_i$-dependent expansion of $\frac{R_+^{(j)}}{R_-^{(j)}}$ is given in \eqref{eq:RpRmexp}. We have also considered that the product ``$\prod_{j=1}^{i-1}$'' is equal to $1$ for the first rod, $i=1$. We define the constants $d_i$ which depend only on the geometry of the rods and are independent of $r_i$ and $\theta_i$.
\begin{equation}
d_1 \equi 1\,,\qquad d_i \equi  \prod_{j=1}^{i-1} \prod_{k=i}^n \left(\dfrac{(z_k^- - z_j^+)(z_k^+ - z_j^-)}{(z_k^+ - z_j^+)(z_k^- - z_j^-)}  \right)^{\alpha_{jk}}\quad \text{when } i=2,\ldots n\,.
\label{eq:diDef}
\end{equation}
Gathering all these expressions, the metric components behave around the $i^\text{th}$ rod as
\begin{equation}
\begin{split}
g_{tt} &\sim-\left( \frac{M_i}{r_i} \right)^{ G^{(0)}_i}\,\prod_{j\neq i} \left(  \frac{R_+^{(j)}}{R_-^{(j)}}\right)^{G^{(0)}_j}\,, \qquad g_{y_ay_a} \sim \left( \frac{M_i}{r_i} \right)^{ G^{(a)}_i}\,\prod_{j\neq i} \left(  \frac{R_+^{(j)}}{R_-^{(j)}}\right)^{G^{(a)}_j}\,,\\
g_{\phi\phi} &\sim  M_i^2\, \sin^2\theta_i\,\left( \frac{r_i}{M_i} \right)^{ 1+ \sum_I G^{(I)}_i}\,\prod_{j\neq i} \left(  \frac{R_+^{(j)}}{R_-^{(j)}}\right)^{-\sum_I G^{(I)}_j} \,.\\
g_{r_i r_i}& \sim  d_i^2 \left(\frac{r_i}{M_i} \right)^{ \sum_I G^{(I)}_i} \,\left(\frac{4 r_i}{M_i \sin^2 \theta_i} \right)^{\alpha_{ii}-1} \,\prod_{j\neq i} \left( \frac{R_+^{(j)}}{R_-^{(j)}}\right)^{-2\alpha_{ij}-\sum_I G^{(I)}_j}   \prod_{j\neq i} \left(\frac{z_j^+ - z_i^-}{z_j^- -z_i^-} \right)^{2\,\text{sign}(j-i) \,\alpha_{ij}}\,, \\
g_{\theta_i \theta_i}& \sim M_i^2\, d_i^2 \left(\frac{r_i}{M_i} \right)^{ 1+\sum_I G^{(I)}_i} \,\left(\frac{4 r_i}{M_i \sin^2 \theta_i} \right)^{\alpha_{ii}-1} \,\prod_{j\neq i} \left( \frac{R_+^{(j)}}{R_-^{(j)}}\right)^{-2\alpha_{ij}-\sum_I G^{(I)}_j}   \prod_{j\neq i} \left(\frac{z_j^+ - z_i^-}{z_j^- -z_i^-} \right)^{2\,\text{sign}(j-i) \,\alpha_{ij}}.
\end{split}
\label{eq:ytphibehaviorBubblerod}
\end{equation}
To avoid singularity along the time and the extra dimensions, it is clear that all $G_i^{(I)} \leq 0$. This necessarily implies that the S$^2$ parametrized by $(\theta_i,\phi)$ must have a finite size at the rod. That is
\begin{equation}
G^{(0)}_i+G^{(1)}_i+G^{(2)}_i =-1 \,,
\end{equation}
which implies $-1 \leq G^{(I)}_i \leq 0$. By studying carefully all the possibilities, we can check that having two $G^{(I)}_i $ different from $0$ leads to a singular metric. Therefore, we have only three possible physical situations for the rod. We will first assume that the rod is disconnected from the others.
\begin{itemize}
\item[•] \underline{$G^{(0)}_i =-1$ and $G^{(1)}_i = G^{(2)}_i=0$: a two-dimensional black brane.}

Having only $G^{(0)}_i$ non-zero makes the time fiber to degenerate while the $y_a$-fibers have a finite size at the rod. More concretely, the $i^\text{th}$ rod corresponds to a horizon where the timelike Killing vector $\partial_t$ shrinks.  Indeed, we have $$ \alpha_{ii} \= 1 \,,\qquad \alpha_{ij} \= - \frac{1}{2} \left(2G_j^{(0)}+G_j^{(1)}+G_j^{(2)} \right)\,,$$
which implies that the $\theta_i$-dependent factors in $g_{tt}$ and $g_{r_i r_i}$ are remarkably the same. The local six-dimensional metric around the i$^\text{th}$ rod is then
\begin{equation}
\begin{split}
ds^2 \bigl|_{r_i = 0} \= &g_{y_1 y_1}(\theta_i) \,dy_1^2+g_{y_2 y_2}(\theta_i) \,dy_2^2 + g_{\theta_i\theta_i}(\theta_i)  \,\left( d\theta_i^2 + \bar{g}_{\phi\phi}(\theta_i) \,\sin^2 \theta_i\, \,d\phi^2\right) \\
&\+ \bar{g}_{r_i r_i}(\theta_i) \left(d\rho_i^2 - \kappa_i^2\,\rho_i^2\,dt^2 \right)\,,
\end{split}
\label{eq:metricatrodBS}
\end{equation}
where $\rho_i^2 \equiv 4 r_i$, $(g_{y_1 y_1}(\theta_i),g_{y_2 y_2}(\theta_i),g_{\theta_i\theta_i}(\theta_i),\bar{g}_{\phi\phi}(\theta_i),\bar{g}_{r_i r_i}(\theta_i))$ are all finite and non-zero for $0\leq \theta_i \leq \pi$  \eqref{eq:ytphibehaviorBubblerod} and the surface gravity, $\kappa_i$, is given by
\begin{equation}
\kappa_i \equi \frac{1}{2d_i\,M_i}\, \prod_{j\neq i} \left(\frac{z_j^+ - z_i^-}{z_j^- -z_i^-} \right)^{\text{sign}(i-j)\, \alpha_{ij}}\,.
\label{eq:surfaceGravVac}
\end{equation}

The metric corresponds then to the horizon of a two-dimensional black brane with a S$^2\times$T$^2$ topology where the T$^2$ is parametrized by $(y_1,y_2)$  and defined the world volume of the black brane. One can relate the surface gravity to the temperature of the black brane by requiring smoothness of the Euclideanized solution. We find
\begin{equation}
\cT_i \= \frac{\kappa_i}{2\pi}\,.
\end{equation}
Note that if we study axisymmetric solutions with multiple black branes in thermal  equilibrium the temperature associated to each black rod must be fixed to be equal.

Moreover, as $g_{y_1y_1}(\theta_i)g_{y_2y_2}(\theta_i)g_{\theta_i \theta_i}(\theta_i)^2\bar{g}_{\phi\phi}(\theta_i)$ is remarkably independent of $\theta_i$, the area of the horizon is simple to derive. We find
\begin{equation}
A_i \= \int_{S^2\times T^2} \sqrt{g_{y_1y_1}(\theta_i)g_{y_2y_2}(\theta_i)g_{\theta_i \theta_i}(\theta_i)g_{\phi\phi}(\theta_i)} \= \frac{(2\pi)^3}{\kappa_i}\,M_i\,R_{y_1}R_{y_2}\,.
\end{equation}
\item[•] \underline{$G^{(1)}_i =-1$ and $G^{(0)}_i = G^{(2)}_i=0$: a species-1 bubble.}

For those values, the time and $y_2$ fibers have a finite size while the $y_1$-fiber degenerates. More concretely, the spacelike Killing vector $\partial_{y_1}$ shrinks on the $i^\text{th}$ rod corresponding to a coordinate singularity of an origin of $\IR^2$ space. We have $$ \alpha_{ii} \= 1 \,,\qquad \alpha_{ij} \= - \frac{1}{2} \left(2G_j^{(1)}+G_j^{(0)}+G_j^{(2)} \right)\,,$$
which implies that the $\theta_i$-dependent factors in $g_{y_1 y_1}$ and $g_{r_i r_i}$ are the same. The local six-dimensional metric around the i$^\text{th}$ rod is then
\begin{equation}
\begin{split}
ds^2 \bigl|_{r_i = 0} \= &-g_{tt}(\theta_i) \,dt^2+g_{y_2 y_2}(\theta_i) \,dy_2^2 + g_{\theta_i\theta_i}(\theta_i)  \,\left( d\theta_i^2 + \bar{g}_{\phi\phi}(\theta_i) \,\sin^2 \theta_i\, \,d\phi^2\right) \\
&\+ \bar{g}_{r_i r_i}(\theta_i) \left(d\rho_i^2 +\frac{\rho_i^2}{{C_i^{(1)}}^2}\,dy_1^2 \right)\,,
\end{split}
\label{eq:metricatrodBu}
\end{equation}
where $(g_{tt}(\theta_i),g_{y_2 y_2}(\theta_i),g_{\theta_i\theta_i}(\theta_i),\bar{g}_{\phi\phi}(\theta_i),\bar{g}_{r_i r_i}(\theta_i))$ can be obtained from \eqref{eq:ytphibehaviorBubblerod} and are all finite and non-zero for $0\leq  \theta_i \leq \pi$. Moreover, we have defined $\rho_i^2 \equiv 4 r_i$ and the constant, $C_i^{(1)}$, is given by
\begin{equation}
C_i^{(1)} \equi 2d_i\,M_i\, \prod_{j\neq i} \left(\frac{z_j^+ - z_i^-}{z_j^- -z_i^-} \right)^{\text{sign}(j-i)\, \alpha_{ij}}\,.
\end{equation}
The two dimensional subspace $(\rho_i,y_1)$ describes a smooth origin of $\IR^2$ or a smooth discrete quotient $\IR^2/\mathbb{Z}_{k_i}$ if the parameters are fixed according to the radius of the $y_1$-circle
\begin{equation}
\begin{split}
R_{y_1} = \frac{C^{(1)}_i}{k_i}\,, \qquad k_i \in \mathbb{N}.
\end{split}
\label{eq:condRymultibubble}
\end{equation}
To conclude, the time slices of the five-dimensional space at the i$^\text{th}$ rod is a bolt described by a warped S$^2\times$S$^{y_2}$ fibration over an origin of a $\IR^2/\mathbb{Z}_{k_i}$ space. 

Moreover, one can check that $g_{y_2y_2}(\theta_i)g_{\theta_i \theta_i}(\theta_i)^2\bar{g}_{\phi\phi}(\theta_i)$ is remarkably independent of $\theta_i$ if there is no black rods in the configuration. Therefore, one can easily derive the area of the S$^2\times$S$^{y_2}$ bubble for such configurations and we find
\begin{equation}
A_{\text{B}i} \= \int_{S^2\times S^{y_2}} \sqrt{g_{y_2y_2}(\theta_i)g_{\theta_i \theta_i}(\theta_i)g_{\phi\phi}(\theta_i)} \= (2\pi)^2\,k_i\,M_i\,R_{y_1}R_{y_2}\,.
\label{eq:AreaBubbleApp}
\end{equation}

\item[•] \underline{$G^{(2)}_i =-1$ and $G^{(0)}_i = G^{(1)}_i=0$: a species-2 bubble.}

This situation is similar to the previous one but the role of the $y_1$ and $y_2$ fibers are inverted. On such a rod, the spacelike Killing vector $\partial_{y_2}$ shrinks defining a bolt. The bolt corresponds to a warped S$^2\times$S$^{y_1}$ at the origin of a $\IR^2/\mathbb{Z}_{k_i}$ space defined by $(\rho_i,y_2)$. The orbifold parameter $k_i\in \mathbb{N}$ relates the radius of the $y_2$-circle to the parameter of the solution such as
\begin{equation}
R_{y_2} = \frac{C^{(2)}_i}{k_i}\,,\qquad C_i^{(2)} \equi 2d_i\,M_i\, \prod_{j\neq i} \left(\frac{z_j^+ - z_i^-}{z_j^- -z_i^-} \right)^{\text{sign}(j-i)\, \alpha_{ij}}\,.
\end{equation}

\end{itemize}

If we assume now that the rod is connected, we still have the three same choices of weights but the local topology might change. For instance, we consider that the $i^\text{th}$ rod is a black rod given by $G^{(0)}_i =-1$ and $G^{(1)}_i = G^{(2)}_i=0$. If the rod is connected from above to a species-1 bubble rod and not connected from below, then the local metric at the horizon is still given \eqref{eq:metricatrodBS}. However, $g_{y_1 y_1}(\theta_i)$ and $\bar{g}_{\phi \phi} (\theta_i)$ are not finite for $0 \leq \theta_i \leq \pi$ anymore. More precisely, we have $g_{y_1 y_1}(\theta_i) \sim 0$ and $\bar{g}_{\phi \phi} (\theta_i) \sin^2 \theta_i $ finite around $\theta_i \to 0$. Therefore, the $y_1$-circle pinches off at the north pole and the horizon has a S$^3\times$S$^1$ topology corresponding to a black string. Note that the surface gravity computed in \eqref{eq:surfaceGravVac} is still the same if the rod is connected and is still well-defined. If the rod is now connected from above and below to two species-1 bubble rods, we have a S$^2\times$T$^2$ horizon again but the S$^2$ is now described by $(\theta_i,y_1)$. Similar scenarios happen if the $i^\text{th}$ rod is a bubble rod: we can have either an S$^2\times$S$^1$ bubble or a S$^3$ bubble depending on what is surrounding the rod.

To conclude, we have three types of rods that can source physically our solutions on the axis. For each type of rods the $\phi$-circle has a finite size. The different rods and their physics has been depicted in Fig.\ref{fig:RodCategories}.

\subsubsection{On the $z$-axis and out of the rods}
\label{App:OutoftheRod}

We now study the behavior of the solutions on the $z$-axis, $\rho \to 0$, and out of the rods where the $\phi$-circle can shrink to zero size. On these segments, each $R^{(i)}_\pm$ \eqref{eq:Rpmdef} is non-zero and finite
\begin{equation}
R^{(i)}_\pm = 2 |z-z_i| \pm M_i\,.
\end{equation}
Thus, all $U_I$ are also non-zero and finite there. The regularity reduces to the study of the three-dimensional subspace $(\rho,z,\phi)$,
\begin{equation}
ds_3^2 \= e^{2 \nu} \left(d\rho^2 +dz^2 \right) +\rho^2 d\phi^2\,.
\end{equation}
At $\rho=0$ and out of the rods, we want this space to correspond to the cylindrical coordinate degeneracy. First we have
\begin{equation}
e^{\nu_{jk}} \sim \begin{cases}
1 \qquad &\text{if } j\leq k  \text{ and } z \not\in [z_j + \frac{M_j}{2} , z_k - \frac{M_k}{2} ]  \\
\dfrac{(z_k^- - z_j^+)^2(z_k^+ - z_j^-)^2}{(z_k^+ - z_j^+)^2(z_k^- - z_j^-)^2} \qquad &\text{if } j<k   \text{ and } z \in ] z_j + \frac{M_j}{2} , z_k - \frac{M_k}{2} [
\end{cases}\,.
\end{equation}
Therefore, we get
\begin{equation}
e^{2 \nu} \sim \begin{cases}
~~ 1  &\text{if } z < a_1 - \frac{M_1}{2}  \text{ and } z > a_n + \frac{M_n}{2} \\
~~ d_i^2\qquad  &\text{if } z \in \, ]z_{i-1} + \frac{M_{i-1}}{2} , z_i - \frac{M_i}{2} [\,,\quad \forall i 
\end{cases}
\end{equation}
where $d_i$ is given by the coordinate of the rod endpoints in \eqref{eq:diDef}. First we notice that asymptotically, $z > z_n^+$ and $z<z_1^-$, the base space is directly flat $\IR^3$ without conical singularity. Now, in between two rods, we have two possibilities if they are connected or disconnected.

\begin{itemize}
\item[•] \underline{Disconnected rods:}
\end{itemize}

We consider the segment in between the disconnected $(i-1)^\text{th}$ rod and $i^\text{th}$ rods, $z_{i-1} + \frac{M_{i-1}}{2} <  z_i - \frac{M_i}{2}$. The three-dimensional base is then given by the metric
 \begin{equation}
ds_3^2 \sim d_i^2 \left(d\rho^2 +dz^2 +\frac{\rho^2}{d_i^2} d\phi^2 \right) \,.
\end{equation}
The segment corresponds to a $\IR^3$ base with the local cylindrical angle $\phi_i \equi \frac{\phi}{d_i}$. Moreover, note that, for the three species of physical rod sources, we have
\begin{equation}
\dfrac{(z_k^- - z_j^+)(z_k^+ - z_j^-)}{(z_k^+ - z_j^+)(z_k^- - z_j^-)}  = \frac{(z_k-z_j)^2 - \frac{1}{4} (M_k+M_j)^2}{(z_k-z_j)^2 - \frac{1}{4} (M_k-M_j)^2} < 1 \,,\qquad \alpha_{jk} \= 1 \text{ or } \frac{1}{2}>0\,.
\label{eq:simpleform}
\end{equation}
Then, we necessarily have $d_i^2 <1$. Thus, the segment has a \emph{conical excess} and the period of the local angle, $\phi_i= \frac{\phi}{d_i}$, is $\frac{2\pi}{d_i}>2\pi$. This manifests itself as a string with negative tension, or strut, between the two rods. The strut exerts the necessary repulsion so that the whole structure does not collapse. We can calculate the stress tensor and the energy of the strut using the method described in section \ref{App:StrutEnergy}, and we will similarly find that the energy is given by 
\begin{equation}
E \=- \frac{1-d_i}{4 G_4}\,,\qquad G_4 \= \frac{G_6}{(2\pi)^2 R_{y_1}R_{y_2}}\,.
\end{equation}

As we aim to construct regular solution, this situation must be prohibited and we will then treat the case where the rods are connected. 

\begin{itemize}
\item[•] \underline{Connected rods:}
\end{itemize}

We consider the intersection between the connected $(i-1)^\text{th}$ rod and $i^\text{th}$ rods, $z_{i-1} + \frac{M_{i-1}}{2} =  z_i - \frac{M_i}{2}$. The intersection then consists of a point with coordinates$(\rho,z)\=(0,z_{i-1}^+)\=(0,z_{i}^-)\,.$
We first define local spherical coordinates as follows
\begin{align}
\mathfrak{r}_i &\equi  \sqrt{\rho^2 +\left(z-z_{i}^-\right)^2}\,,\qquad \cos \tau_i \equi \frac{z -z_{i}^- }{\mathfrak{r}_i }\,, \\
\mbox{that is} \qquad
\rho &= \mathfrak{r}_i \,\sin \tau_i \,,\qquad z =  \mathfrak{r}_i \cos \tau_i + z_i^-\,. 
\end{align}
The two-dimensional base transforms to
\begin{equation}
d\rho^2+dz^2 = d {\mathfrak{r}_i^+}^2+  {\mathfrak{r}_i^+}^2\,d{\tau_i^+}^2\,.
\end{equation}
At $\mathfrak{r}_i \rightarrow 0$ we have
\begin{equation}
\begin{split}
\frac{R_+^{(i-1)}}{R_-^{(i-1)}} &\sim \frac{2 M_{i-1}}{\left(1+\cos \tau_i\right)  \mathfrak{r}_i}\,,\quad \frac{R_+^{(i)}}{R_-^{(i)}} \sim \frac{2 M_{i}}{\left(1-\cos \tau_i\right)  \mathfrak{r}_i}\,,\quad \frac{R_+^{(j)}}{R_-^{(j)}} \sim  \left(\frac{z_j^+ -z_i^+}{z_j^--z_i^+} \right)^{\text{sign}(j-i)}\,,\quad j \neq i-1,i\,, \\
e^{\nu_{i-1\,i-1}}& \sim \frac{(1+\cos \tau_i)^2}{4}\,,\qquad e^{\nu_{ii}} \sim \frac{(1-\cos \tau_i)^2}{4}\,,\qquad e^{\nu_{i -1\,i}} \sim \frac{(z^+_{i}-z^-_{i-1})^2}{(z^+_{i}-z^+_{i-1})^2(z^-_{i}-z^-_{i-1})^2}\,{\mathfrak{r}_i}^2\\
e^{\nu_{jk}} &\sim \begin{cases}
\dfrac{(z_k^- - z_j^+)^2(z_k^+ - z_j^-)^2}{(z_k^+ - z_j^+)^2(z_k^- - z_j^-)^2} \qquad &\text{if } j<i-1 \text{ and } i <k  \\
1 \qquad &\text{otherwise } 
\end{cases}\,, \\
e^{\nu_{i-1\,j}} &\sim \begin{cases} 
\dfrac{(z_j^+ - z_{i-1}^-)^2\left(z_j^- - z_{i-1}^+\right)^2}{(z_j^- - z_{i-1}^-)^2\left(z_j^+ - z_{i-1}^+\right)^2} \qquad &\text{if } j>i,  \\
1 \qquad &\text{if } j<i-1, 
\end{cases}\,, \\
 e^{\nu_{ij}} &\sim \begin{cases} 
1 \qquad &\text{if } j>i,  \\
\dfrac{(z_{i}^+ - z_j^-)^2\left(z_{i}^- - z_j^+\right)^2}{(z_{i}^- - z_j^-)^2\left(z_{i}^+ - z_j^+\right)^2} \qquad &\text{if } j<i-1, 
\end{cases}\,,
\end{split}
\end{equation}
We can directly see that for any types of $i^\text{th}$ and $(i-1)^\text{th}$ rods, the $\phi$-circle keeps a finite size at the intersection unlike the disconnected case. This means that the intersection is protected from the conical excess associated to the degeneracy of the $\phi$-circle in between two rods.

In order to determine the local topology, a distinction must be made between different types of rods. Being connected, the $i^\text{th}$ and $(i-1)^\text{th}$ rods are necessarily of a different nature.\footnote{Otherwise, they form a single rod.}  Consequently, we have three possible scenario: an intersection between a black rod and a species-1 bubble rod, between a black rod and a species-2 bubble rod and between a species-1 bubble rod and a species-2 bubble rod. The two former gives the same topology by inverting the role of the $y_1$ and $y_2$ fibers. From the above expressions we find that the local six-dimensional metric is given by
\begin{equation}\begin{split}
ds^2 \bigl|_{\mathfrak{r}_i=0} \propto \alpha_i \,d\phi^2 + \beta_i\,dy_2^2+  \frac{d\mathfrak{r}_i^2}{\mathfrak{r}_i} + \mathfrak{r}_i \left( d\tau_i^2 - 2 \kappa_i^2\,(1-\cos \tau_i)\,dt^2+ 2 (1+\cos \tau_i)\,\frac{dy_1^2}{R_{y_1}^2\,{k_{i-1}}^2}\right)\,,
\end{split}
\label{eq:intersectionBB&Bu}
\end{equation}
where $\alpha_i$ and $\beta_i$ are irrelevant finite constants and the surface gravity of the black rod $\kappa_i$ and the orbifold parameter of the species-1 bubble rod $k_{i-1}$ are given in \eqref{eq:surfaceGravVac} and \eqref{eq:condRymultibubble}. This corresponds to the usual metric at the north pole of a horizon. At this type of loci, a circle composing the surface of the horizon is degenerating which is here $y_1$. In the present case, it degenerates with a conical defect, parametrized by $k_{i-1}$, related to the species-1 bubble that is connected to the black brane. 

Now, if the two connected rods consist of a species-1 bubble rod and a species-2 bubble rod, the local metric is
\begin{equation}\begin{split}
ds^2 \bigl|_{\mathfrak{r}_i=0} \propto-\beta_i \,dt^2+ \alpha_i \,d\phi^2+  \frac{d\mathfrak{r}_i^2}{\mathfrak{r}_i} + \mathfrak{r}_i \left( d\tau_i^2 + 2 (1-\cos \tau_i)\,\frac{dy_2^2}{R_{y_2}^2\,{k_{i}}^2}+ 2 (1+\cos \tau_i)\,\frac{dy_1^2}{R_{y_1}^2\,{k_{i-1}}^2}\right)\,,
\label{eq:intersectionBu&Bu}
\end{split}
\end{equation}
where $\alpha_i$ and $\beta_i$ are irrelevant finite constants and $(k_{i-1},k_{i})$ are the orbifold parameters of the connected species-1 bubble  and species-2 bubble rods. This corresponds to the metric of the origin of an orbifolded $\IR^{4}$ parametrized by $(\mathfrak{r}_i,\tau_i,y_1,y_2)$. The two angles have the same conical defects as the connected bubbles but the local topology is free from struts and conical excess. Moreover, if $ k_{i-1} = k_{i}=1$, the time slices of the metric corresponds to the origin of a $\IR^{4}$ with a S$^\phi$ fibration and the local spacetime is entirely smooth.

\section{Charged Weyl solutions in six dimensions}
\label{App:Weylcharged}

In this section, we construct in detail physical Weyl solutions of the six-dimensional Einstein-Maxwell theory \eqref{eq:Action6d}. We recall that the ansatz for the metric is
\begin{align}
ds_{6}^2 = &\frac{1}{Z_1} \left[- W_0\,dt^2 + \frac{ dy_1^2}{W_0} \right] + \frac{Z_1}{Z_0}\, \left(dy_2 +H_0 \,d\phi\right)^2 + Z_0 Z_1\,\left[e^{2\nu} \left(d\rho^2 + dz^2 \right) +\rho^2 d\phi^2\right] ,\nonumber\\
 F_3 \= & d\left[ H_1 \,d\phi \wedge dy_2 \+ T_1 \,dt \wedge dy_1 \right]\, \label{eq:WeylSol2circleApp}
\end{align}
The equations of motion obtained from the Maxwell equations and Einstein equations by assuming an electromagnetic duality $d(T_1 \,dt \wedge dy_1 ) \= q \,\star_6 d(H_1 \,d\phi \wedge dy_2)$ can be decomposed into three almost-linear layers:
\begin{align}
&\text{\underline{Vacuum layer:}} \quad \cL \log W_0  \= 0\,,\nonumber\\
&\text{\underline{Maxwell layer:}} \quad  \cL \log Z_I \= \- \frac{\gamma_I^{-2}}{\rho\, {Z_I}^2}\,\left[ (\partial_\rho H_I)^2 + (\partial_z H_I)^2 \right] \,,\quad \left(\gamma_0,\gamma_1 \right)\=\left( 1, \sqrt{\frac{2}{1+q^2}}\right),\nonumber\\
&\hspace{2.90cm} \partial_\rho \left( \frac{1}{\rho\,{Z_I}^2}\,\partial_\rho H_I \right)\+\partial_z \left( \frac{1}{\rho\,{Z_I}^2}\,\partial_z H_I\right)  \=0\,,\nonumber\\
&\text{\underline{Base layer:}} \nonumber \\
& \partial_z \nu \= \frac{\rho}{2}\, \partial_\rho \log W_0\,\partial_z \log W_0 \+ \rho \,\partial_\rho \log Z_1\,\partial_z \log Z_1 \+ \frac{\rho}{2} \,\partial_\rho \log Z_0\,\partial_z \log Z_0 \nonumber\\
& \hspace{1.3cm} \+ \frac{1}{2 \rho \,Z_0^2} \, \partial_\rho H_0 \partial_z H_0\+ \frac{1+q^2}{2\rho \,Z_1^2} \,\partial_\rho H_1 \partial_z H_1 \,, \label{eq:EOMWeylApp}\\
& \partial_\rho \nu \= \frac{\rho}{4}\, \left( (\partial_\rho \log W_0)^2- (\partial_z \log W_0 )^2\right)\+ \frac{\rho}{2} \left( ( \partial_\rho \log Z_1)^2- (\partial_z \log Z_1 )^2\right)\nonumber\\
& \hspace{1.3cm} \frac{\rho}{4} \left( (\partial_\rho \log Z_0)^2-(\partial_z \log Z_0)^2\right) + \frac{1}{4 \rho Z_0^2} \left( ( \partial_\rho H_0 )^2-(\partial_z H_0)^2\right)\nonumber \\
& \hspace{1.3cm}+ \frac{1+q^2}{4 \rho Z_1^2} \left( (\partial_\rho H_1)^2-( \partial_z H_1 )^2\right)\,, \nonumber
\end{align}
We consider $n$ distinct rod sources of length $M_i>0$ along the $z$-axis centered on $z=z_i$ such as $z_i < z_j$ for $i<j$. The warp factors and gauge field potentials that solve the equations of motion are given in \eqref{eq:WarpfactorsRods6d} where we have defined a set of weights, $(G_i,P_i^{(0)},P_i^{(1)})$ and 5 gauge field parameters $(a_I,b_I,q)$.

To study the metric far away from the rods, it is appropriate to use spherical coordinates
$
\rho = r\,\sin \theta \,,\,\, z= r \, \cos \theta \,,
$
and we have, at large $r$, 
\begin{equation}
W_0 \,\sim\, 1 \,,\qquad e^{2\nu} \,\sim \, 1\,,\qquad Z_I \sim\frac{\sinh b_I}{a_I}\,.
\end{equation}
Therefore, the solutions are asymptotic to T$^2\times \IR^{1,3}$ if
\begin{equation}
a_I \= \sinh b_I\,, \quad b_I\geq 0.
\end{equation}

It is rather straightforward from the expressions of $(R_\pm^{(i)},\nu_{ij})$, \eqref{eq:Rpmdef}, that $W_0$ and $e^{2\nu}$ are finite and strictly positive out of the $z$-axis. However, one must impose that $Z_I\geq 0$ \eqref{eq:WarpfactorsRods6d} everywhere which restricts to
\begin{equation}
P_i^{(I)} \geq 0 \,,\qquad \forall i\,.
\end{equation}

\subsection{Regularity on the $z$-axis}
\label{App:Regzaxischarged}

As for vacuum solutions, potential singularities arise on the $z$-axis. We first discuss the regularity on the rods before the regularity elsewhere on the axis.

\subsubsection{At the $i^\text{th}$ rod}
\label{App:AttheRodcharged}

Using the local spherical coordinates around the i$^{\text{th}}$ rod, $(r_i , \theta_i )$ \eqref{eq:localcoorApp}, and the expansion of the generating functions \eqref{eq:RpRmexp} and \eqref{eq:EEEEatrod}, we found that\footnote{We have denoted $\widetilde{g}_{\phi \phi}$ the metric component of $\phi$ without the connection in the $y_2$ fiber, that is $\widetilde{g}_{\phi \phi}= \rho^2 Z_0 Z_1$.}
\begin{equation}
\begin{split}
H_I  &\sim \gamma_I \left(-M_i P_i^{(I)}\,\cos \theta_i \+ \sum_{j\neq i} \text{sign}(j-i) M_j P_j^{(I)}\right)\,,\quad T_1 \sim - q\gamma_1 \sinh b_1 \left[ 1 + \cO \left( r_i^{2 P_i^{(1)} \sinh b_I} \right)\right] , \\
g_{tt} &\sim-\frac{2\sinh b_1}{e^{b_1}} \left( \frac{M_i}{r_i} \right)^{ G_i-P_i^{(1)}\sinh b_1}\,\prod_{j\neq i} \left(  \frac{R_+^{(j)}}{R_-^{(j)}}\right)^{G_j-P_j^{(1)}\sinh b_1}\,, \\
 g_{y_1y_1} &\sim \frac{2\sinh b_1}{e^{b_1}} \left( \frac{M_i}{r_i} \right)^{ -G_i-P_i^{(1)}\sinh b_1}\,\prod_{j\neq i} \left(  \frac{R_+^{(j)}}{R_-^{(j)}}\right)^{-G_j-P_j^{(1)}\sinh b_1}\,,\\
  g_{y_2y_2} &\sim\frac{e^{b_0}\sinh b_1}{e^{b_1}\sinh b_0} \left( \frac{M_i}{r_i} \right)^{ P_i^{(1)}\sinh b_1-P_i^{(0)}\sinh b_0}\,\prod_{j\neq i} \left(  \frac{R_+^{(j)}}{R_-^{(j)}}\right)^{P_j^{(1)}\sinh b_1-P_j^{(0)}\sinh b_0}\,,\\
\widetilde{g}_{\phi\phi} &\sim  \frac{M_i^2\,e^{b_0+b_1}}{4\sinh b_0 \,\sinh b_1} \sin^2\theta_i\,\left( \frac{r_i}{M_i} \right)^{ 1-P_i^{(1)}\sinh b_1-P_i^{(0)}\sinh b_0}\,\prod_{j\neq i} \left(  \frac{R_+^{(j)}}{R_-^{(j)}}\right)^{P_j^{(1)}\sinh b_1+P_j^{(0)}\sinh b_0} \,.\\
g_{r_i r_i}& \sim   \frac{d_i^2\,e^{b_0+b_1}}{4\sinh b_0 \,\sinh b_1} \left(\frac{M_i}{r_i} \right)^{P_i^{(1)}\sinh b_1+P_i^{(0)}\sinh b_0} \,\left(\frac{4 r_i}{M_i \sin^2 \theta_i} \right)^{\alpha_{ii}-1} \\
& \hspace{1cm} \times \prod_{j\neq i} \left( \frac{R_+^{(j)}}{R_-^{(j)}}\right)^{-2\alpha_{ij}+P_j^{(1)}\sinh b_1+P_j^{(0)}\sinh b_0}   \prod_{j\neq i} \left(\frac{z_j^+ - z_i^-}{z_j^- -z_i^-} \right)^{2\,\text{sign}(j-i) \,\alpha_{ij}}\,, \\
g_{\theta_i \theta_i}& \sim  \frac{M_i^2 d_i^2\,e^{b_0+b_1}}{4\sinh b_0 \,\sinh b_1} \left( \frac{r_i}{M_i} \right)^{ 1-P_i^{(1)}\sinh b_1-P_i^{(0)}\sinh b_0} \,\left(\frac{4 r_i}{M_i \sin^2 \theta_i} \right)^{\alpha_{ii}-1} \\
& \hspace{1cm} \times \prod_{j\neq i} \left( \frac{R_+^{(j)}}{R_-^{(j)}}\right)^{-2\alpha_{ij}+P_j^{(1)}\sinh b_1+P_j^{(0)}\sinh b_0}   \prod_{j\neq i} \left(\frac{z_j^+ - z_i^-}{z_j^- -z_i^-} \right)^{2\,\text{sign}(j-i) \,\alpha_{ij}}\,.
\end{split}
\label{eq:ytphibehaviorBubblerodcharged}
\end{equation}
where $\frac{R_+^{(j)}}{R_-^{(j)}}$, $j\neq i$, are finite functions of $\theta_i$ \eqref{eq:RpRmexp}, the exponents $\alpha_{jk}$ are defined according to the weights in \eqref{eq:alphaijDef} and $d_i$ is the function of aspect ratio given in \eqref{eq:diDef}. We have also considered that the product ``$\prod_{j=1}^{i-1}$'' is equal to $1$ for the first rod, $i=1$. By investigating each power in $r_i$, we have three possible choice of physical weights. As for the vacuum solutions, we first discuss the rod's topology when the rod is disconnected from the others.
\begin{itemize}
\item[•] \underline{$P_i^{(1)} \sinh b_1 \=P_i^{(0)} \sinh b_0\=-G_i \=\frac{1}{2}$: a three-charge black brane.}

The $i^\text{th}$ rod corresponds to a horizon of a two-dimensional black brane with a S$^2\times$T$^2$ topology. The local six-dimensional metric has the same form as in \eqref{eq:metricatrodBS}, and the surface gravity is now given by
\begin{equation}
\kappa_i \equi \frac{\sinh b_1}{e^{b_1}}\,\sqrt{\frac{2\sinh b_0}{e^{b_0}}}\frac{1}{d_i\,M_i}\, \prod_{j\neq i} \left(\frac{z_j^+ - z_i^-}{z_j^- -z_i^-} \right)^{\text{sign}(i-j)\,\alpha_{ij}}\,.
\label{eq:surfaceGrav}
\end{equation}
The surface gravity can be associate to a temperature of the black brane by requiring smoothness of the Euclideanized solution $\cT_i \= \frac{\kappa_i}{2\pi}$. The gauge fields \eqref{eq:ytphibehaviorBubblerodcharged} are regular at the horizon and carries two magnetic charges (one is coming from the Kaluza-Klein monopole along $y_2$) and an electric charge, that can be derived by integrating the fluxes at the rod. We find
\begin{equation}
\begin{split}
 \cQ_{e\,i}^{(1)}&\=q\,{\cQ_{m\,i}^{(1)}} \=  \frac{q}{4\sqrt{2\pi(1+q^2) G_4}}\,\frac{M_i}{\sinh b_1} \,, \qquad \cQ_{m\,i}^{(0)}  \= \frac{M_i }{8\sqrt{\pi G_4 }\,\sinh b_0}\,.
 \label{eq:ChargesBBcharged}
\end{split}
\end{equation}
Moreover, the area of the horizon is simple to derive and we find
\begin{equation}
A_i \= \int_{S^2\times T^2} \sqrt{g_{y_1y_1}(\theta_i)g_{y_2y_2}(\theta_i)g_{\theta_i \theta_i}(\theta_i)g_{\phi\phi}(\theta_i)} \= \frac{(2\pi)^3}{\kappa_i}\,M_i\,R_{y_1}R_{y_2}\,.
\end{equation}
Finally, one can compute the contribution of the $i^\text{th}$ rod to the four-dimensional ADM mass following the procedure \eqref{eq:AsymptoticExpGen} and \eqref{eq:conservedchargesGen} and isolating the relevant term. We find 
\begin{equation}
\cM_i \= \frac{M_i}{8 G_4}\,\left(1+ \coth b_0 + 2\coth b_1 \right)\,,
\label{eq:MassBBcharged}
\end{equation}

\item[•] \underline{$P_i^{(1)} \sinh b_1 \=P_i^{(0)} \sinh b_0\=G_i \=\frac{1}{2}$: a three-charge species-1 bubble.}

The $i^\text{th}$ rod corresponds to a S$^{y_2}\times$S$^2$ bubble where the $y_1$-circle degenerates defining an origin of an $\IR^2$ space. The local metric has the same form as in \eqref{eq:metricatrodBu} with a constant $C_i^{(1)}$ given by
\begin{equation}
C_i^{(1)} \equi \frac{e^{b_1}}{\sinh b_1}\sqrt{\frac{e^{b_0}}{2\sinh b_0}}d_i\,M_i\, \prod_{j\neq i} \left(\frac{z_j^+ - z_i^-}{z_j^- -z_i^-} \right)^{\text{sign}(j-i)\,\alpha_{ij}}\,.
\end{equation}
As for vacuum solutions, the regularity (with potential conical defect at the bubble locus) requires
\begin{equation}
\begin{split}
R_{y_1} = \frac{C^{(1)}_i}{k_i}\,, \qquad k_i \in \mathbb{N}.
\end{split}
\label{eq:condRymultibubblecharged}
\end{equation}

Moreover, the component of the gauge field that has a leg along $dy_1$ needs to vanish. One can check from \eqref{eq:ytphibehaviorBubblerodcharged} that we have
\begin{equation}
Z_1 \= -q\gamma_1\sinh b_1 \left[1 \+  \cO\left(\rho_i^2 \right)  \right] \quad \Rightarrow \quad F_3 |_{dy_1} \= \cO\left(\rho_i \right)\,d\rho_i\wedge dt\,.
\label{eq:F4regApp}
\end{equation}
Therefore, the gauge field is regular and carries an electric and two magnetic charges at the bubble given by the same expression as in \eqref{eq:ChargesBBcharged}.

As for neutral solutions, one can easily derive the area of the S$^2\times$S$^{y_2}$ bubble if there is no black rods in the configuration. We find
\begin{equation}
A_{\text{B}i} \= \int_{y_2,\theta_i,\phi} \sqrt{g_{y_2y_2}(\theta_i)g_{\theta_i \theta_i}(\theta_i)g_{\phi\phi}(\theta_i)} \= (2\pi)^2 \, \sqrt{\frac{e^{b_1}}{2 \sinh b_1}}\,k_i\,M_i\,R_{y_1}R_{y_2}\,.
\label{eq:AreaBubbleAppcharged}
\end{equation}
The contribution of the rod to the four-dimensional ADM mass is given by
\begin{equation}
\cM_i \= \frac{M_i}{8 G_4}\,\left(-1+ \coth b_0 + 2\coth b_1 \right)\,,
\end{equation}

\item[•] \underline{$G_i = P_i^{(1)} =0$ and $P^{(0)}_i \sinh b_0 = 1$: a one-charge species-2 bubble.}

This situation is similar to the previous one by inverting the role of the $y_1$ and $y_2$ fibers. On such a rod, the spacelike Killing vector $\partial_{y_2}$ shrinks corresponding to a  S$^{y_1}\times$S$^2$ at an origin of a $\IR^2/\mathbb{Z}_{k_i}$. The orbifold parameter $k_i\in \mathbb{N}$ relates the radius of the $y_2$-circle to the parameter of the solution such as
\begin{equation}
R_{y_2} = \frac{C^{(2)}_i}{k_i}\,,\qquad C_i^{(2)} \equi \frac{e^{b_0}}{\sinh b_0}\,d_i\,M_i\, \prod_{j\neq i} \left(\frac{z_j^+ - z_i^-}{z_j^- -z_i^-} \right)^{\text{sign}(j-i)\,\alpha_{ij}}\,.
\end{equation}
Because $P_i^{(1)}=0$, $H_1$ and $T_1$ are trivial on the rod and the rod is not charged under the gauge field. However, the rod carries a magnetic charge for the KK gauge field $H_0 d\phi$:
\begin{equation}
 \cQ_{e\,i}^{(1)}\={\cQ_{m\,i}^{(1)}} \= 0 \,, \qquad \cQ_{m\,i}^{(0)}  \= \frac{M_i }{4\sqrt{\pi G_4 }\,\sinh b_0}\,.
\end{equation}
The area of the bubble is also computable when there is no black rods and give the same result as for a species-1 bubble \eqref{eq:AreaBubbleAppcharged}.
\end{itemize}

When the rod is connected to other rods, the regularity conditions, the expressions of the conserved charges are identical but the local topology can change. We refer the reader to the end of the section \ref{App:AttheRod} since the discussion is identical to the vacuum case.

\subsubsection{On the $z$-axis and out of the rods}
\label{App:OutoftheRodcharged}

The regularity of the solutions on the $z$-axis out of the rods is identical to the vacuum Weyl solutions, and we refer the reader to section \ref{App:OutoftheRod} for all the details.

On the semi-infinite segments of the $z$-axis above and below the rod configuration, the $\phi$-circle shrinks smoothly as the usual coordinate singularity of the cylindrical coordinate system.

The solutions have a strut at each section in between two disconnected rods. The strut manifests itself as a conical excess on the three-dimensional base where the $\phi$-circle shrinks:
 \begin{equation}
ds_3^2 \sim d_i^2 \left(d\rho^2 +dz^2 +\frac{\rho^2}{d_i^2} d\phi^2 \right) \,.
\end{equation}
where $d_i$ is given in \eqref{eq:diDef} and is smaller than one.

If two rods of different nature are connected, the $\phi$-circle does not shrink at the intersection point and the local geometry is regular. For instance, the local geometry at the intersection of two connected species-1 and species-2 bubble corresponds to the origin of a $\IR^4$ given by \eqref{eq:intersectionBu&Bu}. The intersection between a black brane and a bubble gives the usual metric at a pole of the horizon given by \eqref{eq:intersectionBB&Bu}.

Charged Weyl solutions that are built from connected species-1 and species-2 smooth bubbles are therefore smooth everywhere on the $z$-axis and elsewhere.



\bibliographystyle{utphys}      

\bibliography{microstates}       

\providecommand{\href}[2]{#2}\begingroup\raggedright\begin{thebibliography}{10}

\bibitem{Tod:1983pm}
K.~p. Tod, ``{All Metrics Admitting Supercovariantly Constant Spinors},''
  \href{http://dx.doi.org/10.1016/0370-2693(83)90797-9}{{\em Phys. Lett. B}
  {\bfseries 121} (1983) 241--244}.

\bibitem{Sabra:1997yd}
W.~A. Sabra, ``{General BPS black holes in five-dimensions},''
  \href{http://dx.doi.org/10.1142/S0217732398000309}{{\em Mod. Phys. Lett. A}
  {\bfseries 13} (1998) 239--251},
  \href{http://arxiv.org/abs/hep-th/9708103}{{\ttfamily arXiv:hep-th/9708103}}.

\bibitem{Bena:2004de}
I.~Bena and N.~P. Warner, ``{One ring to rule them all ... and in the darkness
  bind them?},'' {\em Adv. Theor. Math. Phys.} {\bfseries 9} (2005) 667--701,
\href{http://arxiv.org/abs/hep-th/0408106}{{\ttfamily arXiv:hep-th/0408106}}.

\bibitem{Bena:2006kb}
I.~Bena, C.-W. Wang, and N.~P. Warner, ``{Mergers and Typical Black Hole
  Microstates},'' \href{http://dx.doi.org/10.1088/1126-6708/2006/11/042}{{\em
  JHEP} {\bfseries 11} (2006) 042},
\href{http://arxiv.org/abs/hep-th/0608217}{{\ttfamily arXiv:hep-th/0608217}}.

\bibitem{Bena:2007qc}
I.~Bena, C.-W. Wang, and N.~P. Warner, ``{Plumbing the Abyss: Black Ring
  Microstates},'' \href{http://dx.doi.org/10.1088/1126-6708/2008/07/019}{{\em
  JHEP} {\bfseries 07} (2008) 019},
\href{http://arxiv.org/abs/0706.3786}{{\ttfamily arXiv:0706.3786 [hep-th]}}.

\bibitem{Bena:2007kg}
I.~Bena and N.~P. Warner, ``{Black holes, black rings and their microstates},''
  \href{http://dx.doi.org/10.1007/978-3-540-79523-0}{{\em Lect. Notes Phys.}
  {\bfseries 755} (2008) 1--92},
\href{http://arxiv.org/abs/hep-th/0701216}{{\ttfamily arXiv:hep-th/0701216}}.

\bibitem{Bena:2015bea}
I.~Bena, S.~Giusto, R.~Russo, M.~Shigemori, and N.~P. Warner, ``{Habemus
  Superstratum! A constructive proof of the existence of superstrata},''
  \href{http://dx.doi.org/10.1007/JHEP05(2015)110}{{\em JHEP} {\bfseries 05}
  (2015) 110},
\href{http://arxiv.org/abs/1503.01463}{{\ttfamily arXiv:1503.01463 [hep-th]}}.

\bibitem{Heidmann:2017cxt}
P.~Heidmann, ``{Four-center bubbled BPS solutions with a Gibbons-Hawking
  base},'' \href{http://dx.doi.org/10.1007/JHEP10(2017)009}{{\em JHEP}
  {\bfseries 10} (2017) 009},
\href{http://arxiv.org/abs/1703.10095}{{\ttfamily arXiv:1703.10095 [hep-th]}}.

\bibitem{Bena:2017fvm}
I.~Bena, P.~Heidmann, and P.~F. Ramirez, ``{A systematic construction of
  microstate geometries with low angular momentum},''
  \href{http://dx.doi.org/10.1007/JHEP10(2017)217}{{\em JHEP} {\bfseries 10}
  (2017) 217},
\href{http://arxiv.org/abs/1709.02812}{{\ttfamily arXiv:1709.02812 [hep-th]}}.

\bibitem{Stephani:2003tm}
H.~Stephani, D.~Kramer, M.~A.~H. MacCallum, C.~Hoenselaers, and E.~Herlt,
  \href{http://dx.doi.org/10.1017/CBO9780511535185}{{\em {Exact solutions of
  Einstein's field equations}}}.
\newblock Cambridge Monographs on Mathematical Physics. Cambridge Univ. Press,
  Cambridge, 2003.

\bibitem{Bonnor}
W.~B. Bonnor, ``{Physical Interpretation of Vacuum Solutions of Einstein's
  Equations. Part I. Time-independent Solutions},''
  \href{http://dx.doi.org/10.1007/BF00760137}{{\em Gen. Rel. Grav.} {\bfseries
  24} (1992) 551}.

\bibitem{Weyl:book}
H.~Weyl {\em Ann. Phys. (Leipzig)} {\bfseries 54} (1917) 117.

\bibitem{Emparan:2001wk}
R.~Emparan and H.~S. Reall, ``{Generalized Weyl solutions},''
  \href{http://dx.doi.org/10.1103/PhysRevD.65.084025}{{\em Phys. Rev. D}
  {\bfseries 65} (2002) 084025},
  \href{http://arxiv.org/abs/hep-th/0110258}{{\ttfamily arXiv:hep-th/0110258}}.

\bibitem{Elvang:2002br}
H.~Elvang and G.~T. Horowitz, ``{When black holes meet Kaluza-Klein bubbles},''
  \href{http://dx.doi.org/10.1103/PhysRevD.67.044015}{{\em Phys. Rev. D}
  {\bfseries 67} (2003) 044015},
  \href{http://arxiv.org/abs/hep-th/0210303}{{\ttfamily arXiv:hep-th/0210303}}.

\bibitem{Witten:1981gj}
E.~Witten, ``{Instability of the Kaluza-Klein Vacuum},''
  \href{http://dx.doi.org/10.1016/0550-3213(82)90007-4}{{\em Nucl. Phys. B}
  {\bfseries 195} (1982) 481--492}.

\bibitem{Stotyn:2011tv}
S.~Stotyn and R.~B. Mann, ``{Magnetic charge can locally stabilize
  Kaluza\textendash{}Klein bubbles},''
  \href{http://dx.doi.org/10.1016/j.physletb.2011.10.015}{{\em Phys. Lett. B}
  {\bfseries 705} (2011) 269--272},
  \href{http://arxiv.org/abs/1105.1854}{{\ttfamily arXiv:1105.1854 [hep-th]}}.

\bibitem{Bah:2020pdz}
I.~Bah and P.~Heidmann, ``{Topological Stars, Black holes and Generalized
  Charged Weyl Solutions},'' \href{http://arxiv.org/abs/2012.13407}{{\ttfamily
  arXiv:2012.13407 [hep-th]}}.

\bibitem{Costa:2000kf}
M.~S. Costa and M.~J. Perry, ``{Interacting black holes},''
  \href{http://dx.doi.org/10.1016/S0550-3213(00)00577-0}{{\em Nucl. Phys. B}
  {\bfseries 591} (2000) 469--487},
  \href{http://arxiv.org/abs/hep-th/0008106}{{\ttfamily arXiv:hep-th/0008106}}.

\bibitem{bubblebagend}
I.~Bah and P.~Heidmann, ``{Bubble Bag End: A Bubbly Resolution of Curvature
  Singularity},'' \href{http://arxiv.org/abs/2107.13551}{{\ttfamily
  arXiv:2107.13551 [hep-th]}}.

\bibitem{Israel1964}
W.~Israel and K.~A. Khan, ``{Collinear Particles and Bondi Dipoles in General
  Relativity},'' \href{http://dx.doi.org/10.1007/BF02750196}{{\em Nuovo Cim.}
  {\bfseries 33} (1964) 331}.

\bibitem{Gibbons:1979nf}
G.~Gibbons and M.~J. Perry, ``{New Gravitational Instantons and Their
  Interactions},'' \href{http://dx.doi.org/10.1103/PhysRevD.22.313}{{\em Phys.
  Rev. D} {\bfseries 22} (1980) 313}.

\bibitem{Emparan:2008eg}
R.~Emparan and H.~S. Reall, ``{Black Holes in Higher Dimensions},''
  \href{http://dx.doi.org/10.12942/lrr-2008-6}{{\em Living Rev. Rel.}
  {\bfseries 11} (2008) 6}, \href{http://arxiv.org/abs/0801.3471}{{\ttfamily
  arXiv:0801.3471 [hep-th]}}.

\bibitem{Charmousis:2003wm}
C.~Charmousis and R.~Gregory, ``{Axisymmetric metrics in arbitrary
  dimensions},'' \href{http://dx.doi.org/10.1088/0264-9381/21/2/016}{{\em
  Class. Quant. Grav.} {\bfseries 21} (2004) 527--554},
  \href{http://arxiv.org/abs/gr-qc/0306069}{{\ttfamily arXiv:gr-qc/0306069}}.

\bibitem{Papapetrou:1953zz}
A.~Papapetrou, ``{Eine rotationssymmetrische losung in der allgemeinen
  relativitatstheorie},'' {\em Annals Phys.} {\bfseries 12} (1953) 309--315.

\bibitem{Regge:1961px}
T.~Regge, ``{General Relativity without coordinates},''
  \href{http://dx.doi.org/10.1007/BF02733251}{{\em Nuovo Cim.} {\bfseries 19}
  (1961) 558--571}.

\bibitem{Bah:2020ogh}
I.~Bah and P.~Heidmann, ``{Topological Stars and Black Holes},''
  \href{http://arxiv.org/abs/2011.08851}{{\ttfamily arXiv:2011.08851
  [hep-th]}}.

\bibitem{Myers:1986un}
R.~C. Myers and M.~Perry, ``{Black Holes in Higher Dimensional Space-Times},''
\href{http://dx.doi.org/10.1016/0003-4916(86)90186-7}{{\em Annals Phys.}
  {\bfseries 172} (1986) 304}.

\bibitem{1975ApJ}
D.~M. {Eardley}, ``{Observable effects of a scalar gravitational field in a
  binary pulsar.},'' \href{http://dx.doi.org/10.1086/181744}{{\em Astrophys. J.
  Lett.} {\bfseries 196} (Mar., 1975) 59--62}.

\bibitem{Damour:1992we}
T.~Damour and G.~Esposito-Farese, ``{Tensor multiscalar theories of
  gravitation},'' \href{http://dx.doi.org/10.1088/0264-9381/9/9/015}{{\em
  Class. Quant. Grav.} {\bfseries 9} (1992) 2093--2176}.

\bibitem{Mirshekari:2013vb}
S.~Mirshekari and C.~M. Will, ``{Compact binary systems in scalar-tensor
  gravity: Equations of motion to 2.5 post-Newtonian order},''
  \href{http://dx.doi.org/10.1103/PhysRevD.87.084070}{{\em Phys. Rev. D}
  {\bfseries 87} no.~8, (2013) 084070},
  \href{http://arxiv.org/abs/1301.4680}{{\ttfamily arXiv:1301.4680 [gr-qc]}}.

\bibitem{Julie:2017rpw}
F.-L. Juli\'e, ``{On the motion of hairy black holes in
  Einstein-Maxwell-dilaton theories},''
  \href{http://dx.doi.org/10.1088/1475-7516/2018/01/026}{{\em JCAP} {\bfseries
  01} (2018) 026}, \href{http://arxiv.org/abs/1711.10769}{{\ttfamily
  arXiv:1711.10769 [gr-qc]}}.

\bibitem{Denef:2000nb}
F.~Denef, ``{Supergravity flows and D-brane stability},''
  \href{http://dx.doi.org/10.1088/1126-6708/2000/08/050}{{\em JHEP} {\bfseries
  0008} (2000) 050},
\href{http://arxiv.org/abs/hep-th/0005049}{{\ttfamily arXiv:hep-th/0005049
  [hep-th]}}.

\bibitem{Denef:2002ru}
F.~Denef, ``{Quantum quivers and Hall / hole halos},''
  \href{http://dx.doi.org/10.1088/1126-6708/2002/10/023}{{\em JHEP} {\bfseries
  0210} (2002) 023},
\href{http://arxiv.org/abs/hep-th/0206072}{{\ttfamily arXiv:hep-th/0206072
  [hep-th]}}.

\bibitem{Bates:2003vx}
B.~Bates and F.~Denef, ``{Exact solutions for supersymmetric stationary black
  hole composites},'' \href{http://dx.doi.org/10.1007/JHEP11(2011)127}{{\em
  JHEP} {\bfseries 1111} (2011) 127},
\href{http://arxiv.org/abs/hep-th/0304094}{{\ttfamily arXiv:hep-th/0304094
  [hep-th]}}.

\bibitem{GIBBONS1978430}
G.~Gibbons and S.~Hawking, ``Gravitational multi-instantons,''
  \href{http://dx.doi.org/https://doi.org/10.1016/0370-2693(78)90478-1}{{\em
  Physics Letters B} {\bfseries 78} no.~4, (1978) 430 -- 432}.

\bibitem{Jejjala:2005yu}
V.~Jejjala, O.~Madden, S.~F. Ross, and G.~Titchener, ``{Non-supersymmetric
  smooth geometries and D1-D5-P bound states},''
  \href{http://dx.doi.org/10.1103/PhysRevD.71.124030}{{\em Phys. Rev.}
  {\bfseries D71} (2005) 124030},
\href{http://arxiv.org/abs/hep-th/0504181}{{\ttfamily arXiv:hep-th/0504181}}.

\bibitem{Giusto:2004id}
S.~Giusto, S.~D. Mathur, and A.~Saxena, ``{Dual geometries for a set of
  3-charge microstates},''
  \href{http://dx.doi.org/10.1016/j.nuclphysb.2004.09.001}{{\em Nucl. Phys.}
  {\bfseries B701} (2004) 357--379},
\href{http://arxiv.org/abs/hep-th/0405017}{{\ttfamily arXiv:hep-th/0405017}}.

\bibitem{Giusto:2012yz}
S.~Giusto, O.~Lunin, S.~D. Mathur, and D.~Turton, ``{D1-D5-P microstates at the
  cap},'' \href{http://dx.doi.org/10.1007/JHEP02(2013)050}{{\em JHEP}
  {\bfseries 1302} (2013) 050},
\href{http://arxiv.org/abs/1211.0306}{{\ttfamily arXiv:1211.0306 [hep-th]}}.

\bibitem{Gibbons:1978ji}
G.~W. Gibbons and M.~J. Perry, ``{Quantizing Gravitational Instantons},''
  \href{http://dx.doi.org/10.1016/0550-3213(78)90434-0}{{\em Nucl. Phys. B}
  {\bfseries 146} (1978) 90--108}.

\bibitem{Gross:1982cv}
D.~J. Gross, M.~J. Perry, and L.~G. Yaffe, ``{Instability of Flat Space at
  Finite Temperature},'' \href{http://dx.doi.org/10.1103/PhysRevD.25.330}{{\em
  Phys. Rev. D} {\bfseries 25} (1982) 330--355}.

\bibitem{York:1986it}
J.~W. York, Jr., ``{Black hole thermodynamics and the Euclidean Einstein
  action},'' \href{http://dx.doi.org/10.1103/PhysRevD.33.2092}{{\em Phys. Rev.
  D} {\bfseries 33} (1986) 2092--2099}.

\bibitem{Brown:2014rka}
A.~R. Brown, ``{Decay of hot Kaluza-Klein space},''
  \href{http://dx.doi.org/10.1103/PhysRevD.90.104017}{{\em Phys. Rev. D}
  {\bfseries 90} no.~10, (2014) 104017},
  \href{http://arxiv.org/abs/1408.5903}{{\ttfamily arXiv:1408.5903 [hep-th]}}.

\bibitem{Miyamoto:2006nd}
U.~Miyamoto and H.~Kudoh, ``{New stable phase of non-uniform charged black
  strings},'' \href{http://dx.doi.org/10.1088/1126-6708/2006/12/048}{{\em JHEP}
  {\bfseries 12} (2006) 048},
  \href{http://arxiv.org/abs/gr-qc/0609046}{{\ttfamily arXiv:gr-qc/0609046}}.

\bibitem{Chow:2014cca}
D.~D.~K. Chow and G.~Comp\`ere, ``{Black holes in N=8 supergravity from SO(4,4)
  hidden symmetries},''
  \href{http://dx.doi.org/10.1103/PhysRevD.90.025029}{{\em Phys. Rev. D}
  {\bfseries 90} no.~2, (2014) 025029},
  \href{http://arxiv.org/abs/1404.2602}{{\ttfamily arXiv:1404.2602 [hep-th]}}.

\bibitem{Bena:2009fi}
I.~Bena, S.~Giusto, C.~Ruef, and N.~P. Warner, ``{Supergravity Solutions from
  Floating Branes},'' \href{http://dx.doi.org/10.1007/JHEP03(2010)047}{{\em
  JHEP} {\bfseries 03} (2010) 047},
\href{http://arxiv.org/abs/0910.1860}{{\ttfamily arXiv:0910.1860 [hep-th]}}.

\end{thebibliography}\endgroup


\end{document}